\documentclass{article}

\usepackage{arxiv}
\usepackage{amsfonts}
\usepackage{graphicx, color}
\usepackage{amsmath}
\usepackage{url}
\usepackage[algoruled,boxed,lined]{algorithm2e}
\usepackage{float}

\usepackage{natbib}
\newtheorem{theorem}{Theorem}[section]
\newtheorem{remark}[theorem]{Remark}
\newtheorem{conjecture}{Conjecture}[section]

\newcommand{\ds}{\displaystyle} 
\newcommand{\eps}{\varepsilon}

\newcommand{\bE}{\mathbb{E}}

\newcommand{\cF}{\mathcal{F}}

\newcommand{\bM}{\mathbb{M}}

\newcommand{\LSTM}{\mathbb{L}\mathbb{S}\mathbb{T}\mathbb{M}}
\newcommand{\N}{\mathbb{N}}

\newcommand{\cL}{\mathcal{L}}

\newcommand{\bP}{\mathbb{P}}

\newcommand{\bC}{\mathbb{C}}
\newcommand{\bN}{\mathbb{N}}
\newcommand{\bR}{\mathbb{R}}

\newcommand{\bZ}{\mathbb{Z}}
\def\hyph{-\penalty0\hskip0pt\relax}

\allowdisplaybreaks

\title{PDGM: a Neural Network Approach to Solve Path-Dependent Partial Differential Equations}

\author{
  Yuri F.~Saporito\\
  School of Applied Mathematics\\ \\
  Getulio Vargas Foundation\\
  Rio de Janeiro, Brazil \\
  \texttt{yuri.saporito@fgv.br} \\
   \And
  Zhaoyu Zhang \\
  Department of Industrial Engineering \\
  and Operations Research\\
    Columbia University\\
    New York, USA \\
  \texttt{zz2734@columbia.edu} \\
}

\begin{document}

\maketitle

\renewcommand{\arraystretch}{1.5}

\begin{abstract}

In this paper, we propose a novel numerical method for Path-Dependent Partial Differential Equations (PPDEs). These equations firstly appeared in the seminal work of \cite{fito_dupire}, where the functional It\^o calculus was developed to deal with path-dependent financial derivatives contracts. More specificaly, we generalize the Deep Galerking Method (DGM) of \cite{dgm} to deal with these equations. The method, which we call Path-Dependent DGM (PDGM), consists of using a combination of feed-forward and Long Short-Term Memory architectures to model the solution of the PPDE. We then analyze several numerical examples, many from the Financial Mathematics literature, that show the capabilities of the method under very different situations.

\end{abstract}

% keywords can be removed

\section{Introduction}

Neural networks and their modern computational implementations, generally called deep learning, have been successfully applied in several areas of mathematics and science in recent years. In this paper, we will generalize the methodology proposed in \cite{dgm} to numerically solve PDEs, known as Deep Galerkin Method (DGM), to an infinite dimensional setting. Applications of deep learning to solve PDEs date back to \cite{lee_neural_1990, lagaris_artificial_1998, parisi_solving_2003}. Lately, many articles have dealt with the finite-dimensional PDE problems, see, for instance, \cite{e_deep_2017, han_solving_2017, raissi_physics_2017, dgm_ali_et_all2019}. Many of them examine non-linear PDEs and then consider the Backward Stochastic Differential Equation (BSDE) technique. We will not pursue this approach here.

We propose a numerical method based on neural networks to solve path-dependent partial differential equations (PPDEs) that arises from  the functional calculus framework proposed in \cite{fito_dupire}. This theory was firstly proposed in the aforesaid reference with the goal to extend results available for vanilla derivatives contracts in Financial Mathematics to more general, path-dependent derivatives, as, for instance, Asian, barrier and lookback options. Additionally, non-linear PPDEs appear in the context of stochastic optimal control and differential games, see \cite{fito_saporito_control} and \cite{fito_zhang_zero_sum_game}.

One of the main features of this functional calculus is the fact that all the modelling is non-anticipative, meaning that it does not look into the future of the evolution of the state dynamics. This fact suggests the choice of Long-Short Term Memory (LSTM) networks to model these objects. In fact, we propose a novel architecture that combines LSTM and feed-forward, which we called Path-Dependent Deep Galerking Method (PDGM) architecture, that captures the non-anticipativeness of functionals and deals with the necessary path deformations from this functional calculus.

Recently, in \cite{fouque_zhang_2019}, it has been shown that the LSTM network can be used to numerically solve coupled forward anticipated BSDEs, and effectively approximate the conditional expectation for a non-Markovian process. 

% Moreover, as it was argued in \cite{dgm}, the neural network framework is very suitable to handle high-dimensional input variable, allowing us to deal with realistic models.

There are very few methods available to solve PPDEs; for a discussion about them, see \cite{monotone_ppde} and references therein. In this paper, the authors summarize some numerical methods to deal with PPDE, namely finite difference, trinomial tree, probabilistic schemes. These methods are either Monte Carlo or tree based. Our method differs from all of them by considering the recent neural network approach for differential equations.

The closest work to ours, but different nonetheless, is \cite{deep_ppde}. In this paper, the authors consider the functional framework proposed by \cite{fito_zhang_fractional} that generalizes the functional It\^o calculus to deal with the fractional Brownian motion in a very inventive way. The numerical procedure proposed in \cite{deep_ppde} uses the approach that combines BSDE and deep learning to numerically solve PPDE that arises from the rough Heston model. Our approach could be modified to handle those PPDEs. However, it is outside the scope of this paper.

The paper is organized as follows. In Section \ref{sec:ppde} we introduce the functional It\^o calculus and the main theoretical object of our study, the PPDEs. The algorithm is presented and studied in Section \ref{sec:method}. Finally, we show several numerical examples in Section \ref{sec:numerical_examples}. In order to show the capabilities of the method, we mostly consider cases where closed-form solutions are available.

\section{Path-Dependent Partial Differential Equation}
\label{sec:ppde}

In this section, we review the notion of Path-Dependent Partial Differential Equations (PPDEs) and the theory that created them, the functional It\^o calculus. Proposed in the seminal paper \cite{fito_dupire}, this framework allows us to apply the techniques of differential calculus to functions that depend on the history of the state variable being considered. It was firstly developed in the It\^o's stochastic calculus setting, but this generalization could be obviously applied in the usual, deterministic differential calculus. Below we present the necessary definitions and results to define precisely what is a PPDE.

\subsection{Functional It\^o Calculus}

We start by fixing a time horizon $T > 0$. Denote $\Lambda_t$ the space of c\`adl\`ag paths in $[0,t]$ taking values in $\bR^n$ and define $\Lambda = \bigcup_{t \in [0,T]} \Lambda_t$. Capital letters will denote elements of $\Lambda$ (i.e. paths) and lower-case letters will denote spot value of paths. In symbols, $Y_t \in \Lambda$ means $Y_t \in \Lambda_t$ and $y_s = Y_t(s)$, for $s \leq t$.

\begin{figure}[h!]
\centering
  \begin{minipage}[b]{0.45\linewidth}
    \centering
    \includegraphics[width=0.5\linewidth]{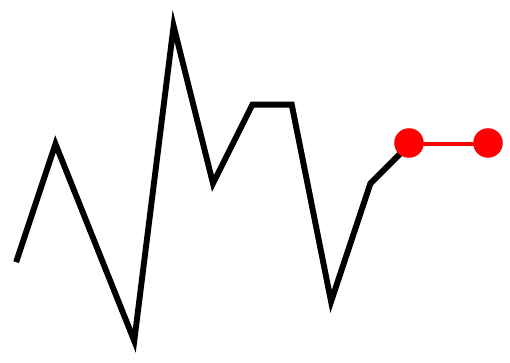}
    \caption{Flat extension of a path.}
    \label{fig:flat}
  \end{minipage}
  \begin{minipage}[b]{0.45\linewidth}
    \centering
    \includegraphics[width=0.5\linewidth]{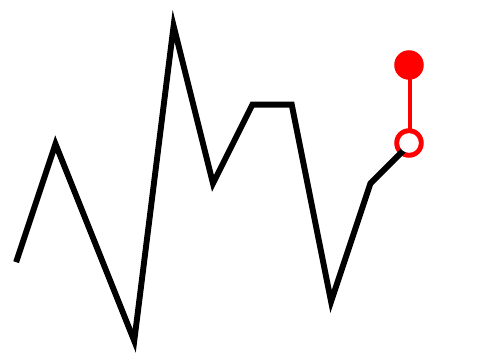}
    \caption{Bumped path.}
    \label{fig:bump}
  \end{minipage}
\end{figure}

A functional is any function $f: \Lambda \longrightarrow \bR$. For such objects, we define, when the limits exist, the \textit{time} and \textit{space functional derivatives}, respectively, as
\begin{align}
\Delta_t f(Y_t) &= \lim_{\delta t \to 0^+} \frac{f(Y_{t,\delta t}) - f(Y_t)}{\delta t}, \label{eq:time_deriv}\\
\Delta_x f(Y_t) &= \lim_{h \to 0} \frac{f(Y_t^h) - f(Y_t)}{h}, \label{eq:space_deriv}
\end{align}
where
\begin{align*}
Y_{t,\delta t}(u) &= \left\{
\begin{array}{ll}
  y_u,  &\mbox{ if } \quad 0 \leq u \leq t, \\
  y_t,  &\mbox{ if } \quad t \leq u \leq t + \delta t,
\end{array}
\right. \\
Y_t^h(u) &= \left\{
\begin{array}{ll}
  y_u,      &\mbox{ if } \quad 0 \leq u < t, \\
  y_t + h,  &\mbox{ if } \quad u = t,
\end{array}
\right.
\end{align*}
see Figures \ref{fig:flat} and \ref{fig:bump}. In the case when the path $Y_t$ lies in a multidimensional space, the path deformations above are understood as follows: the flat extension is applied to all dimension jointly and equally and the bump is applied to each dimension individually.

We consider here continuity of functionals as the usual continuity in metric spaces with respect to the metric:
\begin{align*}
d_{\Lambda}(Y_t,Z_s) = \| Y_{t,s-t} - Z_s\|_{\infty} + |s -t|,
\end{align*}
where, without loss of generality, we are assuming $s \geq t$, and
\begin{align*}
\|Y_t\|_{\infty} = \sup_{u \in [0,t]} |y_u|.
\end{align*}
The norm $| \cdot |$ is the usual Euclidean norm in the appropriate Euclidean space, depending on the dimension of the path being considered. This continuity notion could be relaxed, see, for instance, \cite{fito_extension_ito_formula}.

Moreover, we say a functional $f$ is \textit{boundedness- preserving} if, for every compact set $K \subset \bR^n$, there exists a constant $C$ such that $|f(Y_t)| \leq C$, for every path $Y_t$ satisfying $Y_t([0,t]) = \{y \in \bR^n \ ; \ Y_t(s) = y \mbox{ for some } s \in [0,t]\} \subset K$, see \cite{fito_cont_change}.

A functional $f: \Lambda \longrightarrow \bR$ is said to belong to $\bC^{1,2}$ if it is $\Lambda$\hyph continuous, bounded\-ness\hyph preserving and it has $\Lambda$\hyph continuous, boundedness\hyph preserving derivatives $\Delta_t f$, $\Delta_x f$ and $\Delta_{xx} f$. Here, clearly, $\Delta_{xx} = \Delta_x \Delta_x$.

Our numerical method is based on the following approximation of the functional derivatives: for a smooth functional $f \in \bC^{1,2}$, we use
\begin{equation}\label{eq:finite_diff}
\begin{aligned}
\Delta_t f(Y_t) &= \frac{f(Y_{t,\delta t}) - f(Y_t)}{\delta t} + o(\delta t),\\
\Delta_x f(Y_t) &= \frac{f(Y_t^h) - f(Y_t)}{h} + o(h),\\
\Delta_{xx} f(Y_t) &= \frac{f(Y_t^h) -2 f(Y_t) + f(Y_t^{-h})}{h^2} + o(h^2).
\end{aligned}
\end{equation}
Additionally, one could obviously consider
$$\Delta_x f(Y_t) = \frac{f(Y_t^h) - f(Y_t^{-h})}{2h} + o(h^2).$$

% The It\^o formula can be generalized to this framework. The proof can be found in \cite{fito_dupire}. We start by fixing a probability space $(\Omega, \cF, \bP)$.
% \begin{theorem}[Functional It\^o Formula; \cite{fito_dupire}]\label{thm:fif}
% Let $x$ be a continuous semimartingale and $f \in \bC^{1,2}$. Then, for any $t \in [0,T]$,
% \begin{align*}
% f(X_t) = f(X_0) + \int_0^t \Delta_t f(X_s) ds + \int_0^t \Delta_x f(X_s) dx_s + \frac{1}{2} \int_0^t \Delta_{xx} f(X_s) d\langle x\rangle_s \quad \bP\mbox{-a.s.}
% \end{align*}
% \end{theorem}

\subsection{PPDEs}

For any $s \leq t$ in $[0,T]$, denote by $\Lambda_{s,t}$ the space of $\bR^n$-valued c\`adl\`ag paths on $[s,t]$. Now define the operator $(\cdot \ \otimes \ \cdot) : \Lambda_{s,t} \times \Lambda_{t,T} \longrightarrow \Lambda_{s,T}$, the \textit{concatenation} of paths, by
$$(Y \otimes Z)(u) = \left\{
\begin{array}{ll}
  y_u,  &\mbox{ if } s \leq u < t, \\
  z_u - z_t + y_t, &\mbox{ if }  t \leq u \leq T,
\end{array}
\right.$$
which is a paste of $Y$ and $Z$.

Given functionals $\mu$ and $\sigma$ and fixing a probability space $(\Omega, \cF, \bP)$, we consider a process $x$ given by the stochastic differential equation (SDE)
\begin{align}
dx_s = \mu(X_s)ds + \sigma(X_s)dw_s, \label{eq:sde}
\end{align}
with $s \geq t$ and $X_t = Y_t$. The process $(w_s)_{s \in [0,T]}$ denotes a standard Brownian motion in $(\Omega, \cF, \bP)$ and we assume $\mu$ and $\sigma$ are such that there exists a unique strong solution for the SDE (\ref{eq:sde}). This unique solution will be denoted by $x_s^{Y_t}$ and the path solution from $t$ to $T$ by $X_{t,T}^{Y_t}$. We forward the reader, for instance, to \cite{rogerswilliams} for results on SDEs with functional coefficients.

Finally, we define the \textit{conditioned expectation} as
\begin{align}
\bE[g(X_T) \ | \ Y_t] = \bE[g(Y_t \otimes X_{t,T}^{Y_t})], \label{eq:conditioned_expec}
\end{align}
for any $Y_t \in \Lambda$. The path $Y_t \otimes X_{t,T}^{Y_t} \in \Lambda_T$ is equal to the path $Y_t$ up to $t$ and follows the dynamics of the SDE (\ref{eq:sde}) from $t$ to $T$ with initial path $Y_t$. Moreover, if we define the filtration $\cF_t^x$ generated by $\{x_s \ ; \ s \leq t\}$, one may prove
$$\bE[g(X_T) \ | \ X_t(\omega)] = \bE[g(X_T) \ | \ \cF_t^x](\omega) \quad \bP\mbox{-a.s.}$$
where the expectation on the left-hand side is the one discussed above and the one on the right-hand side is the usual \textit{conditional expectation}.

The Feynman-Kac formula in the classical stochastic calculus is a very important result that relates conditional expectations of functions of diffusions and PDEs. It turns out that a functional extension of this result is available.

\begin{theorem}[Functional Feynman-Kac Formula; \cite{fito_dupire}]\label{thm:feynman-kac}

Let $x$ be a process given by the SDE (\ref{eq:sde}). Consider functionals $g :\Lambda_T \longrightarrow \bR$, $\lambda: \Lambda \longrightarrow \bR$ and $k:\Lambda \longrightarrow \bR$ and define the functional $f$ as
$$f(Y_t) = \bE\left[\left.e^{-\int_t^T \lambda(X_u) du} g(X_T) + \int_t^T e^{-\int_t^s \lambda(X_u) du} k(X_s) ds \ \right| \ Y_t \right],$$
for any path $Y_t \in \Lambda$, $t \in [0,T]$. Thus, if $f \in \bC^{1,2}$ and $k$, $\lambda$, $\mu$ and $\sigma$ are $\Lambda$-continuous, then $f$ satisfies the (linear) Path-dependent Partial Differential Equation (PPDE):
\begin{align}\label{eq:feynman-kac-equation}
\Delta_t f(Y_t) + \mu(Y_t) \Delta_x f(Y_t)  + \frac{1}{2} \sigma^2(Y_t) \Delta_{xx} f(Y_t) - \lambda(Y_t)f(Y_t) + k(Y_t) = 0,
\end{align}
with $f(Y_T) = g(Y_T)$, for any $Y_t$ in the topological support of the stochastic process process $x$.

\end{theorem}

\begin{remark}\label{rmk:stroock_varadhan}
In diffusion models ($\mu(Y_t) = \mu(t,y_t)$ and $\sigma(Y_t) = \sigma(t,y_t)$), under mild assumptions on $\mu$ and $\sigma$, the Stroock-Varadhan Support Theorem states that the topological support of $x$ is the space of continuous paths starting at $x_0$, see for instance \cite[Chapter 2]{pinsky95}. So, under these assumptions, the PPDE (\ref{eq:feynman-kac-equation}) will hold for any continuous path. See \cite{fito_greeks} for a discussion on this type of result in the case of SDEs with functional coefficients. For instance, the arithmetic and geometric Brownian motions have full support on the space of continuous path, with the GBM having a restriction for positive range for the paths.
\end{remark}

\begin{remark}
Existence and uniqueness of classical (in the functional sense) of solution of PPDEs of the form (\ref{eq:feynman-kac-equation}) was studied in \cite{ppde_existence}, for instance. We forward the reader to the aforesaid reference for conditions of the functional parameters to ensure this result. Furthermore, several results have been developed to study non-linear versions of such PPDE and the existence and uniqueness of viscosity solutions, see \cite{fito_zhang_1}, \cite{fito_touzi_ppde1, fito_touzi_ppde2}.
\end{remark}

\section{Path-Dependent Deep Galerkin Method (PDGM)}
\label{sec:method}

In this section, we will present our algorithm to numerically solve a vast class of PPDEs. The main idea of algorithm is to apply the DGM methodology of \cite{dgm} to the PPDE framework. By DGM methodology we mean to approximate the solution of the equation by finding an neural network that approximately solve the equation in a given sense and any other additional conditions. In order to achieve this we need to consider a neural network architecture that correctly models the functionals that appear in PPDEs. Since functionals are non-anticipative, their value at $t$ does not depend on state values after $t$, for any given time $t$. Because of this characteristic we consider a combination of feed-forward and the LSTM networks. 

Another difference between our setting and the DGM is the space where the equation is defined. In their case, the domain of the PDE is some subset of an Euclidean space. In our case, it is a subset of the space of paths $\Lambda$ or of the space of continuous paths.

\subsection{Long Short-Term Memory (LSTM) arquitecture}

We start by stating some useful definitions. A set of layers $\bM_{d, k}^{\rho}$ with input  $x \in \bR^d$ in a feed-forward neural network can be defined as
\begin{equation}
 \bM_{d, k}^\rho :=  \{M: \bR^d \to \bR^k \ ; \   M(x) =  \rho(Ax + b), A \in \bR^{k\times d}, b \in \bR^k\},
\end{equation}
where $\rho$ is some activation function such as $\rho_{\tanh} (x) := \tanh(x)$, $\rho_s(x):= \frac{1}{1+e^{-x}}$ and $\rho_{Id}(x) := x$.
Then the set of feed-forward neural networks with $\ell$ hidden layers is defined as  a composition of layers:
\begin{align*}
\bN\bN_{d_1, d_2}^\ell = \{\widetilde{M}: \bR^{d_1} \to \bR^{d_2} \ ;& \ \widetilde{M} =  M_{\ell} \circ \cdots \circ M_1 \circ M_0, \\ 
&M_0 \in  \bM^{\rho}_{d_{1}, k_1},  M_{\ell} \in  \bM^{\rho}_{k_{l}, d_2}, M_i \in \bM^{\rho}_{k_{i}, k_{i+1}}, k_i \in \bZ^+, i = 1, \dots, \ell-1\}.
\end{align*}

Instead of all the inputs being not ordered  as in the feed-forward neural network, we often encounter sequential information as input, which is the case of our application. Additionally, for instance, in natural language processing, one of the main topics is sentimental analysis, where given paragraphs of texts, one classify them into different categories. In such cases, recurrent neural networks (RNN) come into play, which stores information so far, and uses them to perform computations in the next step. However, it was shown in \cite{vanish-gradient} that plain RNNs suffer from exploding or vanishing gradient problems. The LSTM network in \cite{LSTM} is designed to tackle this problem, in which the inputs and outputs are controlled by gates inside each LSTM cell. This architecture is powerful for capturing long-range dependence of the data. Each LSTM cell is composed of a cell state, which contains the information, and three gates, which regulate the flow of information. Mathematically, the rule inside the $i$th cell follows, for $x_i \in \bR^d$,

\begin{equation}\label{gates}
\begin{aligned}
\Gamma_{F_i}(x_i, a_{i-1}) = & \rho_{s} (A_F x_i + U_F a_{i-1} + b_F), \\
\Gamma_{I_i}(x_i, a_{i-1}) = & \rho_{s}(A_I x_i + U_I a_{i-1} + b_I), \\
\Gamma_{O_i}(x_i, a_{i-1}) = & \rho_{s} (A_O x_i + U_O a_{i-1} + b_O), \\
c_i = & \Gamma_{F_i} \odot c_{i-1} + \Gamma_{I_i} \odot  \rho_{\tanh} (A_C x_i + U_C a_{i-1} + b_C), \\
a_i = & \Gamma_{O_i} \odot  \rho_{\tanh}(c_i), \\
\end{aligned}
\end{equation}
where the operator $\odot $ denotes the element-wise product. Additionally, $a_i \in \bR^k$ is known as the output vector with initial value $a_{-1} = 0$, and $c_i \in \bR^k$ is called the cell state vector with initial value $c_{-1} = 0$; $k$ refers to the number of hidden units. Moreover, $A_\cdot \in \bR^{k \times d}$, $U_{\cdot} \in \bR^{k \times k}$ are weight matrices, and $b_{\cdot} \in \bR^k$ is the bias vector. These parameters are learned during training via a stochastic gradient descent algorithm combined with backpropagation to compute the gradients.

The set of LSTM network up to time $i$ is defined as
\begin{multline}
\LSTM_{i, d, k} = \bigg\{M: (\bR^d)^i \times \bR^k \times \bR^k \to \bR^k \times \bR^k \  ; \ M(x_{[0, i]}, a_{-1}, c_{-1}) = (a_i, c_i), \\
c_i = \Gamma_{F_i} \odot c_{i-1} + \Gamma_{I_i} \odot  \rho_{\tanh} (A_C x_i + U_C a_{i-1} + b_C), a_i = \Gamma_{O_i} \odot  \rho_{\tanh}(c_i), a_{-1} = c_{-1} = 0  \bigg\},
\end{multline}
where $\Gamma_{F_\cdot}, \Gamma_{I_\cdot}, \Gamma_{O_\cdot}$ are defined in \eqref{gates} and $x_{[0, i]} = [x_{0}, \dots, x_i]$.

\subsection{PDGM architecture}

\begin{figure}[htbp]
\begin{center}
  \includegraphics[width = 1\textwidth, height = 6cm]{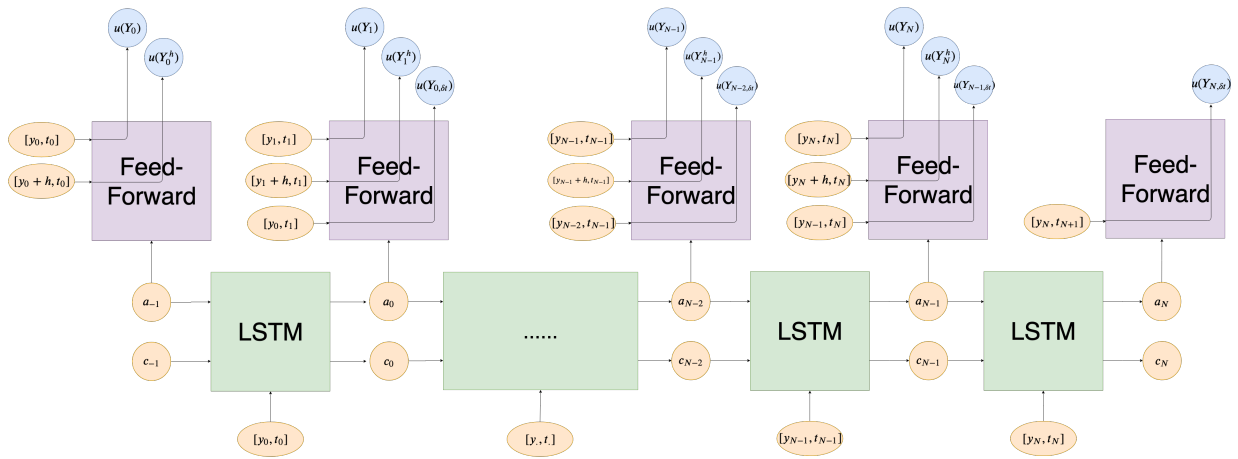}
  \caption{PDGM architecture.}
  \label{fig:NN}
    \end{center}
\end{figure}

In order to model the objects from the functional It\^o calculus, we propose a novel neural network architecture that combines LSTM and feed-forward networks in order to guarantee non-anticipativeness and to deal with the necessary path deformations from the functional It\^o calculus. We call such architecture Path-Dependent Deep Galerking Method (DPGM). The network structure is displayed in Figure \ref{fig:NN}. 

The PDGM architecture approximates a functional $f$ as follows. We start by considering a time discretization $\{t_i\}_{i=1,\ldots,N}$, with $\delta t = t_i - t_{i-1}$. We then approximate $f(Y_t)$ by a feed-forward neural network $u(Y_{t_i};\theta) = \varphi(t_i, y_{t_i}, a_{t_{i-1}}; \theta^f)$, where $t_i \leq t < t_{i+1}$. Here $\varphi \in \N\N_{k+2, 1}^\ell$, where $a$ is an output vector from an LSTM network, i.e. $a_{t_{i-1}} = \psi(y_{t_0}, \ldots, y_{t_{i-1}}; \theta^r)$, for some $\psi \in \LSTM_{i-1, 1, k}$.  $\theta = [\theta^f, \theta^r]$ are the neural network's parameters. The spatial and time extensions can be properly obtained by assigning correct inputs to the feed-forward neural network. This shows the effectiveness of the PDGM architecture for the functional It\^o calculus setting. Therefore, the functional derivatives can be approximated according to the approximation as in \eqref{eq:finite_diff}:
\begin{equation}\label{eq:finite_diff_NN}
\begin{aligned}
&u(Y_{t_i}^h;\theta) = \varphi(t_i, y_{t_i}+h, a_{t_{i-1}}; \theta^f), \\
&u(Y_{t_i, \delta t};\theta)  = \varphi(t_{i+1}, y_{t_i}, a_{t_{i}}; \theta^f),\\
&\Delta_t^{[\delta t]} u(Y_{t_i};\theta) = \frac{u(Y_{t_i, \delta t};\theta) - u(Y_{t_i};\theta)}{\delta t},\\
&\Delta_x^{[h]}  u(Y_{t_i};\theta) = \frac{u(Y_{t_i}^h;\theta) - u(Y_{t_i};\theta)}{h},\\
&\Delta_{xx}^{[h]} u(Y_{t_i};\theta) = \frac{u(Y_{t_i}^h;\theta) - 2u(Y_{t_i};\theta) + u(Y_{t_i}^{-h};\theta)}{h^2}.\\
\end{aligned}
\end{equation}

\subsection{Algorithm}

Consider the general class of final-value PPDE problem:
\begin{align}\label{eq:ppde_algo}
\begin{cases}
\displaystyle \Delta_t f(Y_t) + \cL f(Y_t) = 0,\\[10pt]
f(Y_T) = g(Y_T),
\end{cases}
\end{align}
where $Y_t \in \Lambda_0 \subset \Lambda$, for some subset of paths $\Lambda_0$. As an illustration, $\cL$ could be given by the linear operator
\begin{align}\label{eq:ppde_algo_linear}
\displaystyle \cL f(Y_t) = \mu(Y_t) \Delta_x f(Y_t)  + \frac{1}{2} \sigma^2(Y_t) \Delta_{xx} f(Y_t) - \lambda(Y_t)f(Y_t) + k(Y_t).
\end{align}

% {\color{blue} $u(Y_{t_i, \delta t};\theta)  = \varphi(t_{i+1}, y_{t_i}, a_{t_{i}}; \theta^f)$ should be correct because the flat extension is a function of (1) future time, which is ($t_{i+1}$), (2) flat extension, which is ($Y_T(t_{i+1}) = y_{t_i}$), (3) past information up to time $t_{i}$, which is ($a_{t_i}$)} \com{This makes sense. Thanks!}\\

We train the neural network to minimize the following objective function:
\begin{align*}
    J(\theta) &=\left\| \Delta_t^{[\delta t]} u(\cdot \ ; \theta) + \cL^{[h]} u(\cdot \ ; \theta) \right\|^2_{\Lambda, \nu_1} + \|u(\cdot \ ; \theta) - g\|^2_{\Lambda_T, \nu_2},
\end{align*}

where $\cL^{[h]}$ is the operator $\cL$ with the finite difference approximation of the functional derivatives and
\begin{align*}
&\|f\|^2_{\Lambda, \nu_1} = \bE_{\nu_1}\left[\int_0^T f^2(X_t) dt \right],\\
&\|g\|^2_{\Lambda_T, \nu_2} = \bE_{\nu_2}\left[g^2(X_T) \right].
\end{align*}

Here, $\nu_1$ and $\nu_2$ are measures in the path space $\Lambda$ and $\Lambda_T$, respectively. The choice of this measures and consequently how we should sample the paths $X_t$ in order to approximate the theoretical loss function $J$ will be discussed below. Additionally, we apply stochastic gradient descent to minimize the loss function over a set of parameters $\theta$. The neural network with optimized parameters $\theta$ delivers an approximation of the solution of the PPDE (\ref{eq:ppde_algo}). Given $M$ simulated paths accordingly to the laws $\nu_1$ and $\nu_2$, time and space discretization parameters $\delta t$ and $h$, the loss $J$ will be approximated by
\begin{align}\label{eq:loss_NM}
J_{N,M}(\theta) = \frac{1}{M} \frac{1}{N}\sum_{j=1}^M \sum_{i=0}^N \left(\Delta_{t}^{[\delta t]} u(Y^{(j)}_{t_i};\theta) + \cL^{[h]} u(Y^{(j)}_{t_i};\theta)\right)^2 +  \frac{1}{M}\sum_{j=1}^M \left(u(Y^{(j)}_{t_N}; \theta) - g(Y_{t_N}^{(j)}) \right)^2.
\end{align}
Moreover, when a closed-form solution is available, we compute the $L^2$-error from this one and our numerical solution approximating the mean squared error $\|\cdot\|^2_{\Lambda, \nu_1}$ defined above.

Our algorithm works as follows:
\begin{algorithm}[htbp]
 \caption{Path-Dependent DGM - PDGM}
 \label{algo:nn_ppde}
 initialize discretization parameter $\delta t$, mini-batch size $M$ and threshold $\epsilon$\\
 \While{$J_{N,M}(\theta) > \epsilon$}{
 generate a mini-batch size of $M$ paths $\{{(Y_{t_i}^{(j)})}_{i = 0, \dots, N}\}_{j = 1, \dots, M}$\\
 \For{$i \in \{1, \dots, N\}$ }{
  calculate $u(Y_{t_i}^{(j)};\theta)$, $\Delta_t^{[\delta t]} u(Y_{t_i}^{(j)};\theta)$, $\Delta_x^{[h]} u(Y_{t_i}^{(j)};\theta)$ and $\Delta_{xx}^{[h]} u(Y_{t_i}^{(j)};\theta)$ according to \eqref{eq:finite_diff_NN}\;
  put them all together to compute $\cL^{[h]} u(Y_{t_i}^{(j)};\theta)$ \;  
 }
calculate the approximated loss function, $J_{N,M}(\theta)$, as in (\ref{eq:loss_NM});\\
to minimize $J_{N,M}(\theta)$, update $\theta$ using stochastic gradient descent.
}
\end{algorithm}

% \begin{enumerate}
%     \item Generate a mini-batch size of $M$ paths $\{{(Y_{t_i}^{(j)})}_{i = 0, \dots, N}\}_{j = 1, \dots, M}$ with $\delta t = t_i - t_{i-1}$ for $i \in {1, \dots, N}$.
%     \item For $i \in \{1, \dots, N\}$, calculate $u(Y_{t_i}^{(j)};\theta)$, $\Delta_t[\delta t] u(Y_{t_i}^{(j)};\theta)$, $\Delta_x[h] u(Y_{t_i}^{(j)};\theta)$ and $\Delta_{xx}[h] u(Y_{t_i}^{(j)};\theta)$ according to \eqref{eq:finite_diff_NN}. Put them all together to compute $\cL[h] u(Y_{t_i}^{(j)};\theta)$. 
%     \item Calculate the approximated loss function, 
% $$J_{N,M}(\theta) = \frac{1}{M} \frac{1}{N}\sum_{j=1}^M \sum_{i=0}^N \left(\Delta_{t} u(Y^{(j)}_{t_i};\theta) + \cL u(Y^{(j)}_{t_i};\theta)\right)^2 +  \frac{1}{M}\sum_{j=1}^M \left(u(Y^{(j)}_{t_N}; \theta) - g(Y_{t_N}^{(j)}) \right)^2.$$
%     \item Minimize $J_{N,M}$ and update $\theta$ according to stochastic gradient descent.
%     \item Repeat until $J_{N,M}$ is below some previously defined threshold.
% \end{enumerate}

\newpage

\subsubsection{Simulation}\label{sec:simulation}

One important ingredient of the method above is the simulation of the paths $Y^{(j)}$. The goal of this step is to select good representatives of the set $\Lambda_0$, the domain of the PPDE. Usually, for problems that arises from the Feynman-Kac formula (even in its non-linear form), it is straightforward to choose the generating process that should be considered for the simulation of these paths. For instance, if we have the path-dependent heat equation (studied in Section \ref{sec:brownian_example}), one should simulate from the Brownian motion. 

However, the simulation does not to have as precise as in the Monte Carlo methods. The reason is that the PPDE itself has the dynamics of the state variable within its formulation. For example, in the Heston model studied in Section \ref{sec:heston}, one could simulate the CIR dynamics simplifying the natural reflecting barrier at 0 (taking the maximum of the simulated value and zero, for instance).

Nonetheless, one should be aware of the choice of simulated paths for the training. Usually, the space $\Lambda_0$ is much bigger than the possible simulated paths (e.g. in the Brownian case, $\Lambda_0$ is the space of continuous paths in $[0,T]$). An interesting exercise is to verify that training from a given set of simulated paths gives the algorithm sufficient knowledge to predict the value of the functional on a different type of path. Numerical experiments showed us that one important aspect is the range of the test paths. If the range is very different from the trained paths, the approximation will not work very well. In the numerical examples below, we consider the exact model coming from the PPDE to simulate the paths for the training sets and test the trained functional in very smooth and very rough paths different from the generating process of the training paths, but respecting their range. The method performs very well in all of them.

\begin{remark}
An idea similar to control variates applied in Monte Carlo methods would be the following. Suppose that $\phi$ is a path-independent functional (i.e. $\phi(Y_t) = \phi(t,y_t)$) such that $\Delta_t \phi + \cL \phi = 0$ with $\phi(T, y_T)$ being somewhat analogous to $g(Y_T)$ (e.g. grows similarly). Then, the functional $\tilde{f}(Y_t) = f(Y_t) - \phi(t,y_t)$ solves the same PPDE and the final condition might be better behaved. We then could apply the algorithm to approximate $\tilde{f}$ and use the formula $f(Y_t) = \tilde{f}(Y_t) + \phi(t,y_t)$ to find an approximation for $f$.
\end{remark}

\subsubsection{Convergence Result}

The derivation of convergence results similar to the ones shown in \cite{dgm} are very challenging in this setting. We leave them for possible future work since it would require some new results from the functional It\^o calculus theory. However, we sketch an approach for the proof of existence of a PDGM network such that the loss function is arbitrarily small.

The argument would be as follows: fix $N$ as the time discretization parameter and consider the approximation of the functional $f$ as $f(Y_t) \approx \phi_N(y_{t_0}, \ldots, y_{t_i}, y_{t})$ where, $t_i < t < t_{i+1}$. If $f$ is smooth, then $\phi_N$ is smooth in the last variable and $\phi_N \to f$ as $N \to +\infty$, where the convergence is of the functional and their derivatives. Now, fundamentally, the PDGM approximates $\phi_N$ and this is very similar to argument presented in \cite{dgm}. The result, under possible additional technical conditions, could be formulated as:

\begin{conjecture}
Assume there exists a classical solution for the PPDE (\ref{eq:ppde_algo}). Then, for any $\eps > 0$, there exists $k, \ell \in \bN$, $\varphi \in \N\N_{k+2, 1}^{\ell}$ and $\psi \in \LSTM_{i-1, 1, k}$ such that $u(Y_{t_i};\theta) = \varphi(t_i, y_{t_i}, a_{t_{i-1}};\theta^f)$ with $a_{t_{i-1}} = \psi(y_{t_0}, \ldots, y_{t_{i-1}}; \theta^r)$ satisfies
$$J(\theta) < \eps.$$
\end{conjecture}

\section{Numerical Examples}
\label{sec:numerical_examples}

In this section we will provide several examples of PPDEs with their closed-form and PDGM solutions. We will consider different dynamics (Brownian motion, geometric Brownian motion and Heston model) and different path-dependent final conditions (running integral, running maximum and running minimum). Moreover, we will also consider a non-linear case. The algorithm could handle more complex problems, as for instance, high-dimensional PPDEs. We decided to choose classic examples for pedagogical reasons: they are well-known to the readers, they have closed-form solutions and demonstrate how powerful the method is. Furthermore, as it was clear from the exposition of the method, the PDGM is able to deal with any path-dependent structure as long as it might be written as a PPDE of the form (\ref{eq:ppde_algo}). Additional conditions, such as boundary and integral conditions, could be added to the loss function similarly to the DGM methodology.

We have used a personal desktop with Intel Core i7, 16GB RAM, and a NVIDIA RTX 2080 graphic card to run these numerical examples. Additionally, we have used the \texttt{TensorFlow}. Each epoch takes approximately 0.4s to 0.8s depending on the complexity of the neural network and the PPDE. The Python code for an illustrative example of the geometric Asian option shown in Section \ref{sec:geo_asian} is available at \url{https://github.com/zhaoyu-zhang/PDGM-Geometric_Asian}.

\subsection{Brownian motion}\label{sec:brownian_example}

In this section, we consider the class of examples
\begin{align}
\begin{cases}
\Delta_t f(Y_t) + \displaystyle \frac{1}{2}\Delta_{xx} f(Y_t) = 0,\\[10pt]
f(Y_T) = g(Y_T),
\end{cases}
\end{align}
where $Y_T$ is any continuous path, see Remark \ref{rmk:stroock_varadhan}. These PPDEs arise from the linear expectations of path-dependent final condition $g$ under a Brownian model. Under smoothness condition on $f$, the PPDE above holds for any continuous path $Y$. These simple examples allow us to provide a very clear introduction to the method and serve as illustrations. Below we will consider five different final conditions $g$: path-independent, linear running integral, quadratic running integral, one high-dimensional case and a strongly path-dependent example.

Training paths in this subsection are sampled from standard Brownian motions paths with $T = 1$ and time discretization $N = 100$. For the path independent, linear running integral, quadratic running integral examples,  we choose mini-batch size $M = 128$ paths.  We use a single layer LSTM network with 64 units connecting with a deep feed-forward neural network which consists of three hidden layers with 64, 128, 64 respectively.  Although we only train our neural network using standard Brownian motions simulated paths, our algorithm is able to provide a good approximation to the true solution for paths other than those. Furthermore, we show the train losses, test losses and MSE after 10,000 epochs in the table below.

\begin{table}[htbp]
\centering
\begin{tabular}{lccc}
Example                    & Train Loss         & Test Loss         & MSE \\
\hline
Path Independent           & $9 \times 10^{-6}$ & $9 \times 10^{-6}$  & $8.7 \times 10^{-4}$ \\
Linear Running Integral    & $2 \times 10^{-5}$ & $2 \times 10^{-5}$  & $4.3 \times 10^{-6}$\\
Quadratic Running Integral & $6 \times 10^{-5}$ & $6 \times 10^{-5}$  & $7.3 \times 10^{-5}$\\
High Dimensional Example with Dimension = 20 & $7 \times 10^{-2}$ & $7 \times 10^{-2}$  & $6.95 \times 10^{-2}$\\
Stopping Time Example & $1.6 \times 10^{-2}$ & $1.8 \times 10^{-2}$  & -- -- \\
\hline
\end{tabular}
\vspace{1mm} 
\caption{Train and test losses for the Brownian case}
\end{table}

For path independent, linear running integral, quadratic running integral examples, three representatives test paths with their corresponding solution and derivatives are plotted in Figures \ref{fig:ind_paths}, \ref{fig:linear_paths}, and \ref{fig:quadratic_paths} respectively. Path 1 is a standard Brownian motion path. Path 2 is the smooth path $y_t = (1-t)^2$ for $t \in [0,1]$. Path 3 is a realization of a sequence of uniform random variables between -1 and 1, i.e., $y_{t_i} \sim U(-1, 1)$, for each  $i \in \{0, \dots, N\}$.

\subsubsection{Path Independent}
As a sanity check, consider the case where $g(Y_T) = \phi(y_T)$, which yields a PDE with solution
$$f(Y_t) = \int_\bR \phi(y_t + z) \frac{e^{-z^2/(2(T-t))}}{\sqrt{2\pi (T-t)}} dz.$$

\begin{figure}[htbp]
\begin{center}
  \includegraphics[width = 0.45\textwidth, height = 5cm]{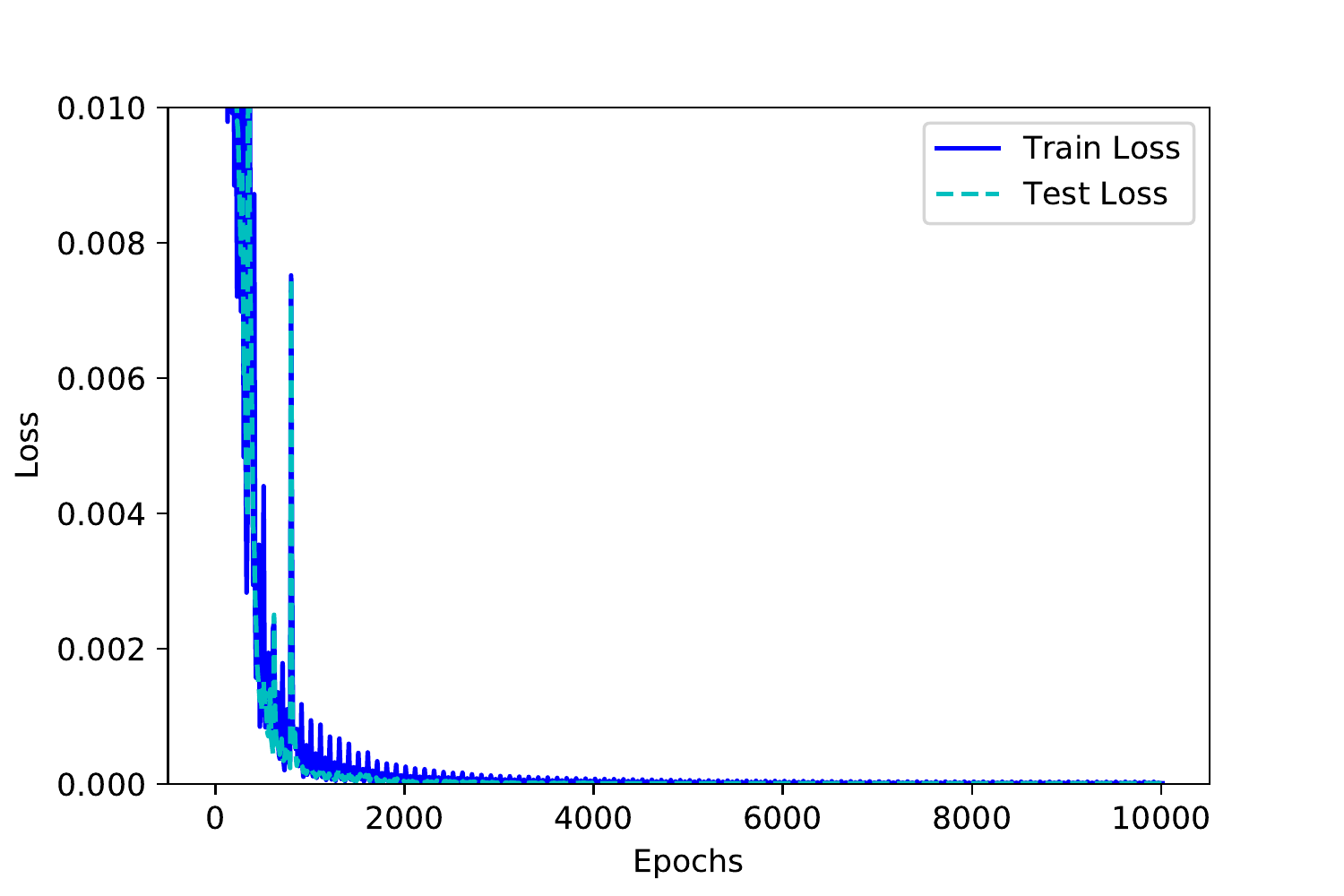}

  \caption{Train and test losses for the path-independent example.}
  \label{fig:ind_loss}
    \end{center}
\end{figure}

\begin{figure}[htbp]
\begin{center}
  \includegraphics[width = 0.45\textwidth, height = 5cm]{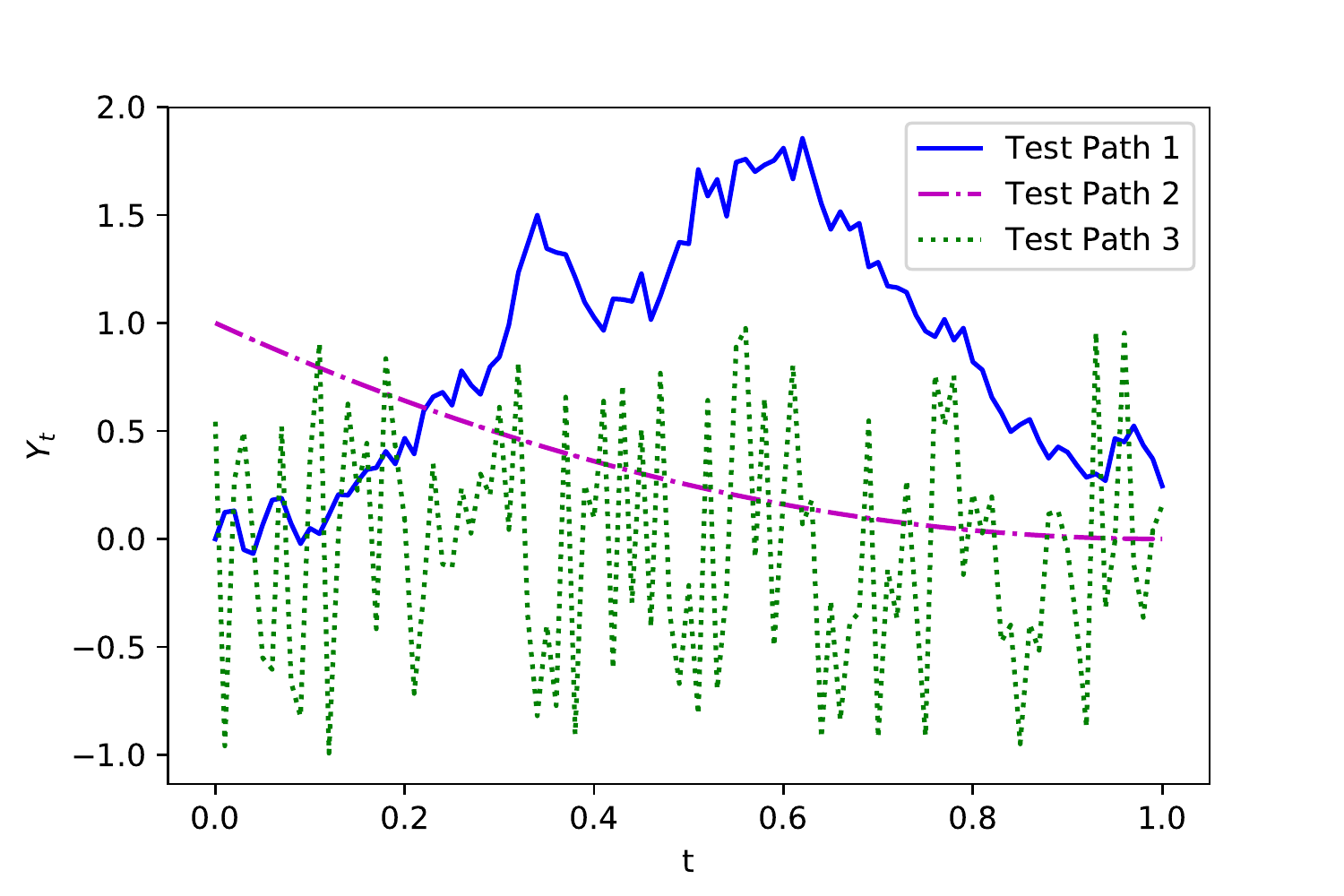}
  \includegraphics[width = 0.45\textwidth, height = 5cm]{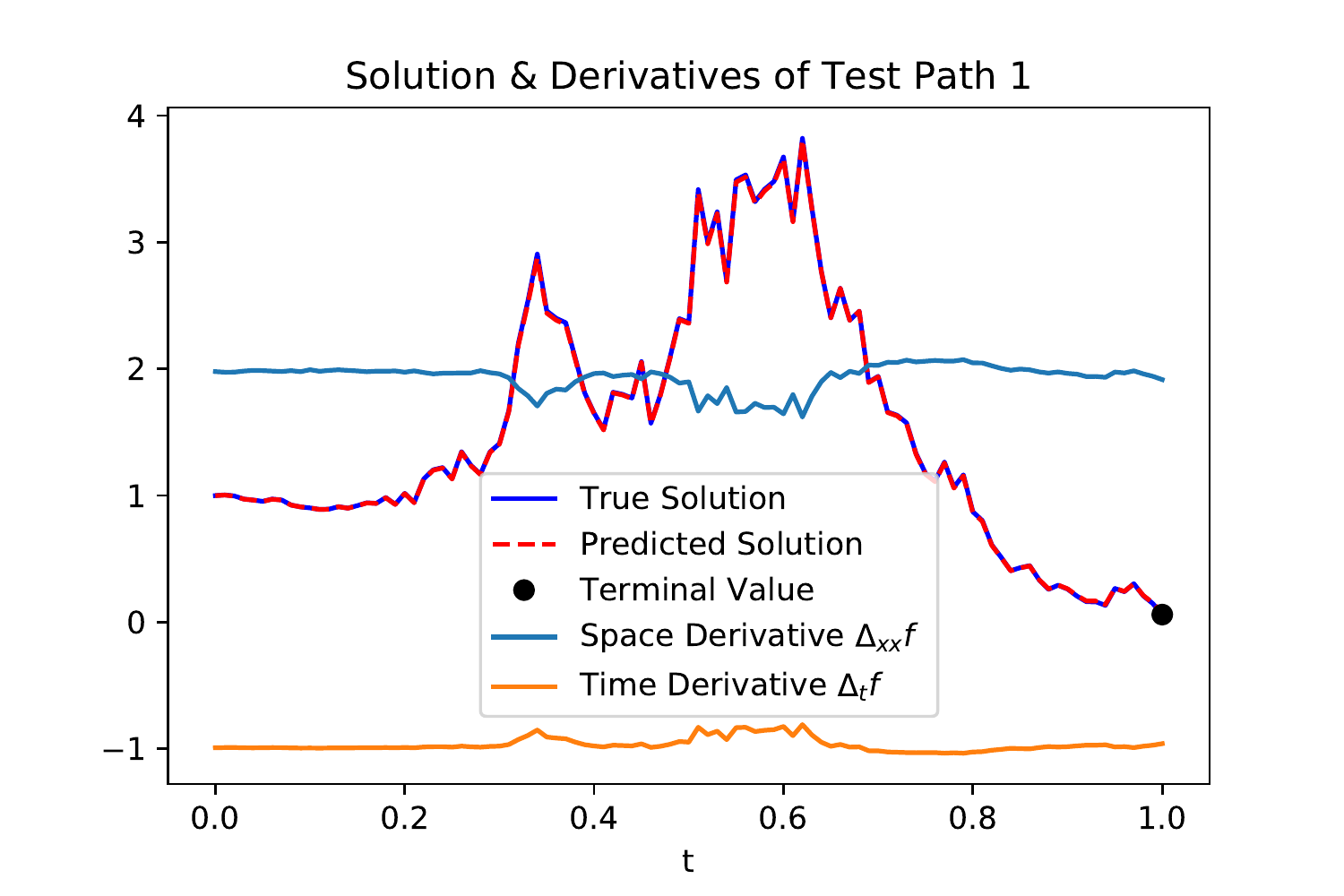}
  \includegraphics[width = 0.45\textwidth, height = 5cm]{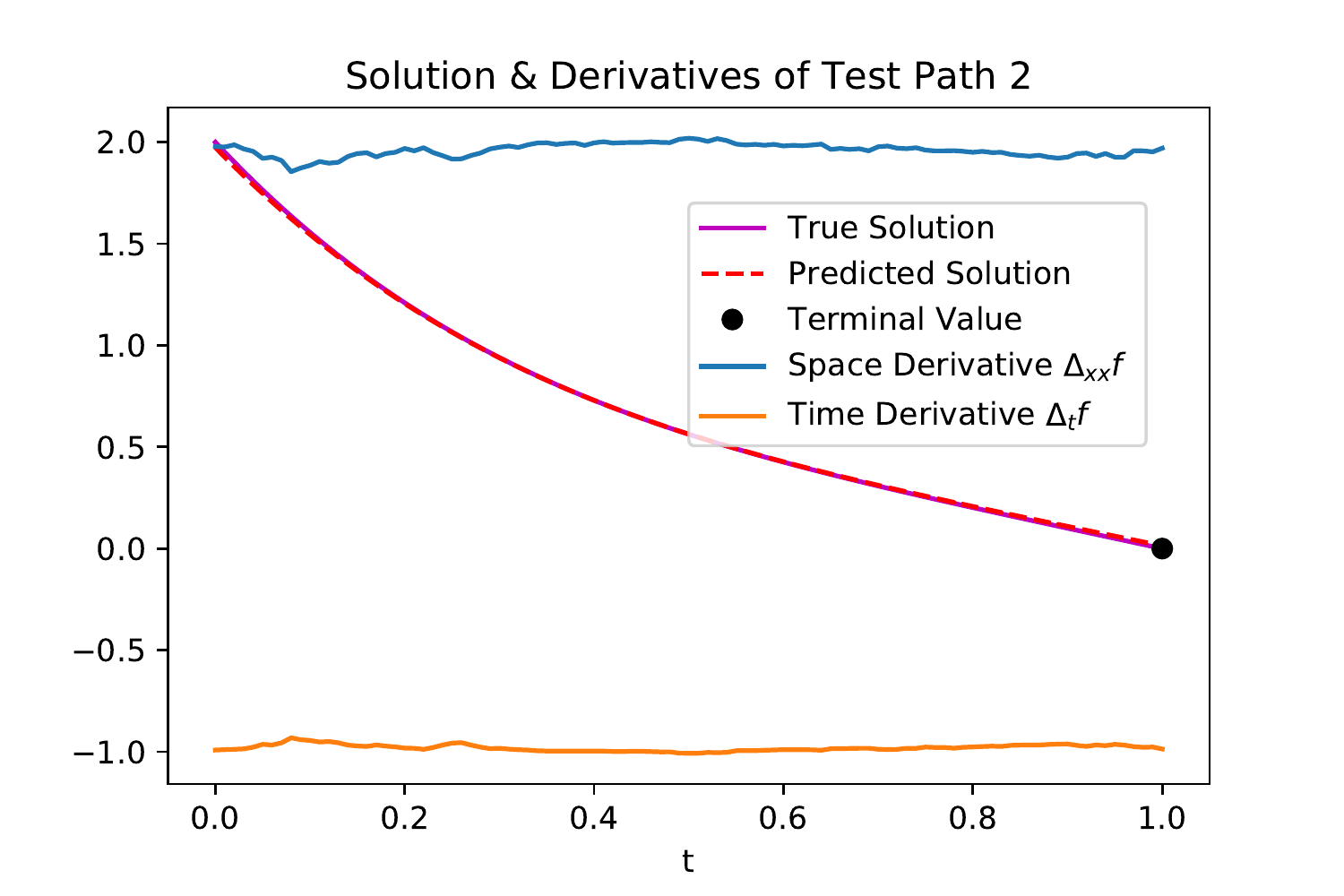}  \includegraphics[width = 0.45\textwidth, height = 5cm]{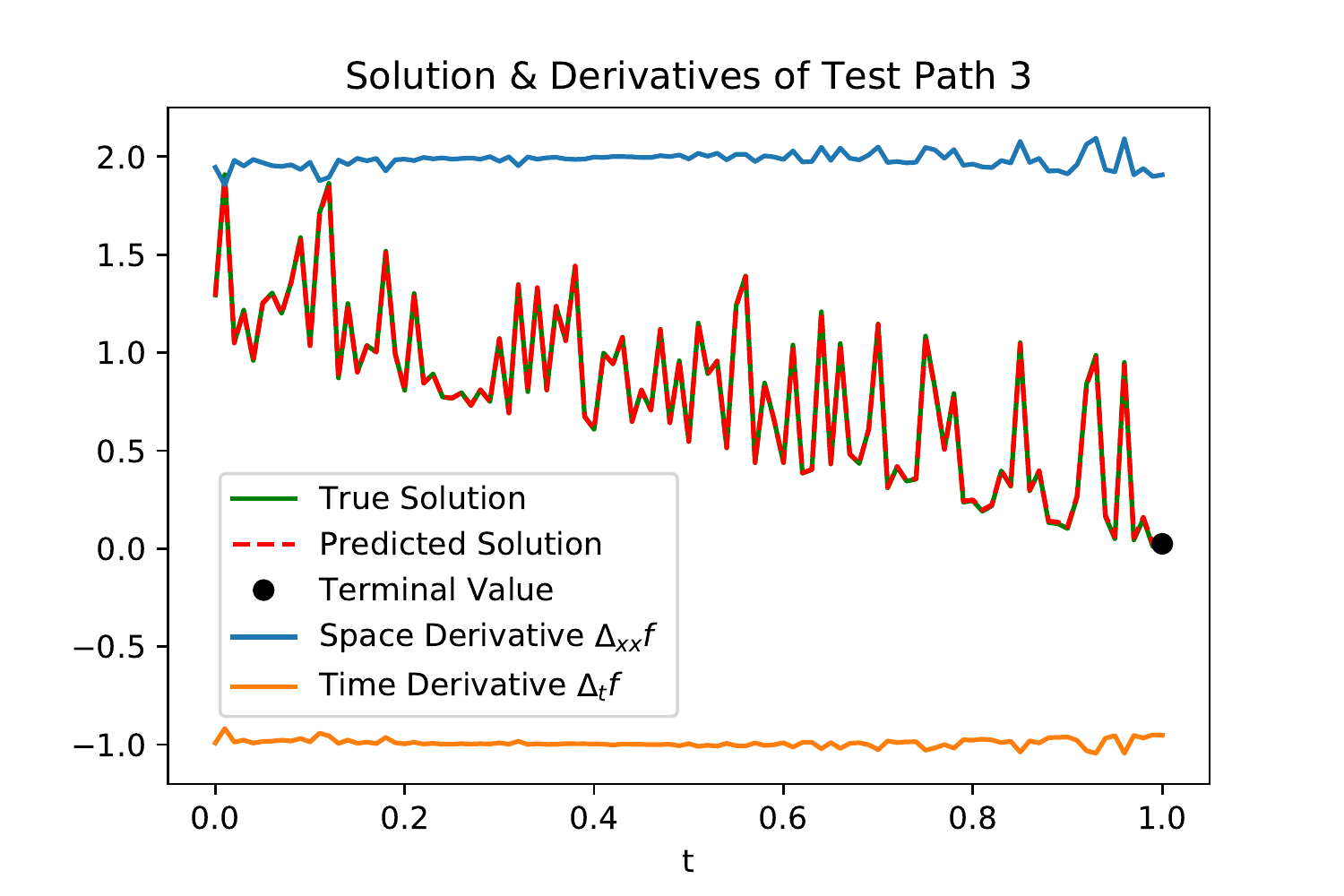}
  \caption{Three representative paths with corresponding solutions and functional derivatives for the path-independent example.}
  \label{fig:ind_paths}
    \end{center}
\end{figure}

As an example, we consider $\phi(y) = y^2$, which gives $f(Y_t) = y_t^2 + T - t$.
Figure \ref{fig:ind_loss} shows the training and testing losses. Three representative paths with their corresponding solution and derivatives are shown in Figure  \ref{fig:ind_paths}. It can be seen that our algorithm provides a good approximation. The functional derivatives for this example are $\Delta_t f(Y_t) = -1$ and $\Delta_{xx} f(Y_t) = 2$, which are also captured by the algorithm.

\subsubsection{Linear Running Integral}

\begin{figure}[htbp]
\begin{center}
  \includegraphics[width = 0.45\textwidth, height = 5cm]{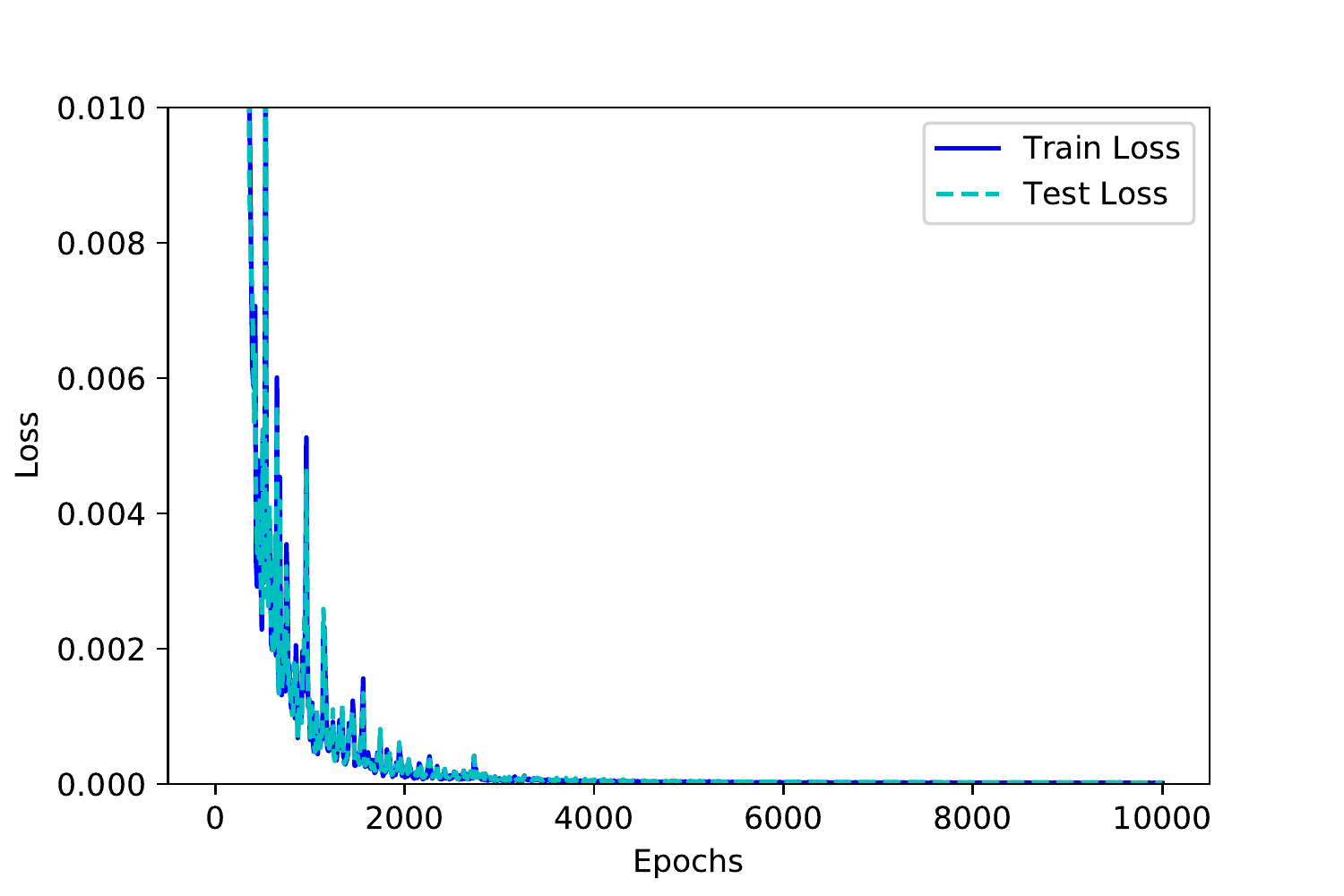}

  \caption{Train and test losses for the linear running integral example.}
  \label{fig:linear_loss}
    \end{center}
\end{figure}

\begin{figure}[htbp]
\begin{center}
  \includegraphics[width = 0.45\textwidth, height = 5cm]{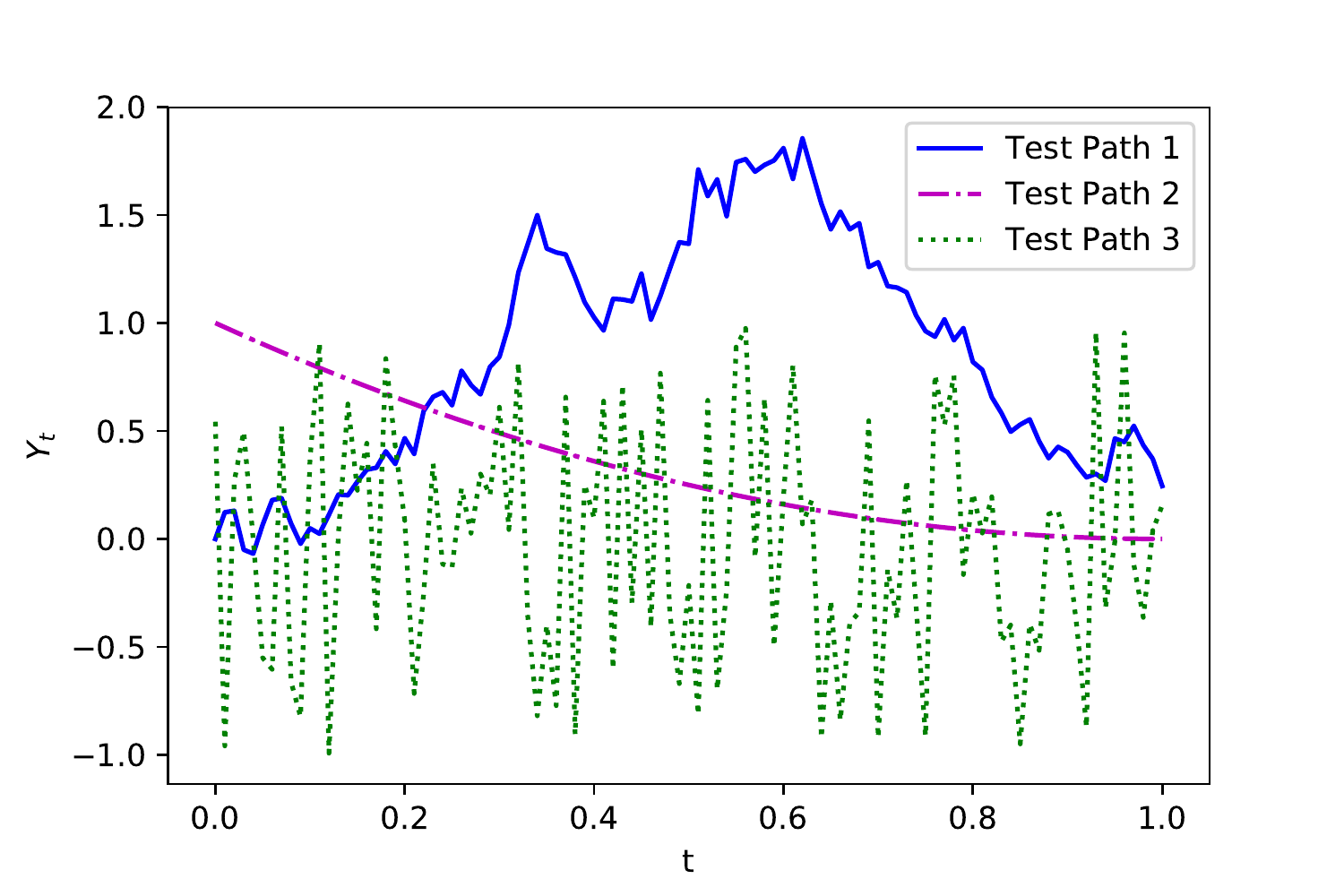}
  \includegraphics[width = 0.45\textwidth, height = 5cm]{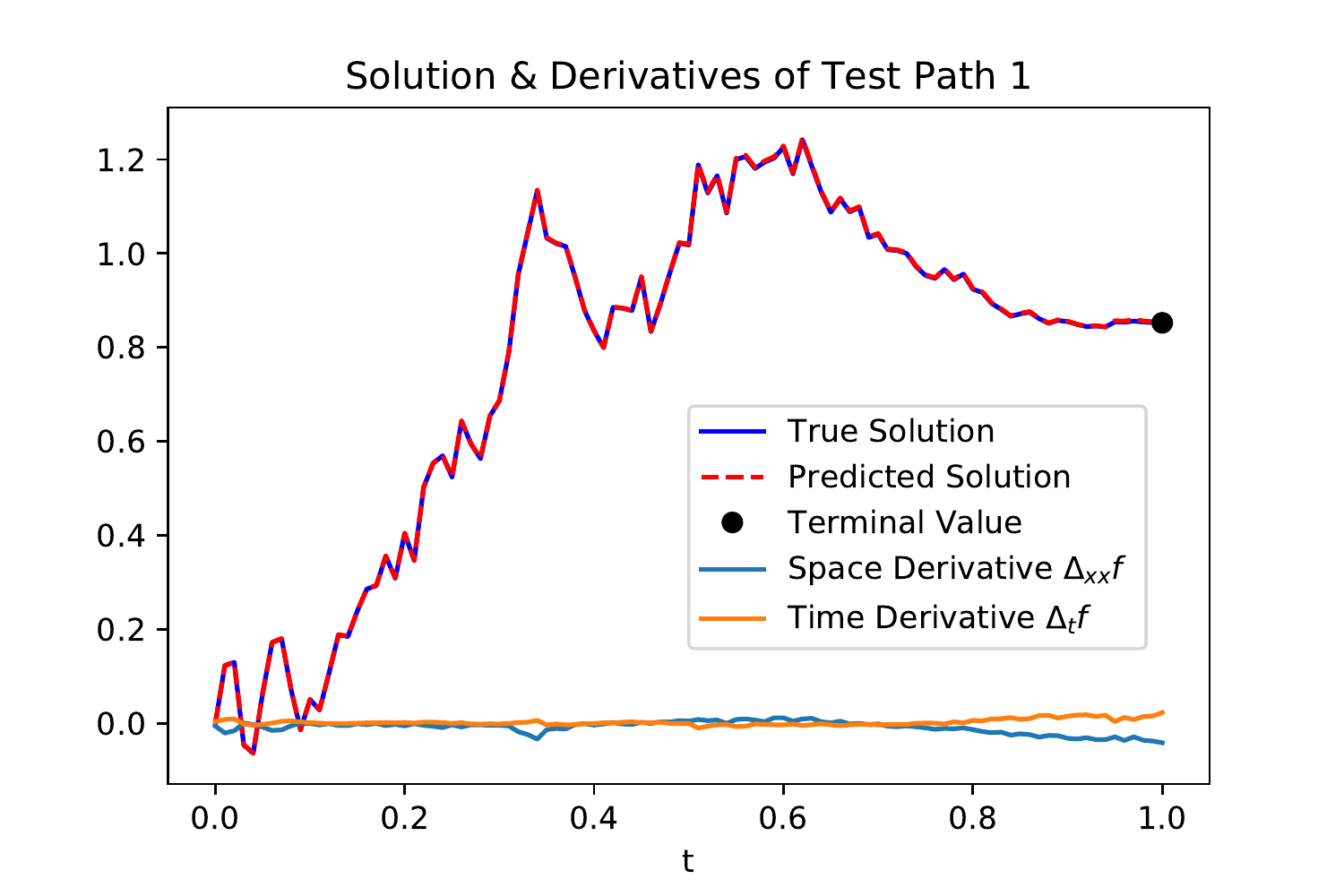}
  \includegraphics[width = 0.45\textwidth, height = 5cm]{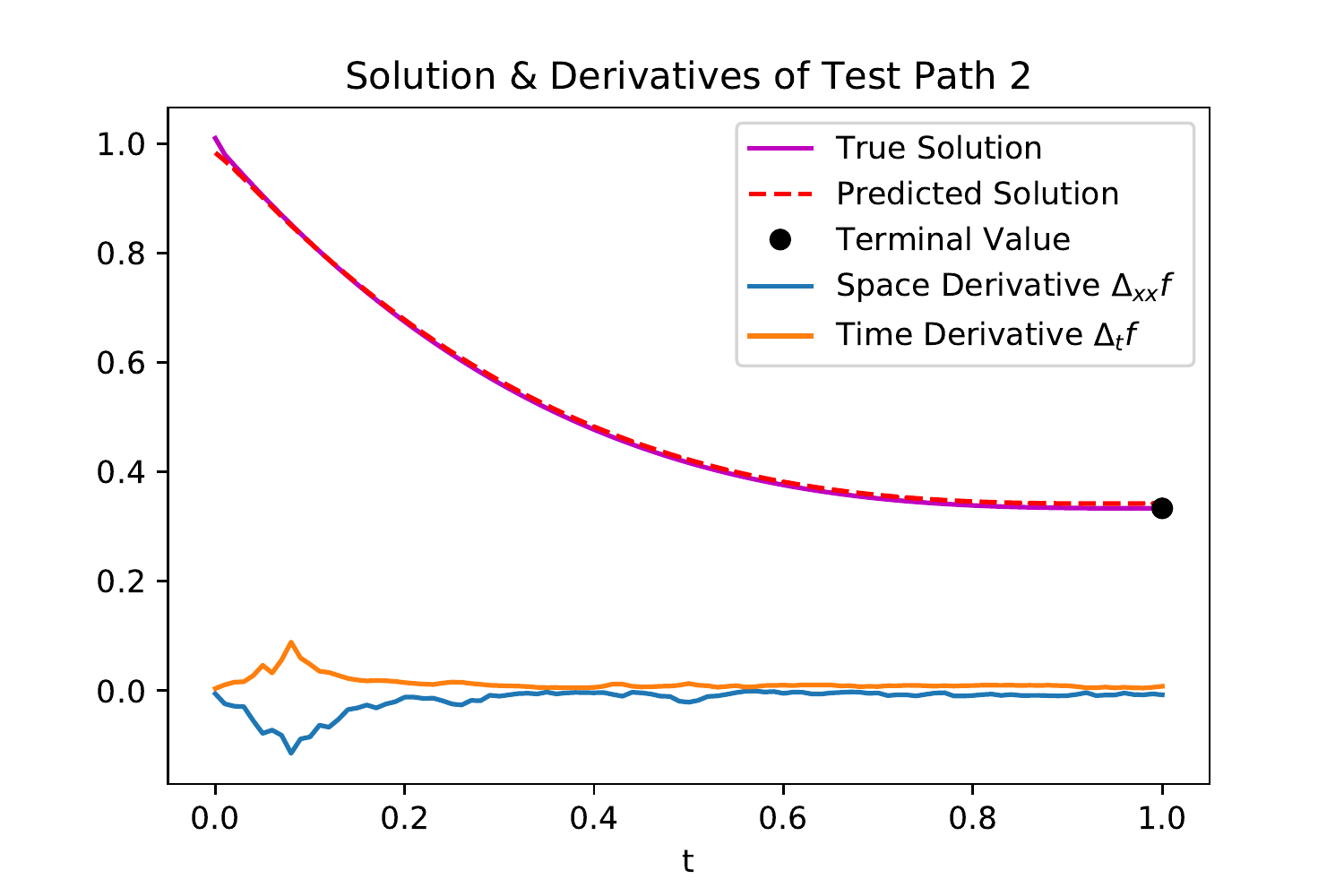}  \includegraphics[width = 0.45\textwidth, height = 5cm]{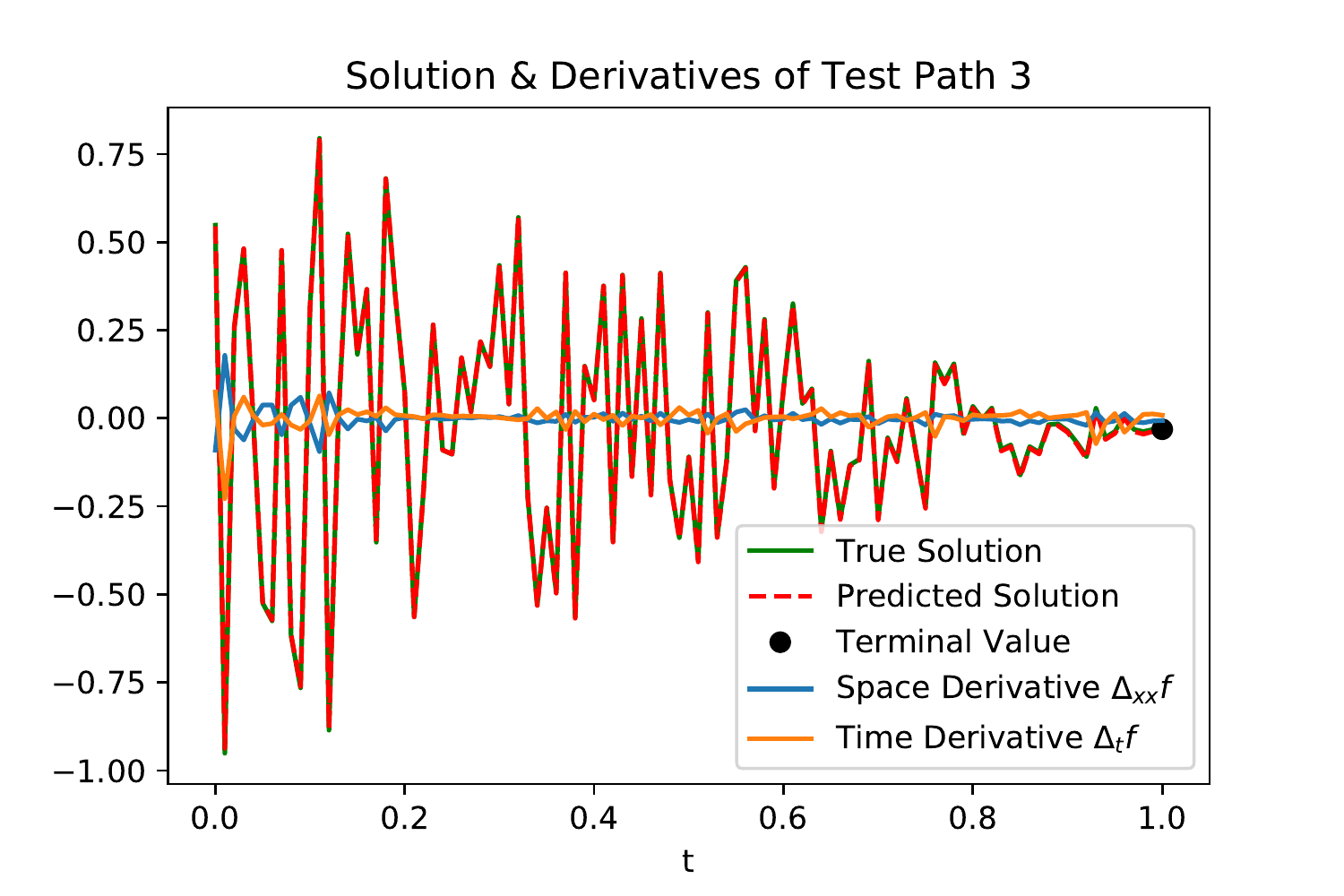}
  \caption{Three representative paths with corresponding solutions and functional derivatives for the linear running integral example.}
  \label{fig:linear_paths}
    \end{center}
\end{figure}

Consider the path-dependent case of $g(Y_T) = \ds \int_0^T y_u du$, which gives the solution
$$f(Y_t) = \int_0^t y_udu + y_t (T-t).$$
In this example, training and test losses reach $2 \times 10^{-5}$ after 10000 epochs, which is shown in Figure \ref{fig:linear_loss}. Figure \ref{fig:linear_paths} plots three representative paths (as in the path-independent example) with their corresponding solution and functional derivatives. The predicted solutions are approximately the same as true solutions. From the plot, the derivatives in this example for both $\Delta_t f(Y_t)$ and $\Delta_{xx} f(Y_t)$ are 0 which is true also by direct computation.

\begin{figure}[htbp]
\begin{center}
  \includegraphics[width = 0.45\textwidth, height = 5cm]{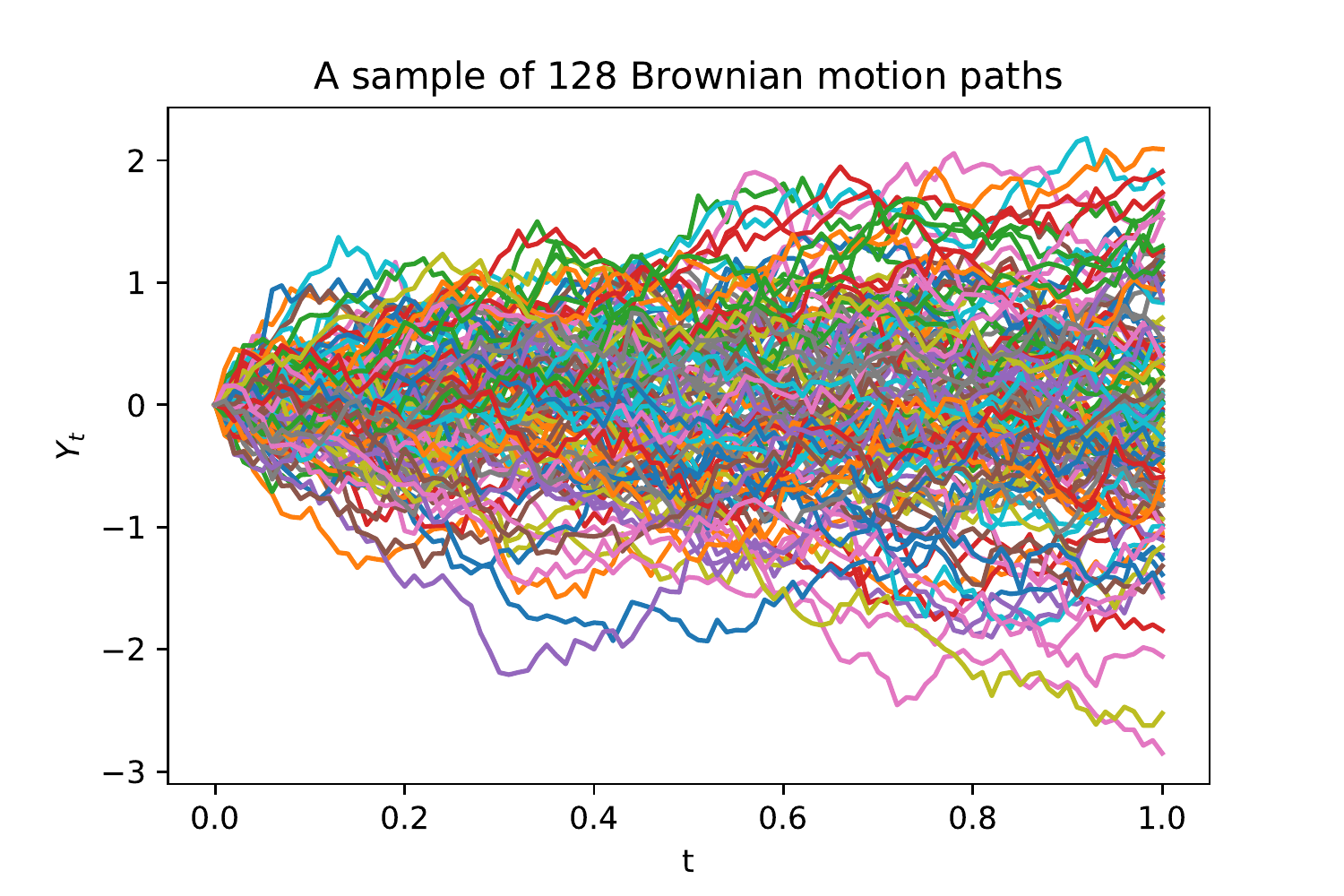}
  \caption{A sample of 128 Brownian motion paths.}
  \label{fig:bm_paths}
    \end{center}
\end{figure}

\begin{figure}[htbp]
\begin{center}
  \includegraphics[width = 0.45\textwidth, height = 5cm]{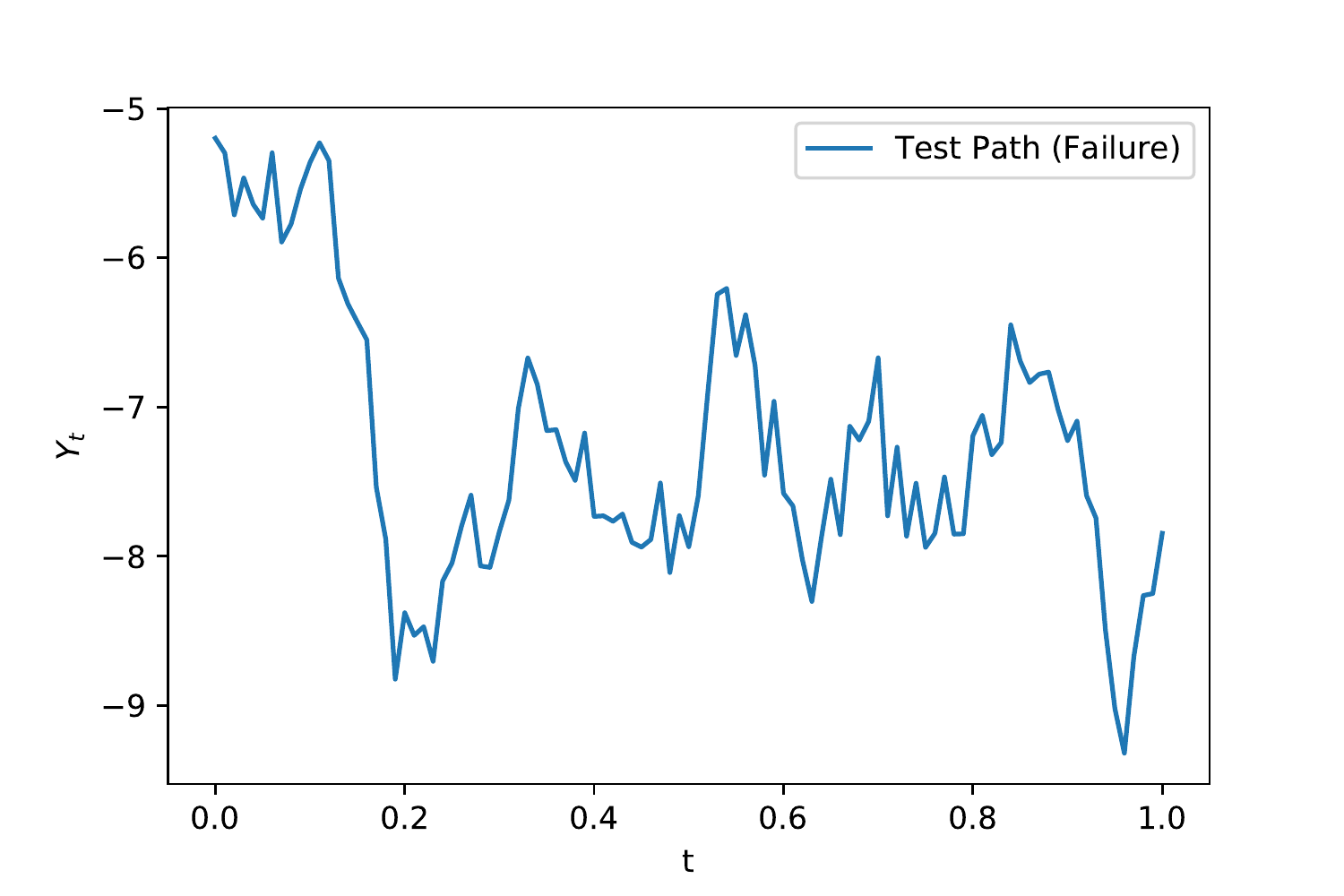}
  \includegraphics[width = 0.45\textwidth, height = 5cm]{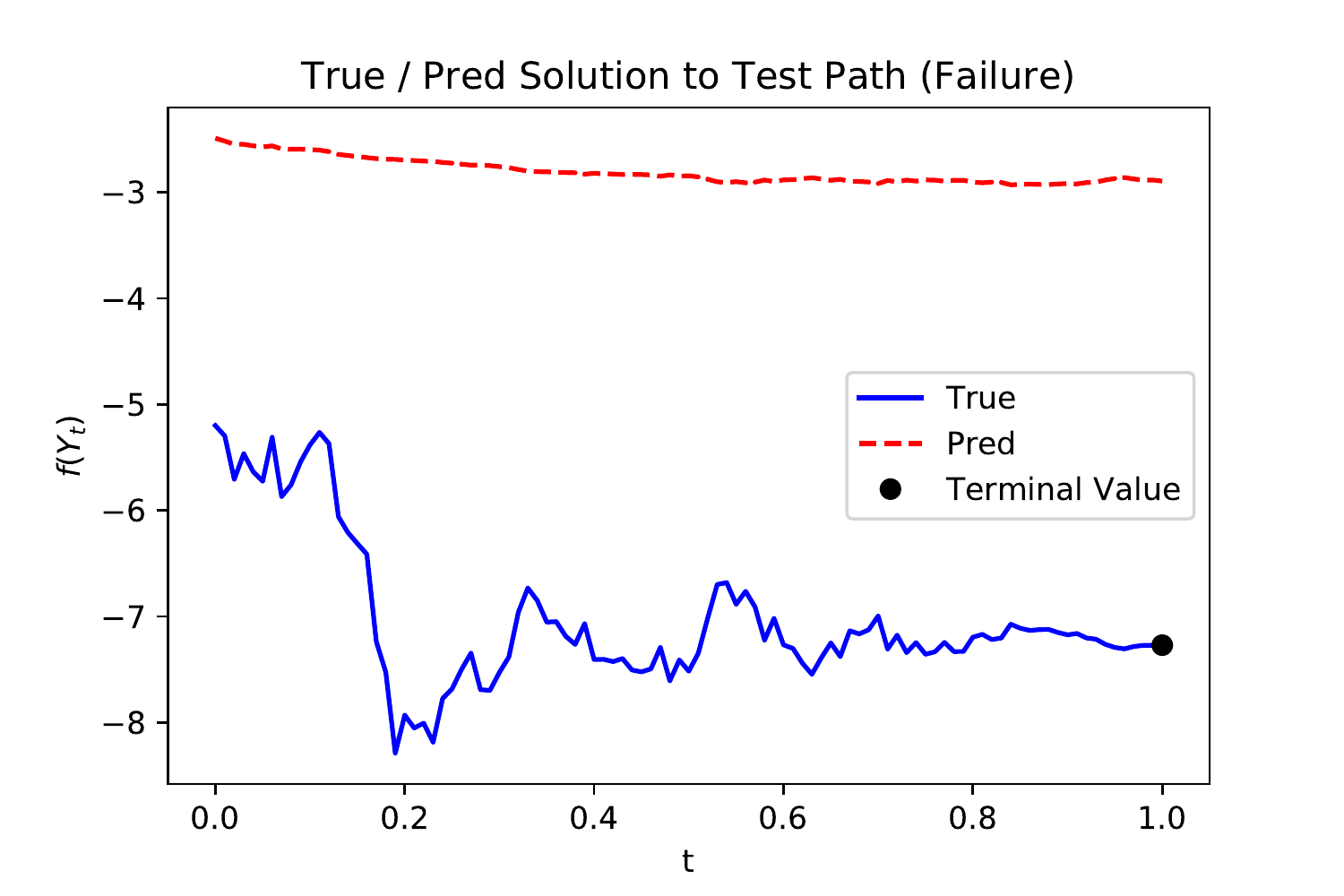}
  \caption{Prediction failure due to the limitation of training domain.}
  \label{fig:linear_paths_fail}
    \end{center}
\end{figure}

Though using Brownian motions as training paths yields a faster convergence with a small number of neurons, one drawback is that when the test path is outside the domain of the trained paths, it would yield a poor prediction, as it was discussed in Section \ref{sec:simulation}. In particular, Figure \ref{fig:bm_paths} plots 128 Brownian paths used for training, showing that the domain is from $-3$ to 2. In Figure \ref{fig:linear_paths_fail}, the neural network is \textit{not} able to find the right solutions to a Brownian path with volatility 4 which starts at $-5.2$.

\begin{figure}[htbp]
\begin{center}
  \includegraphics[width = 0.45\textwidth, height = 5cm]{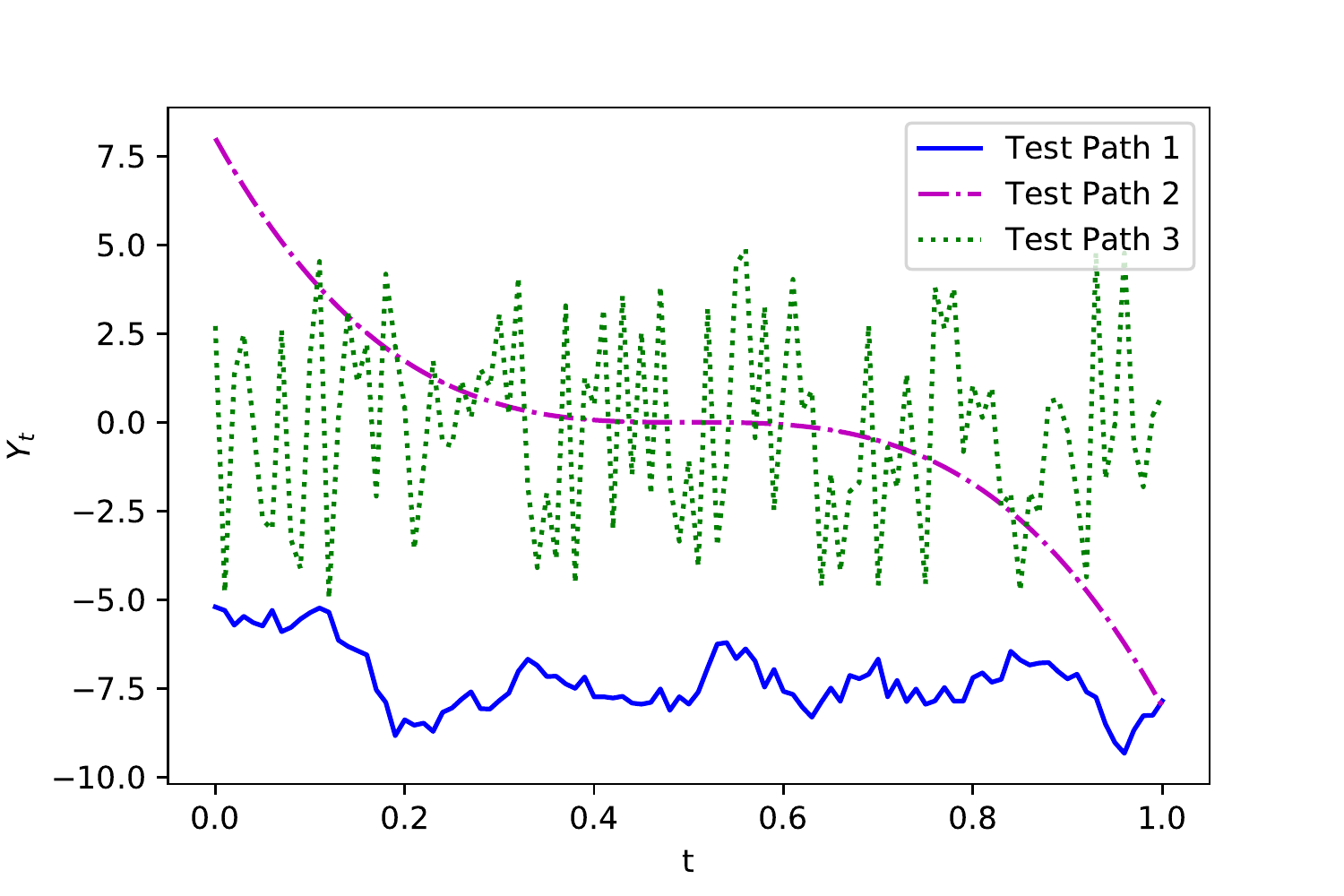}
  \includegraphics[width = 0.45\textwidth, height = 5cm]{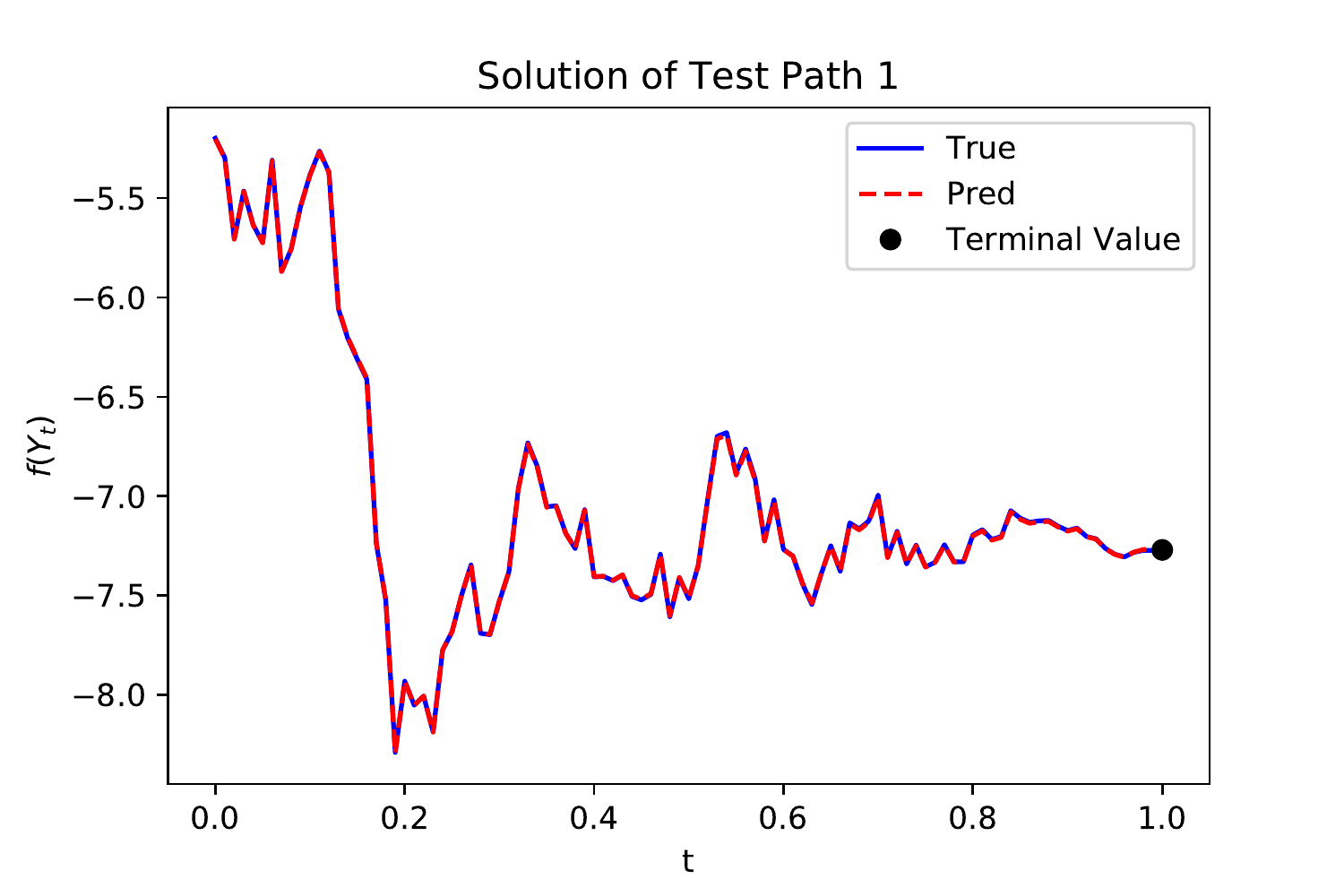}
  \includegraphics[width = 0.45\textwidth, height = 5cm]{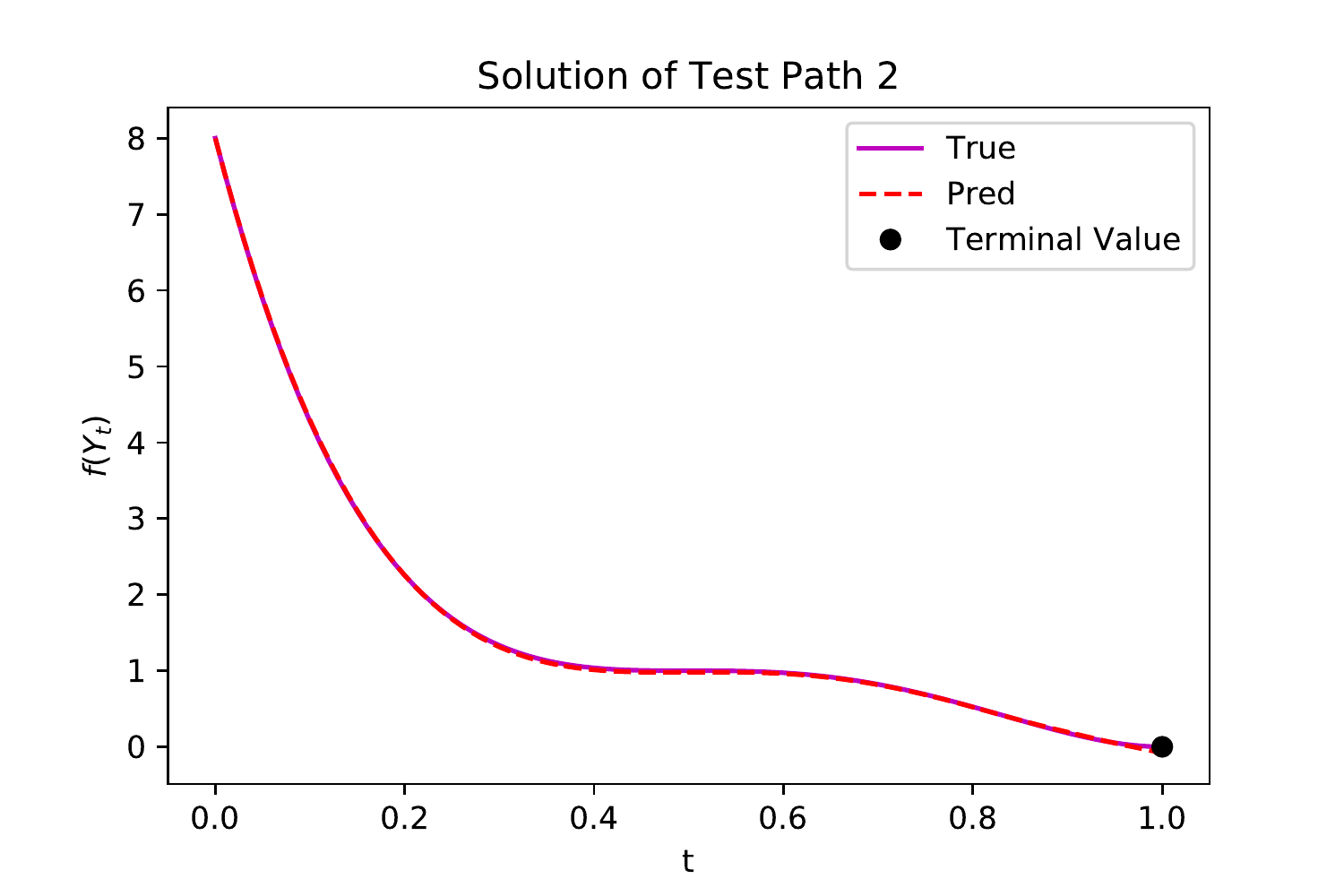}  \includegraphics[width = 0.45\textwidth, height = 5cm]{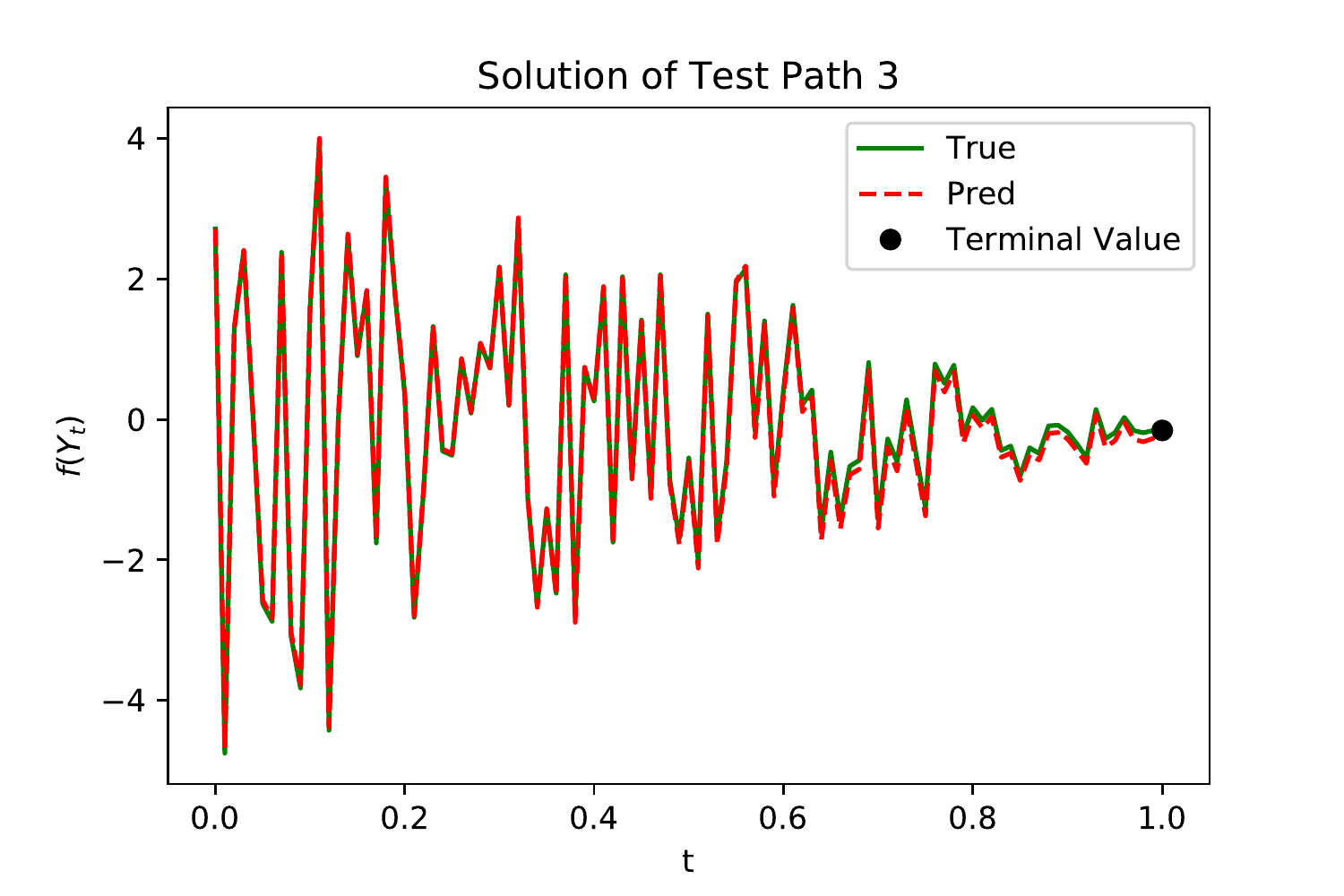}
  \caption{Three representative paths with corresponding solutions and functional derivatives for the linear running integral example.}
  \label{fig:linear_paths_large}
    \end{center}
\end{figure}

One easy remedy for the above problem is to use varying volatility of a Brownian paths with varying initial values. In addition, one may also need to enlarge the neural network. For example, the training paths for Figure \ref{fig:linear_paths_large} are Brownian paths with volatility $\sigma \in \{1,2,3,4\}$ and initial value $x_0 \sim U(-10, 10)$. The single layer LSTM network consists of 128 units, and each of the three layer feed-forward neural networks contains 128 hidden neurons.  For example, in Figure \ref{fig:linear_paths_large}, test path 1 is a Brownian path with volatility 4 and starting at $-5.2$; test path 2 is a function $y_t = (2-4t)^3$; test path 3 is a realization of a sequence of i.i.d. uniform random variables drawn from $-5$ to 5. As a result, according to the above setup, the neural network is capable to predict solutions to the paths with wider domain.

\subsubsection{Quadratic Running Integral}
In order to consider a more complicated case, take $g(Y_T) = \left(\ds \int_0^T y_u du\right)^2$, then
\begin{align*}
f(Y_t) &= \bE\left[\left(\int_0^t y_udu + \int_t^T (y_t + w_u - w_t)du\right)^2\right]\\
&= \bE\left[\left(\int_0^t y_udu + y_t(T-t) + \int_t^T (w_u - w_t)du\right)^2\right]\\
&= \left(\int_0^t y_udu\right)^2 + y_t^2 (T-t)^2 + 2y_t(T-t)\int_0^t y_udu + \bE\left[\left(\int_t^T (w_u - w_t)du\right)^2\right].
\end{align*}
Moreover
\begin{align*}
\bE\left[\left(\int_t^T (w_u - w_t)du\right)^2\right] &= \bE\left[\left(\int_0^{T-t} (w_{u+t} - w_t)du\right)^2\right] \\
&= \bE\left[\left(\int_0^{T-t} w_u du\right)^2\right] = \frac{1}{3}(T-t)^3,
\end{align*}
yielding
\begin{align*}
f(Y_t) = \left(\int_0^t y_udu\right)^2 + y_t^2 (T-t)^2 + 2y_t(T-t)\int_0^t y_udu + \frac{1}{3}(T-t)^3.
\end{align*}

\begin{figure}[htbp]
\begin{center}
  \includegraphics[width = 0.45\textwidth, height = 5cm]{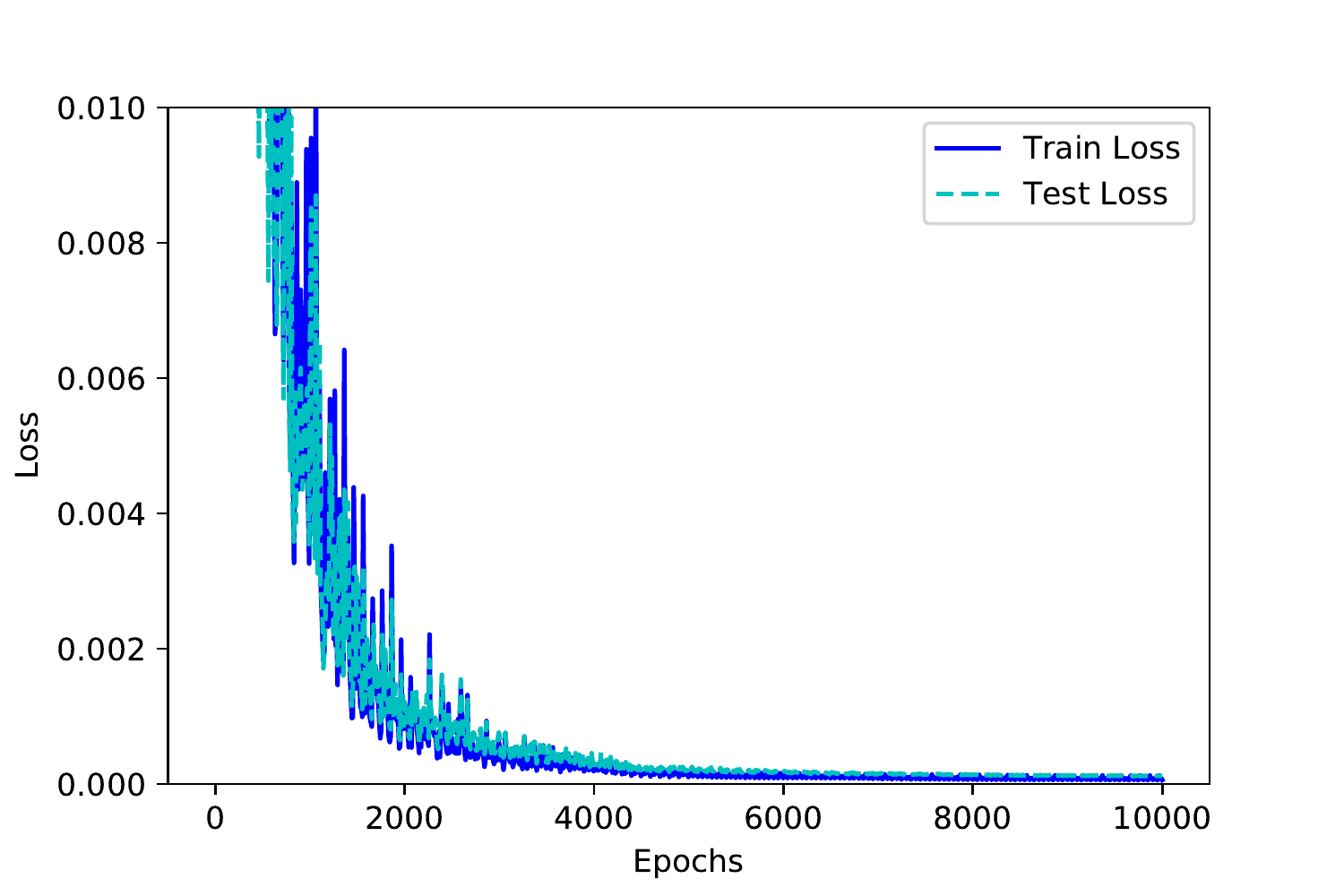}

  \caption{Train and test losses for the quadratic running integral example.}
  \label{fig:quadratic_loss}
    \end{center}
\end{figure}

\begin{figure}[htbp]
\begin{center}
  \includegraphics[width = 0.45\textwidth, height = 5cm]{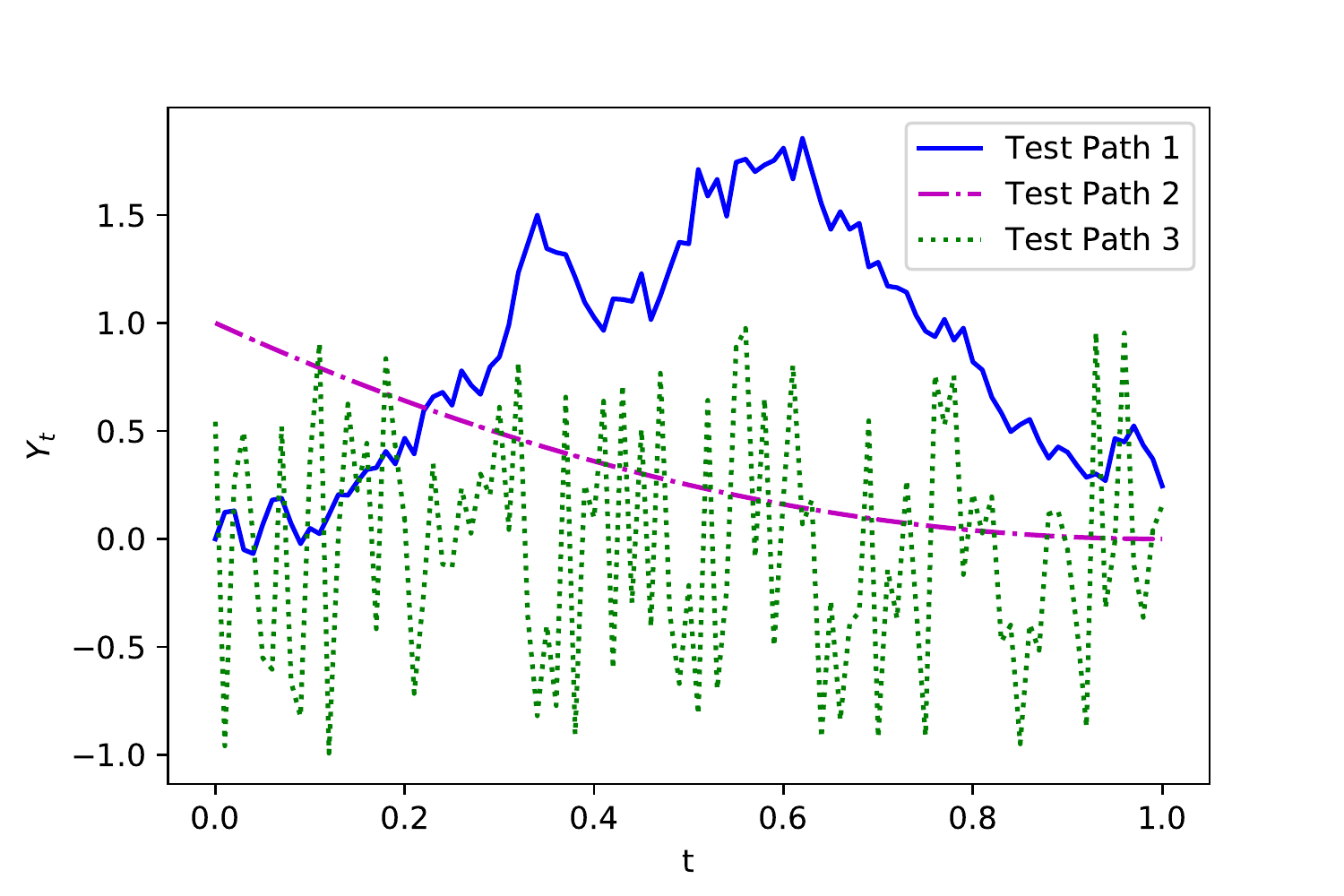}
  \includegraphics[width = 0.45\textwidth, height = 5cm]{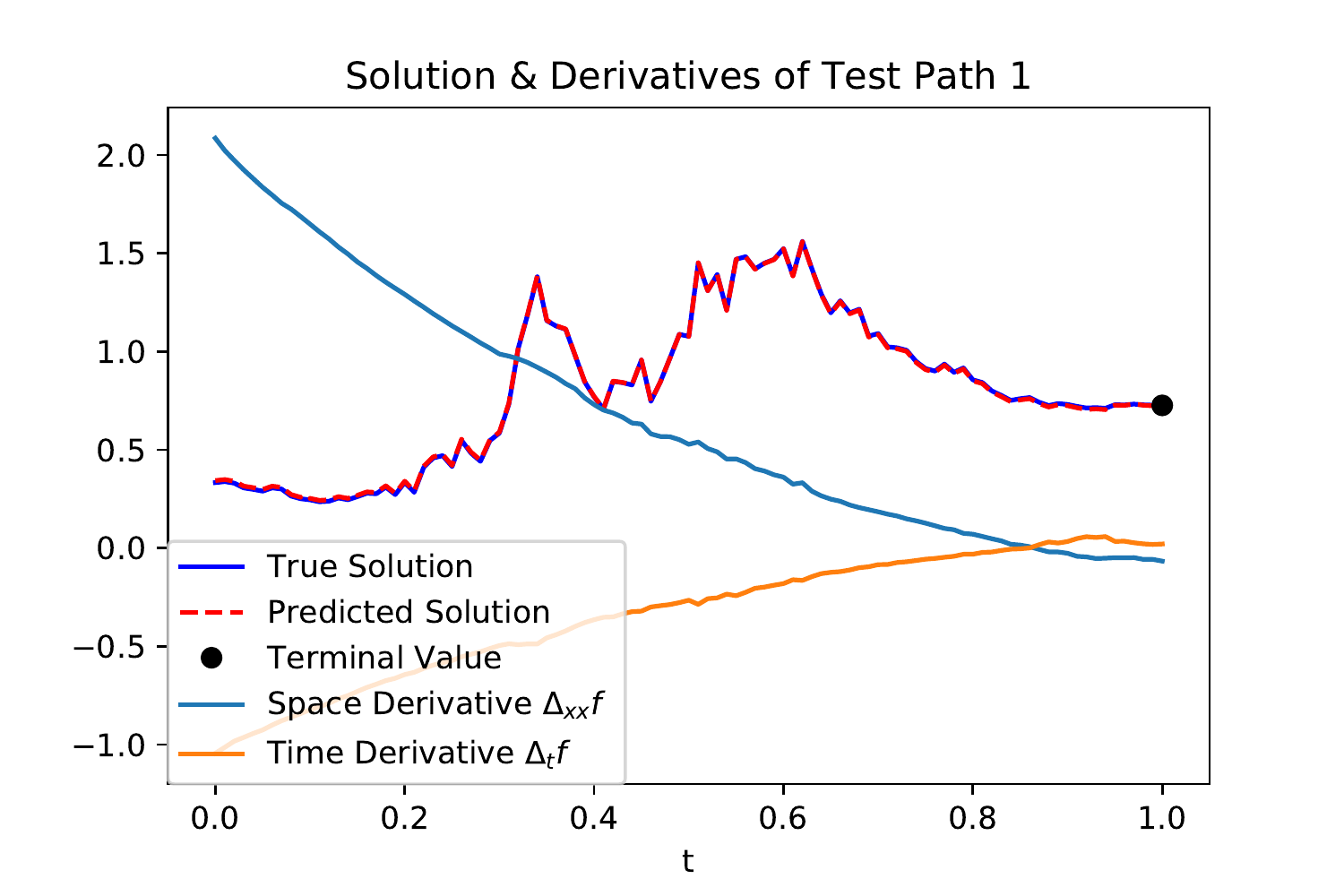}
  \includegraphics[width = 0.45\textwidth, height = 5cm]{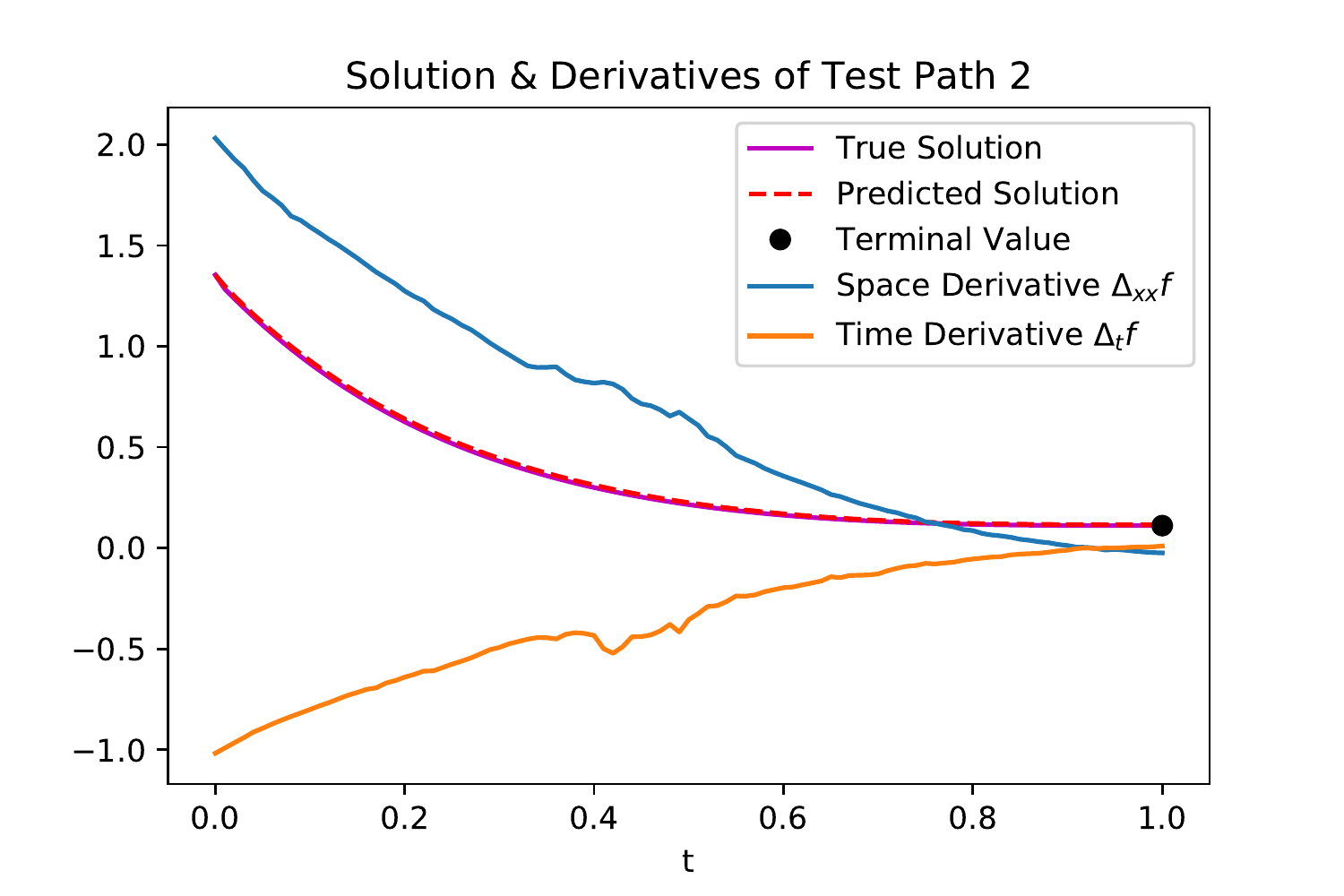}  \includegraphics[width = 0.45\textwidth, height = 5cm]{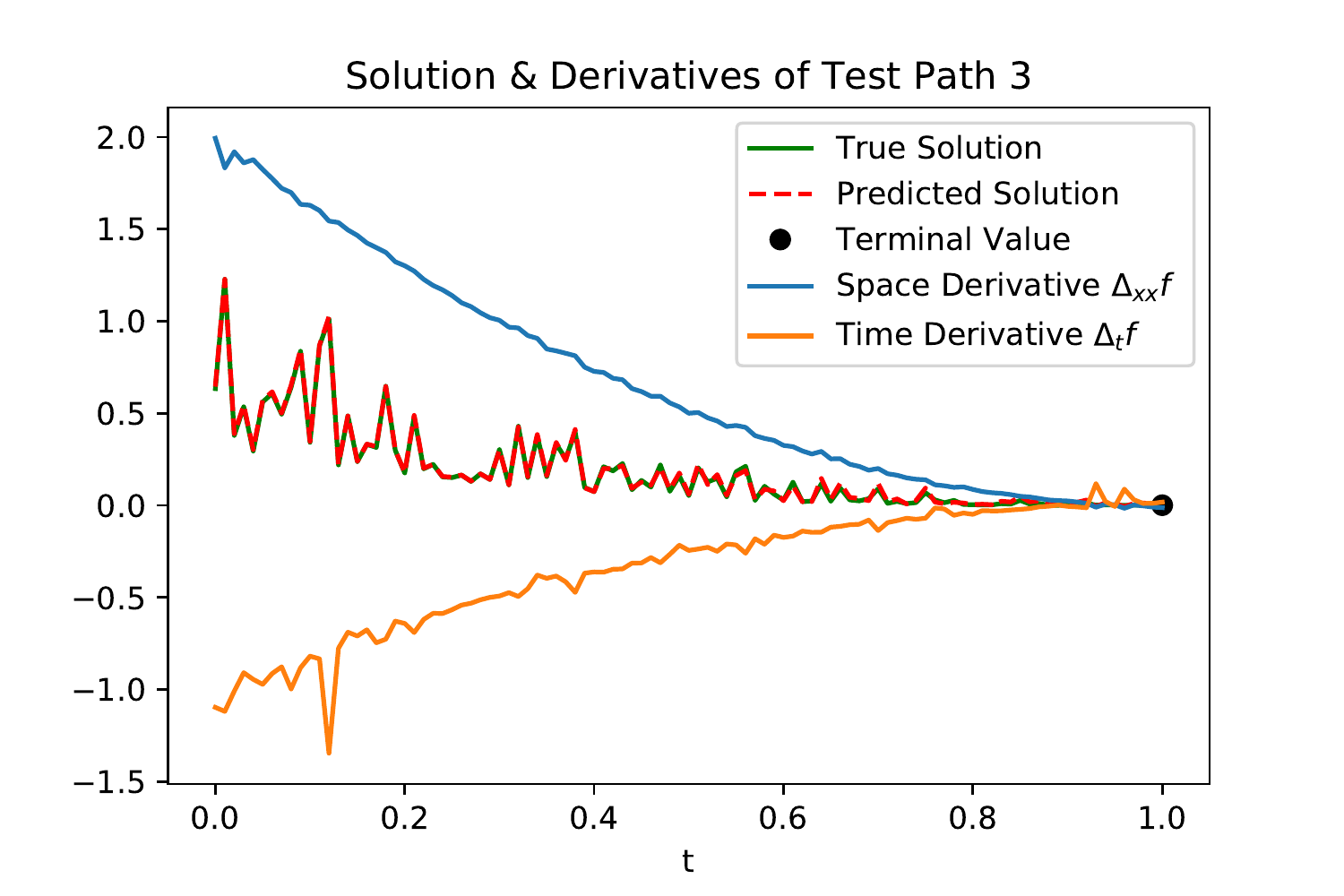}
  \caption{Three representative paths with corresponding solutions and functional derivatives for the quadratic running integral example.}
  \label{fig:quadratic_paths}
    \end{center}
\end{figure}

Similar to the above examples, training and testing loss is around $6 \times 10^{-5}$ after 10000 epochs as in Figure \ref{fig:quadratic_loss}, and three representative paths with their corresponding solution and derivatives are plotted in Figure \ref{fig:quadratic_paths}.

\subsubsection{High-Dimensional Example}

Here we will show that the methodology we have developed could handle high-dimensional PPDEs. Since this is not the main focus of the paper, the example serves more as an illustration of the method under this setting. Several numerical improvements could be introduce following the suggestions outlined in \cite{dgm}.

\begin{figure}[htbp]
\begin{center}
  \includegraphics[width = 0.45\textwidth, height = 5cm]{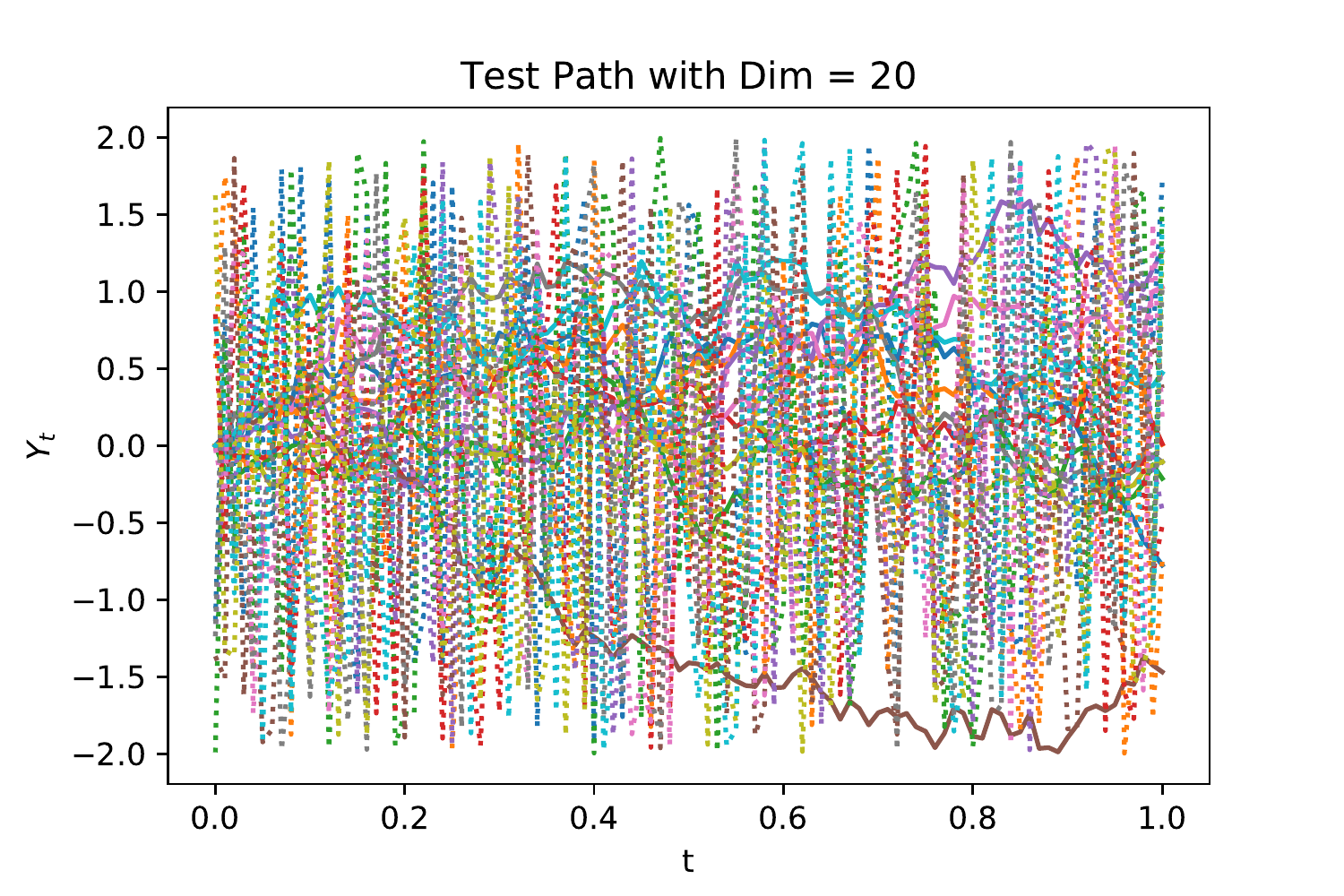}
  \includegraphics[width = 0.45\textwidth, height = 5cm]{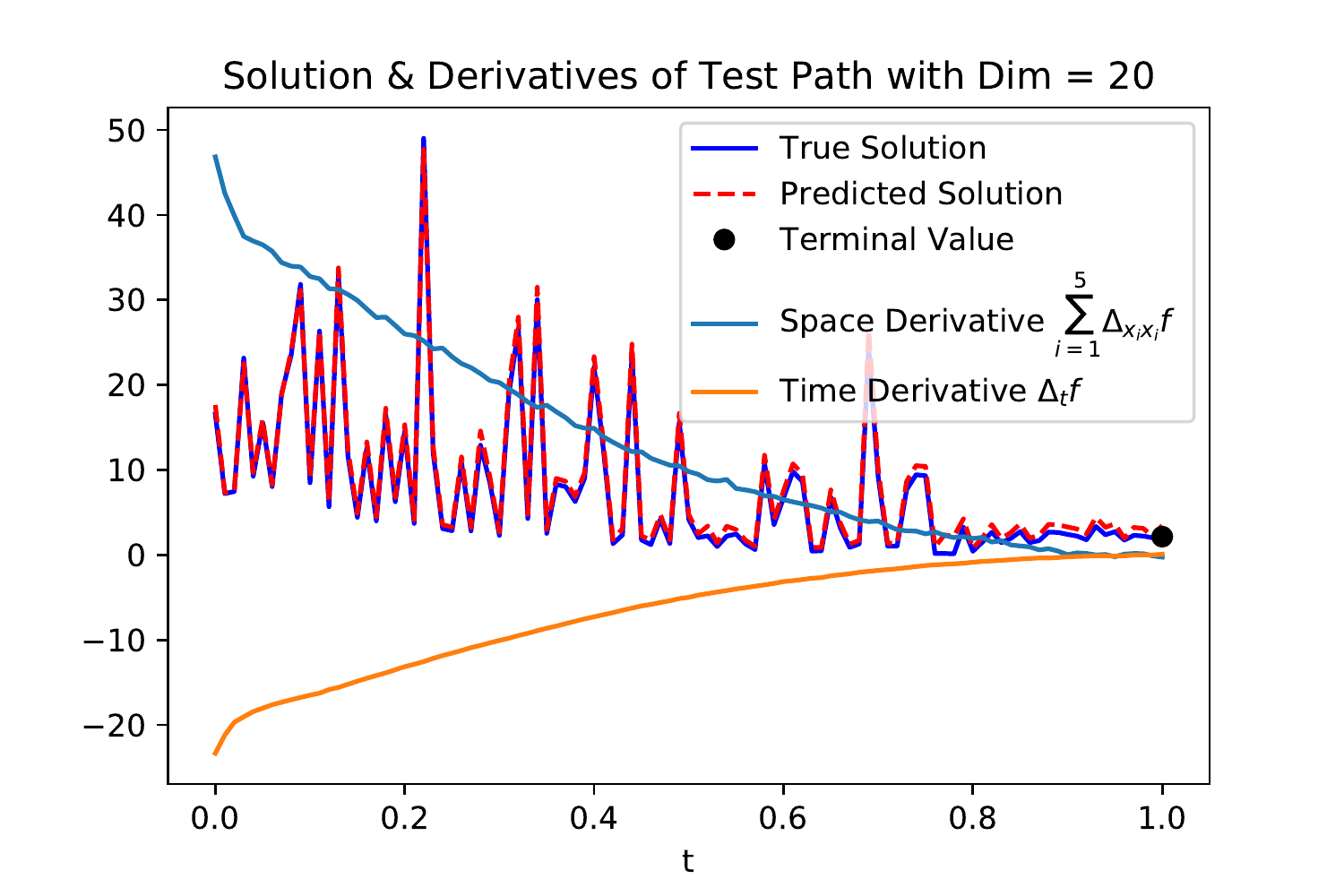}
  \caption{High-Dimensional Example}
  \label{fig:hd_paths}
    \end{center}
\end{figure}

Consider a $d$-dimensional Brownian motion $w_t = (w_t^{(1)}, \ldots, w_t^{(d)})$ and the payoff functional
$$g(Y_T) = \left(\ds \int_0^T \sum_{i=1}^d y_u^{(i)} du\right)^2.$$
It can be straightforwardly shown that
$$f(Y_t) = \left(\ds \int_0^t \sum_{i=1}^d y_u^{(i)} du\right)^2 + 2(T-t) \left(\sum_{i=1}^d y_t^{(i)} \right) \int_0^t \sum_{i=1}^d y_u^{(i)} du + \frac{d}{3}(T-t)^3.$$
Moreover, $f$ satisfies the PPDE:
\begin{align}
\begin{cases}
\Delta_t f(Y_t) + \displaystyle \frac{1}{2} \sum_{i=1}^d \Delta_{x_ix_i} f(Y_t) = 0,\\[10pt]
f(Y_T) = g(Y_T).
\end{cases}
\end{align}
 As an illustration, we choose $d = 20$, $T = 1$ and $\delta = 0.01$. Each dimension of the training paths are sampled from Brownian motions. For the numerical implementation, our algorithm works the same as in other aforementioned examples. On the left of Figure \ref{fig:hd_paths}, we plot the each dimension of a 20-dimensional path separately. The first 10 dimensions are sampled from standard Brownian motion paths. For the remaining 10 paths, points at each time step are sample from an uniform distribution between -2 and 2. The right plot in Figure \ref{fig:hd_paths} compares the true solution and the solution predicted, which are similar.. Time and spatial derivatives can also found on the right plot of Figure \ref{fig:hd_paths}.

\subsubsection{Hitting Time of the Final Value}

Consider the final functional
$$g(Y_T) = \inf\{t \in [0,T] \ ; \ y_t = y_T\},$$
which is the hitting time of the final value of the path $Y_T$. This example presents a stronger type of path-dependence than the ones presented so far. The value of $f(Y_0)$ can be found in closed form. Indeed, notice that in the Brownian case,
\begin{align*}
f(Y_0) &=  \bE[\inf\{t \in [0,T] \ ; \ y_0 + w_t = y_0 + w_T\}]\\
&= \bE[\inf\{t \in [0,T] \ ; \ w_t = w_T\}] = \int_\bR \bE[T_x] f_T(x)dx = 2\int_0^{+\infty} \bE[T_x] f_T(x)dx,
\end{align*}
where $T_x$ is the hitting time of the value $x$ of a Brownian bridge from $0$ to $x$ and $f_T$ is the probability density of $w_T$. Fixing $T=1$ and $x>0$, one might show that the probability density if $T_x$ is given by
$$f_{T_x}(t) = \frac{x}{\sqrt{2\pi t^3(1-t)}} e^{- \frac{1-t}{2t}x^2}.$$
We might then compute
\begin{align*}
\bE[T_x] &= \frac{x}{\sqrt{2\pi}} \int_0^1 t \frac{1}{\sqrt{t^3(1-t)}} e^{- \frac{1-t}{2t}x^2}dt\\
&= \frac{x}{\sqrt{2\pi}} \int_0^1\frac{1}{\sqrt{t(1-t)}} e^{- \frac{1-t}{2t}x^2}dt \\
&=\frac{x}{\sqrt{2\pi}} \pi e^{x^2/2} \mbox{erfc}(x/\sqrt{2}) = x \sqrt{\frac{\pi}{2}} e^{x^2/2} \mbox{erfc}(x/\sqrt{2}).
\end{align*}
Therefore,
\begin{align*}
f(Y_0) &= 2\int_0^{+\infty} x \sqrt{\frac{\pi}{2}} e^{x^2/2} \mbox{erfc}(x/\sqrt{2}) f_T(x)dx\\
&= 2\int_0^{+\infty} x \sqrt{\frac{\pi}{2}} e^{x^2/2} \mbox{erfc}(x/\sqrt{2}) \frac{1}{\sqrt{2\pi}} e^{-x^2/2}dx\\
&= \int_0^{+\infty} x \mbox{erfc}(x/\sqrt{2}) dx = \frac{1}{2}.
\end{align*}

% \com{we could run our code with the $g$ above. Notice that we would have to approximate the hitting time. For a given path, we need to find the first interval $[t_n, t_{n+1}]$ where $y_{t_{n}} < y_T $ and $y_{t_{n+1}} > y_T $. Then we use $t_n$ as an approximation of the final condition. The PPDE is the Brownian one. Since I don't think we can fin $f(Y_t)$ for any $t$, we could plot $f(Y_0)$ for many $Y_0$ and plot an horizontal line in 1/2. What do you think?}

\begin{figure}[htbp]
\begin{center}
  \includegraphics[width = 0.45\textwidth, height = 5cm]{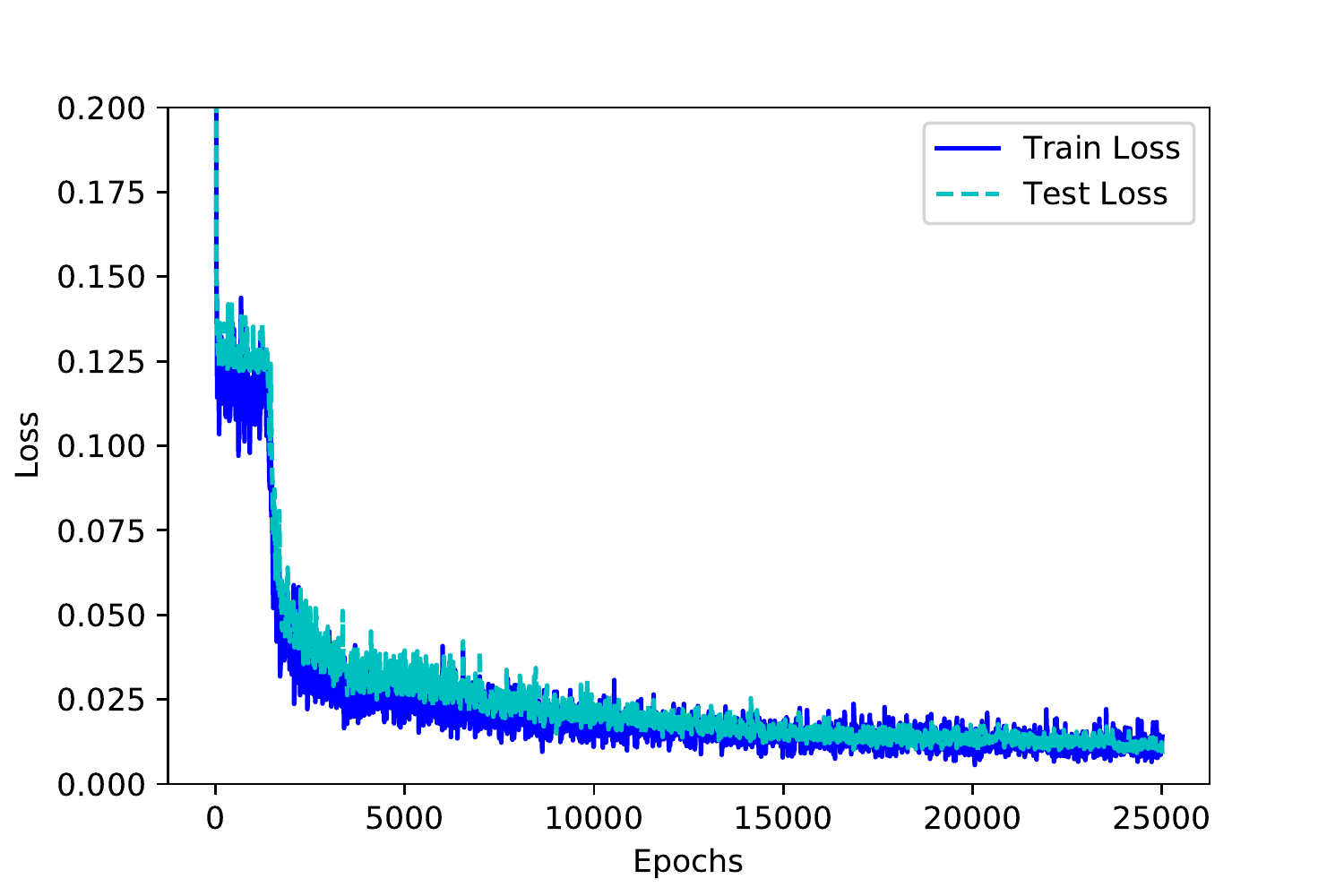}
  \includegraphics[width = 0.45\textwidth, height = 5cm]{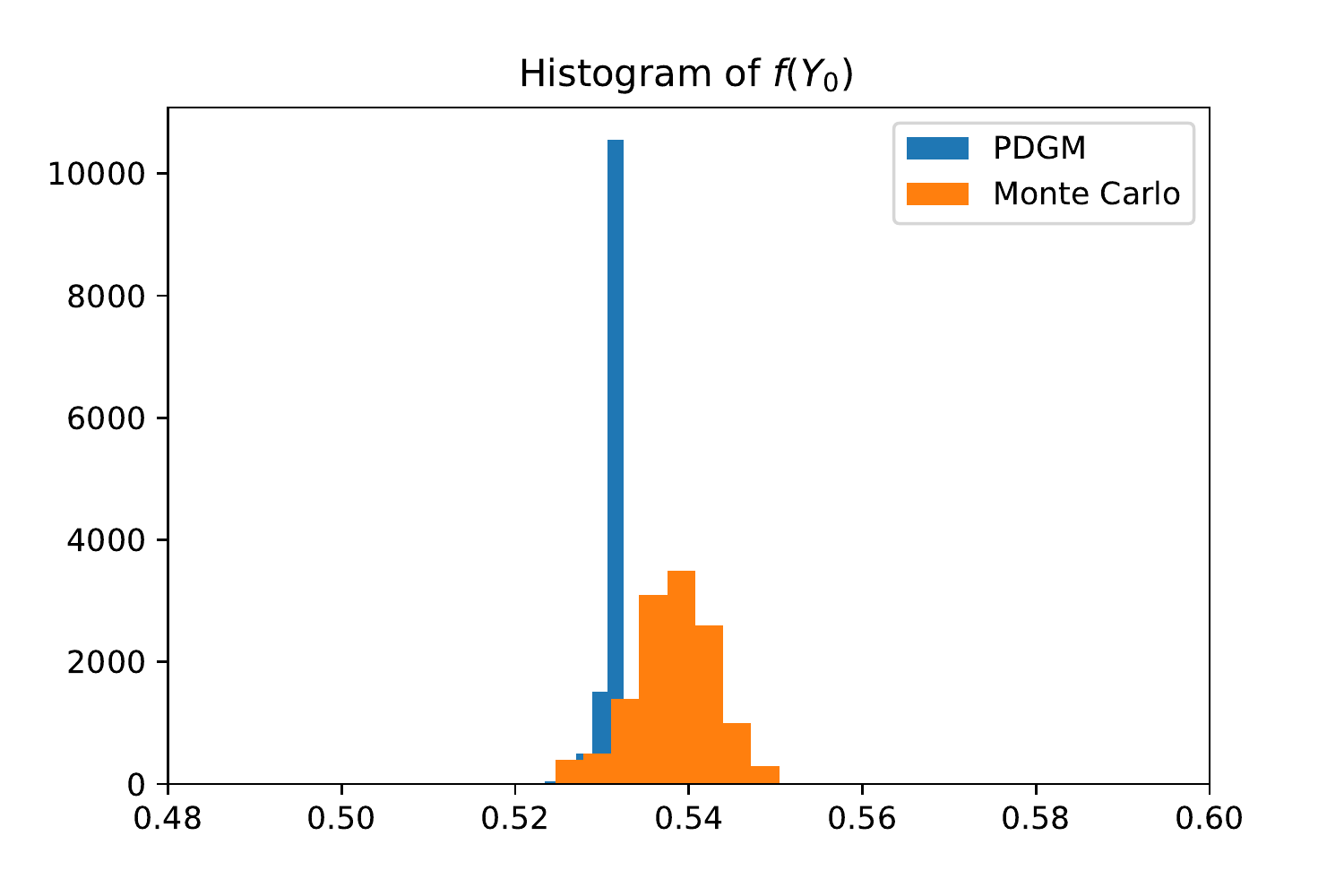}

  \caption{Train and test losses on the left, and the histogram comparison of $f(Y_0)$ between PDGM architecture and Monte Carlo method on the right for the hitting time example.}
  \label{fig:stopping_loss}
    \end{center}
\end{figure}

In this example, we use Brownian motion paths with starting points following a standard normal distribution as training paths. The discretization mesh size is again chosen to be 0.01. The training and testing losses approach to 0.01 after 25,000 epochs as shown on the left side of Figure \ref{fig:stopping_loss}. On the right of Figure \ref{fig:stopping_loss}, we compare $f(Y_0)$ of 12,800 paths between PDGM architecture and Monte Carlo method. In the Monte Carlo method, for each starting position $Y_0$, we simulate 5,000 Brownian motion paths in order to compute the sample mean. The results from Monte Carlo simulation have bell shape with mean around 0.54, but the results from our method are more concentrated at 0.53, and the difference is less than 1\%. This bias comes from discretization of time as also discussed in Remark \ref{rmk:barrier}.

\begin{figure}[htbp]
\begin{center}
  \includegraphics[width = 0.45\textwidth, height = 5cm]{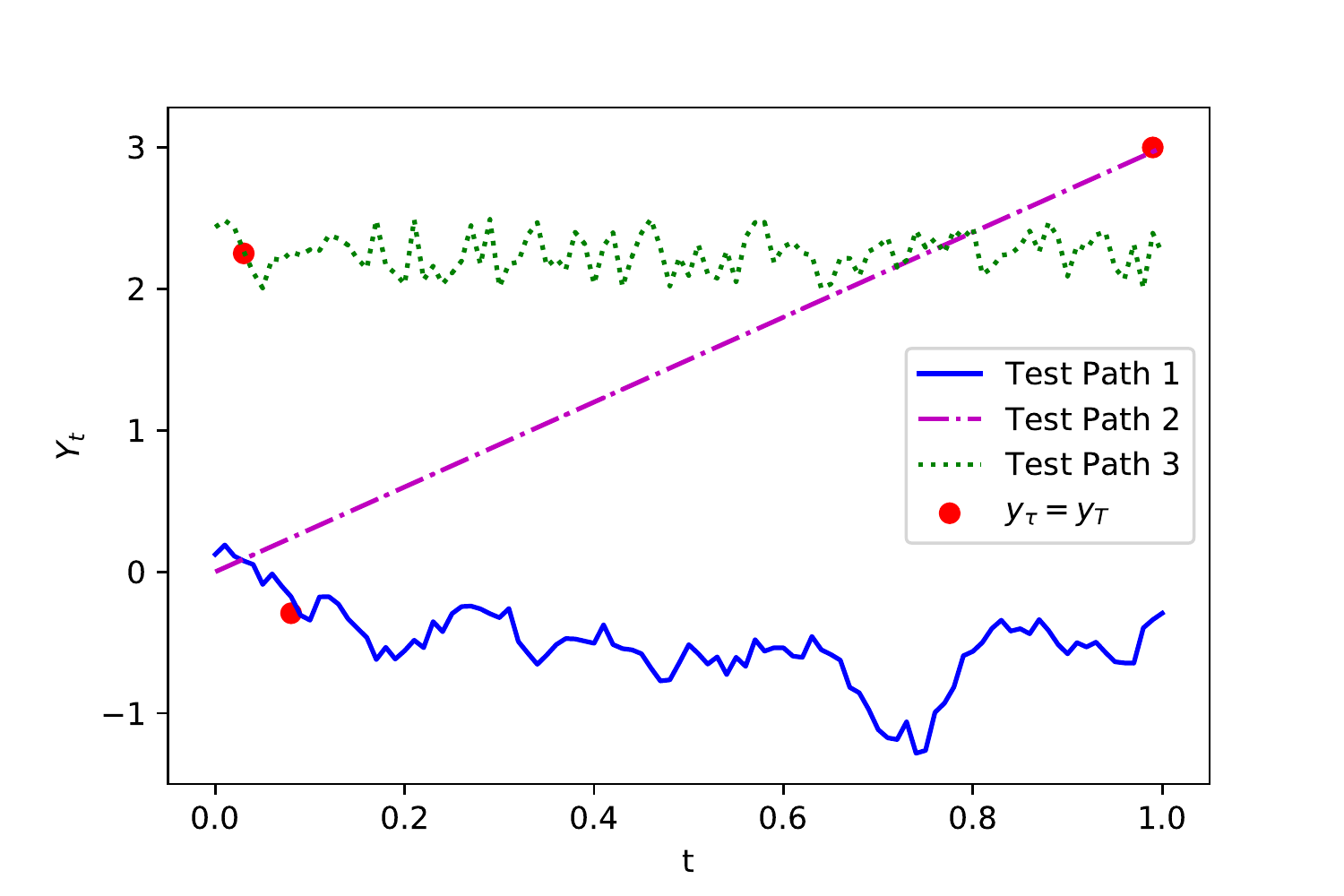}
  \includegraphics[width = 0.45\textwidth, height = 5cm]{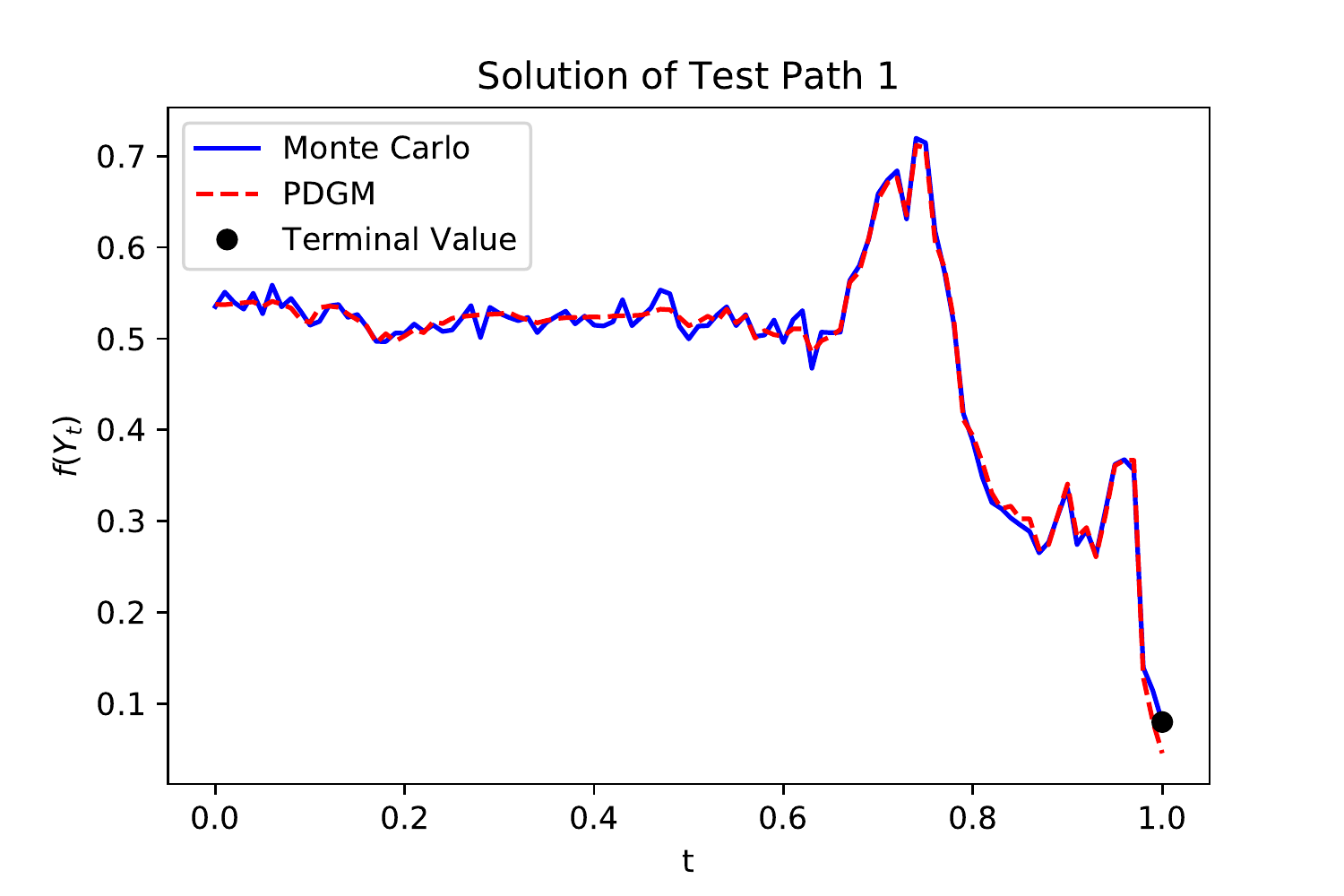}
  \includegraphics[width = 0.45\textwidth, height = 5cm]{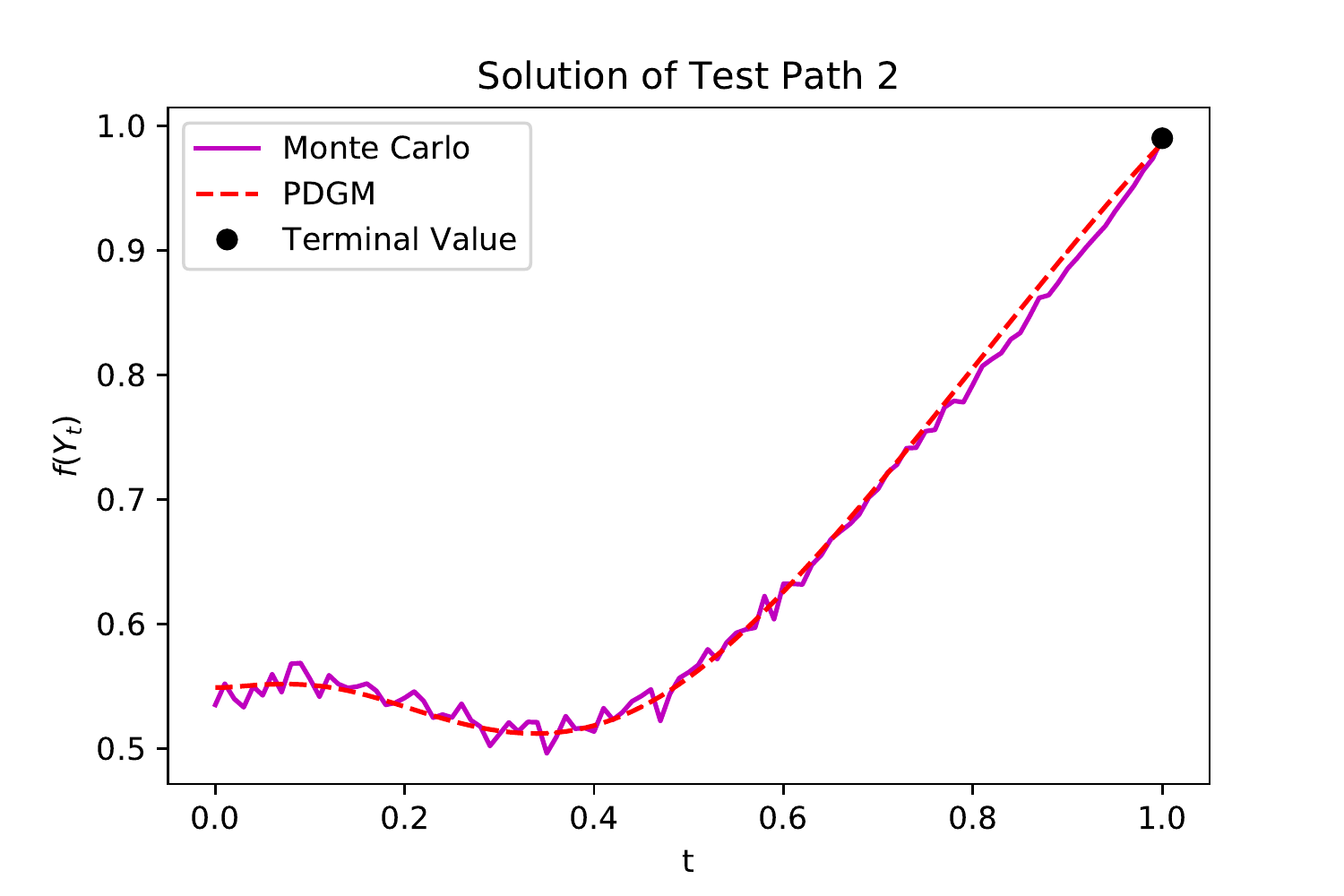}  \includegraphics[width = 0.45\textwidth, height = 5cm]{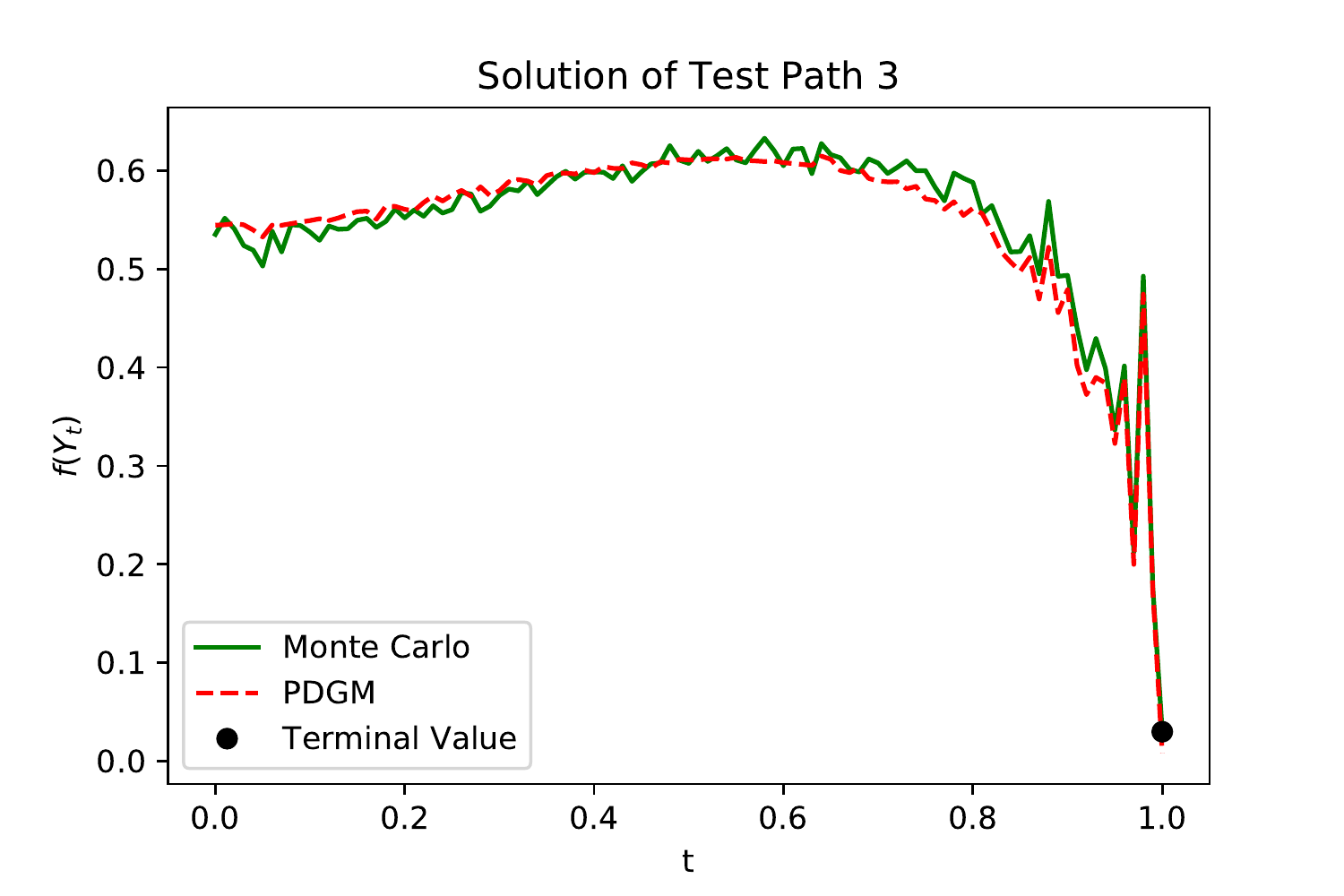}
  \caption{Three representative paths with corresponding solutions for the hitting time example.}
  \label{fig:stopping_paths}
    \end{center}
\end{figure}

Figure \ref{fig:stopping_paths} shows three representative test paths and their solutions from both our method and Monte Carlo simulation. Finding the solution from Monte Carlo simulation for an entire path is quite expensive. At each time step of a given path, we simulate 2,000 Brownian motion concatenated paths with the original path, i.e. we need to simulate 200,000 paths to approximate the pathwise solution. Test path 1 is a Brownian motion path starting at 0.1233; test path 2 is a straight line from 0 to 3; test path 3 is a realization of a sequence of i.i.d. uniform random variables between 2 and 2.5. The solutions are similar, and the solution from the PDGM algorithm tends to be smoother. Our algorithm after properly trained is able to compute path solutions for any path with similar range, however, the Monte Carlo simulation is only capable to compute the solution for each entire path at a time.

\subsection{Applications in Mathematical Finance}

Functional It\^o calculus, and hence PPDEs, was born from the necessity to deal path-dependent financial derivatives in the Mathematical Finance literature. In this section we will consider the classical Black--Scholes model, where the spot value follows a geometric Brownian Motion with constant parameters
$$dx_t = (r - q)x_tdt + \sigma x_t dw_t.$$
Under this model, the price of a general path-dependent financial derivative with maturity $T$ and payoff $g: \Lambda_T \longrightarrow \bR$ solves the PPDE
\begin{align}\label{eqn:PPDE_BS}
\begin{cases}
\Delta_t f(Y_t) + (r - q) y_t \Delta_x f(Y_t) + \displaystyle \frac{1}{2}\sigma^2 y_t^2 \Delta_{xx} f(Y_t) - r f(Y_t) = 0,\\[10pt]
f(Y_T) = g(Y_T),
\end{cases}
\end{align}
for any continuous path $Y_T$ taking positive values, see Remark \ref{rmk:stroock_varadhan}.

We will consider three examples (Geometric Asian, Lookback and Barrier options) where closed-form solutions are available. Moreover, we will consider one path-dependent example with the process $x$ having stochastic volatility. Additionally, one could consider several other path-dependent, exotic derivatives with different dynamics. The PDGM could be applied similarly to these cases requiring possibly more computational power or time.

In the examples below, we consider the PPDE (\ref{eqn:PPDE_BS}) with parameters $x_0 = 1$, $r = 0.03$, $q = 0.01$, $\sigma  = 1$ and $T = 1$. The payoff functional $g$ will vary for each case. We use the geometric Brownian motion with these parameters as training paths for the algorithm with number of batch size of $M = 128$ paths and $N = 100$ time steps. Moreover, there are 128 units in a single layer LSTM cell, and the deep feed-forward neural network consists of three hidden layers with 128 neurons in each. 

For geometric Asian option and the lookback option, Figures \ref{fig:asian} and \ref{fig:lookback} show three representative test paths with corresponding closed-form solutions. Path 1 is a geometric Brownian motion path with the same parameters as above. Path 2 is the smooth path $y_t = (1-t)^2$ for $t \in [0, 1]$. Path 3 is a realization of a sequence of i.i.d. uniform random variables between 1 and 3. 
For the barrier option, we consider a down and out option. Figures \ref{fig:barrier} shows three representative test paths (different from the ones above and defined in Section \ref{sec:barrier}) with corresponding closed-form solutions.

The solutions predicted from our algorithm are approximately the same as the true solutions. Our algorithm is able to predict solutions for any given paths in the domain of training paths regardless of the shape of a path. Furthermore, the losses after 15,000 epochs in these examples, together with the MSE when closed-form solution is available, are given in the table below.

\begin{table}[htbp]
\centering
\begin{tabular}{lccc}
Example                    & Train Loss         & Test Loss       & MSE  \\
\hline
Geometric Asian          & $5.8 \times 10^{-5}$ & $5.4 \times 10^{-5}$  &  $9.9 \times 10^{-5}$ \\
Barrier    & $4.7 \times 10^{-3}$ & $1.2 \times 10^{-2}$ &  $6.9 \times 10^{-3}$\\
Lookback & $1.3 \times 10^{-3}$ & $1.6 \times 10^{-3}$ & $8.8 \times 10^{-4}$ \\
Heston model ($K=0.4, T = 1$) & $1.4 \times 10^{-2}$ & $1.2 \times 10^{-2}$  & -- --\\
Nonlinear & $6 \times 10^{-6}$ & $6 \times 10^{-6}$ & $3.6 \times 10^{-6}$\\
\hline
\end{tabular}
\vspace{1mm}
\caption{Train and test losses for the Mathematical Finance examples}
\end{table}

\subsubsection{Geometric Asian Option}\label{sec:geo_asian}

\begin{figure}[htbp]
\begin{center}
  \includegraphics[width = 0.45\textwidth, height = 5cm]{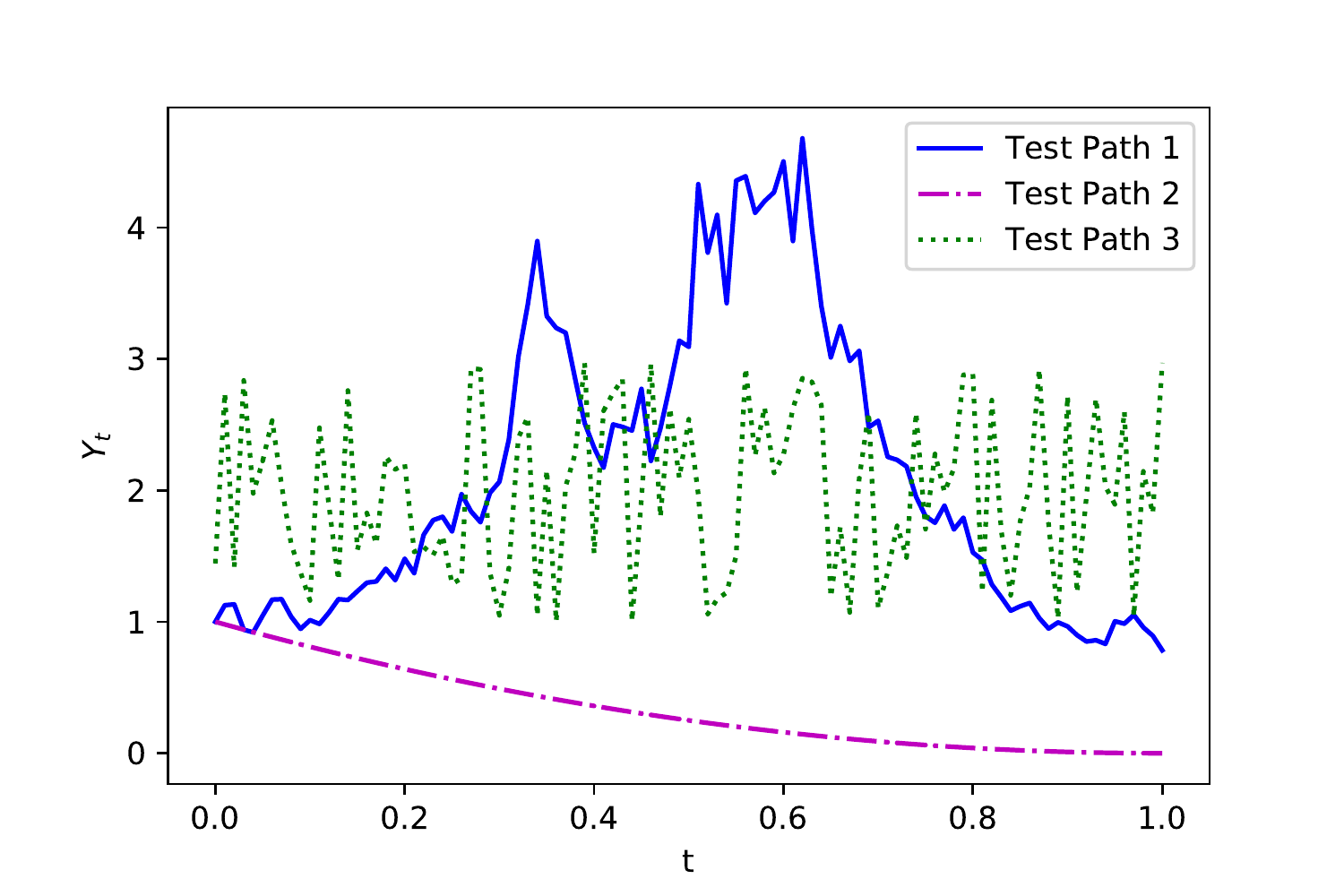}
  \includegraphics[width = 0.45\textwidth, height = 5cm]{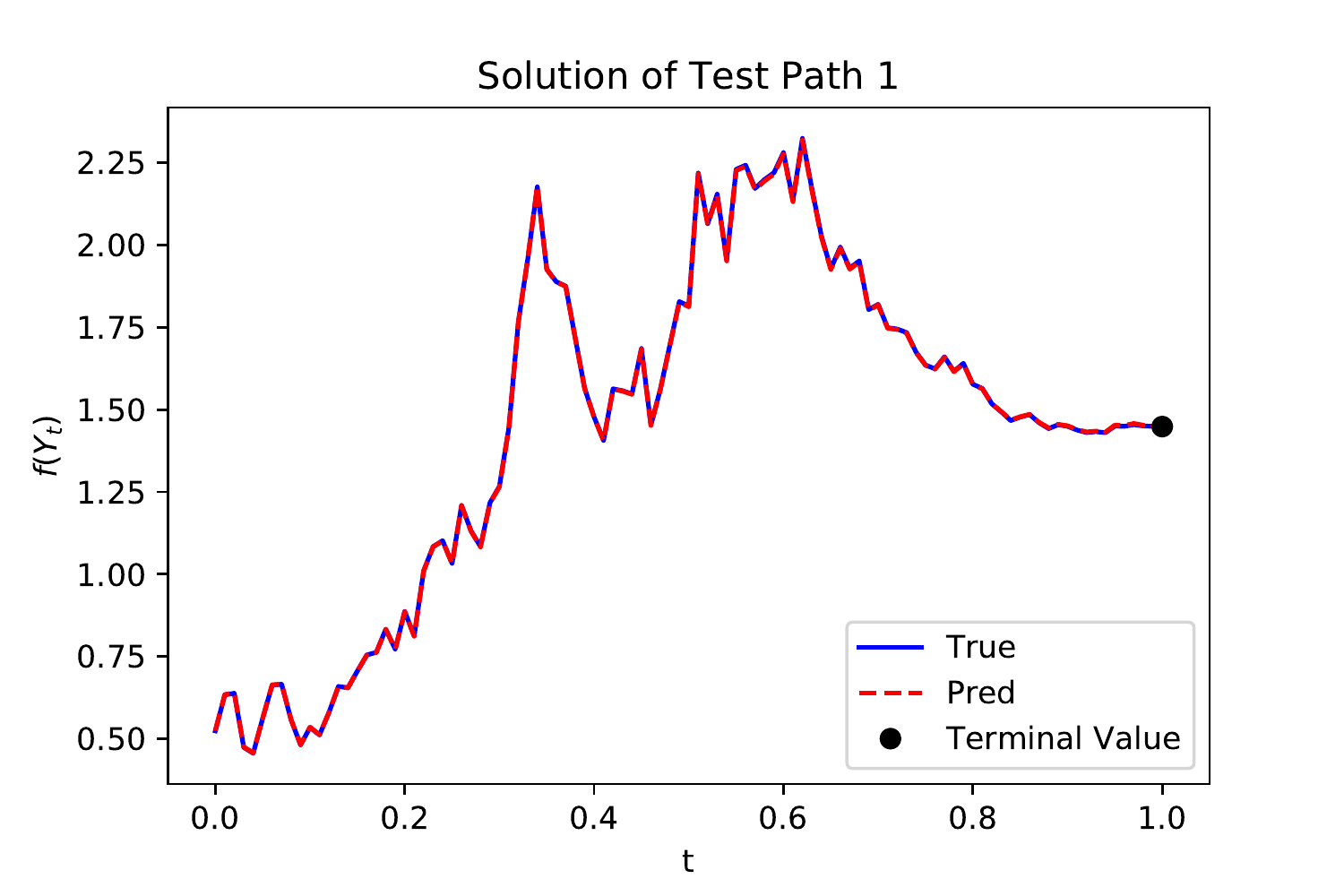}
  \includegraphics[width = 0.45\textwidth, height = 5cm]{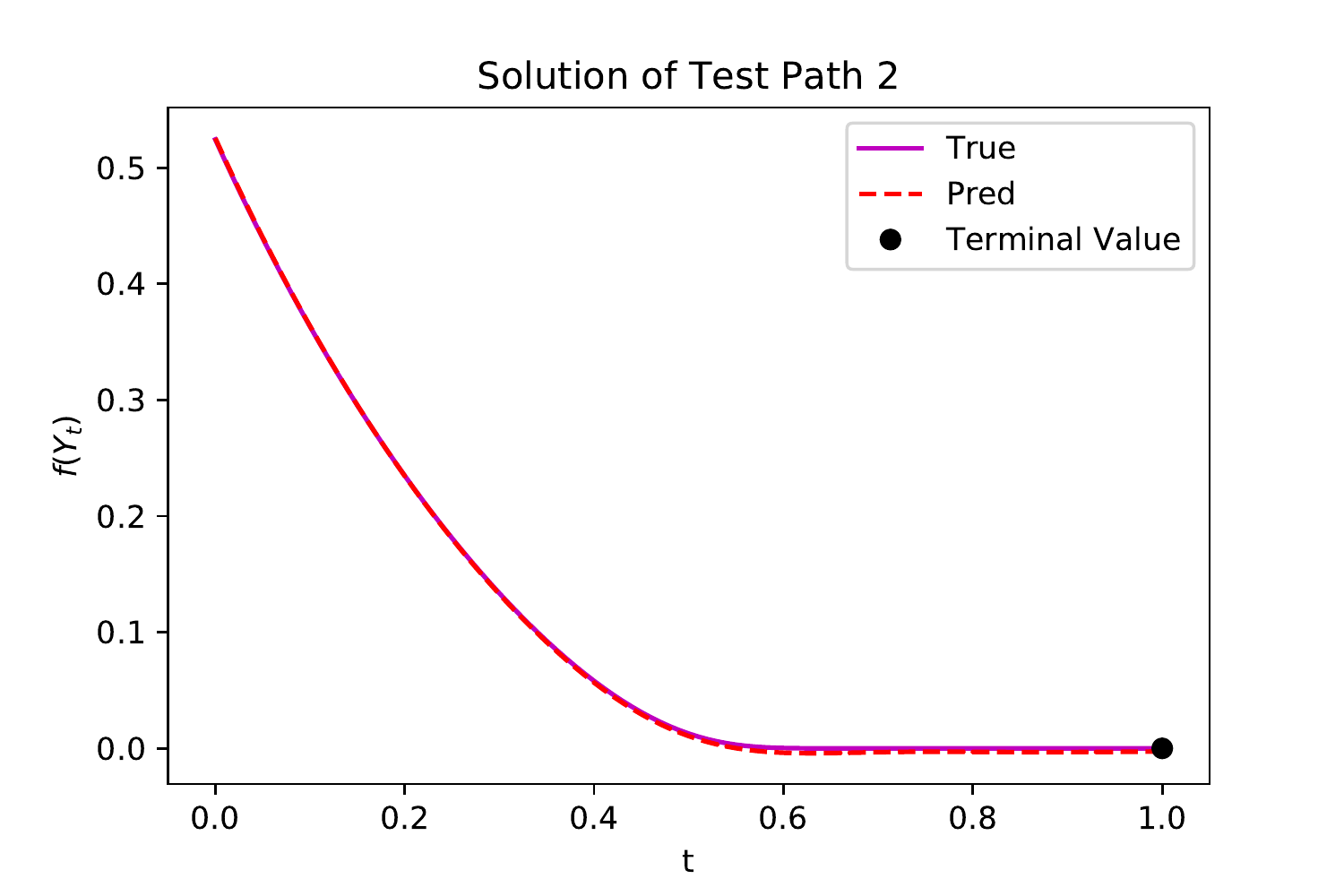}  \includegraphics[width = 0.45\textwidth, height = 5cm]{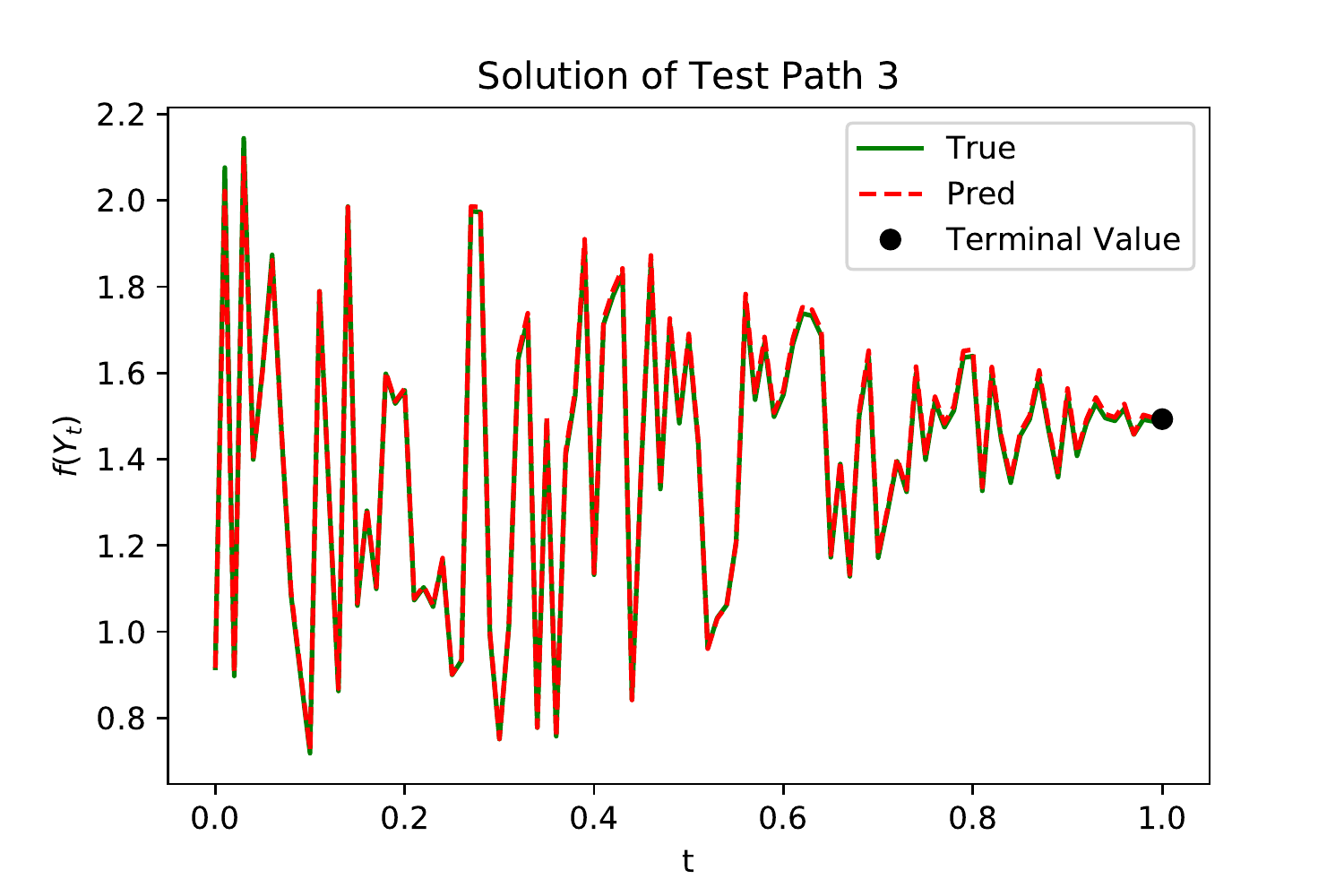}
  \caption{Three representative paths with corresponding solutions for the geometric Asian option.}
  \label{fig:asian}
    \end{center}
\end{figure}

The case of continuously-monitored geometric Asian options with fixed strike is determined by the payoff
$$g(Y_T) = \left(\exp\left\{\frac{1}{T} \int_0^T \log y_t dt \right\} - K\right)^+,$$
where $x^+$ is the positive part of $x$ and $K > 0$ is called the strike.  A closed-form solution is available in this case:
$$f(Y_t) = e^{-r(T-t)} \left(G_t^{t/T}y_t^{1 - t/T} e^{\bar{\mu} + \bar{\sigma}^2/2} \Phi(d_1) - K\Phi(d_2) \right),$$
where $\Phi$ is the cumulative distribution function of the standard normal,
\begin{equation}
\begin{aligned}
G_t &= \exp\left\{\frac{1}{t} \int_0^t \log y_u du \right\}\\
\bar{\mu} &= \left(r - q - \frac{\sigma^2}{2}\right)\frac{1}{2T}(T-t)^2,\\
\bar{\sigma} &= \frac{\sigma}{T} \sqrt{\frac{1}{3}(T-t)^3},\\
d_2 &= \frac{(t/T) \log G_t + \left(1 - t/T\right) \log y_t + \bar{\mu} - \log K}{\bar{\sigma}},\\
d_1 &= d_2 + \bar{\sigma}.
\end{aligned}
\end{equation}
In the numerical examples below, we fix the strike at $K = 0.4$. Three representative test paths with corresponding closed-form solutions are given in Figure \ref{fig:asian}.

\subsubsection{Lookback option}

\begin{figure}[htbp]
\begin{center}
  \includegraphics[width = 0.45\textwidth, height = 5cm]{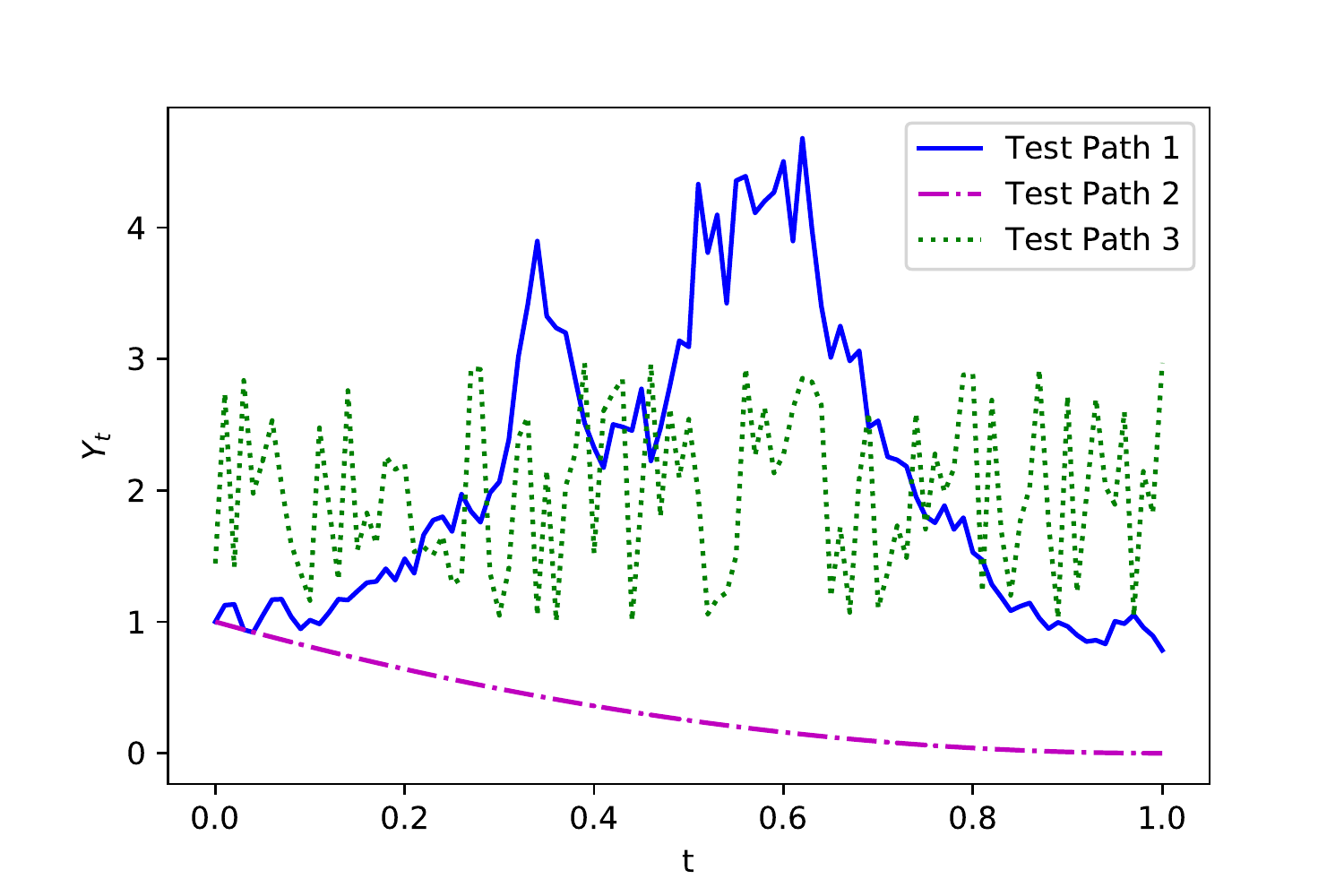}
  \includegraphics[width = 0.45\textwidth, height = 5cm]{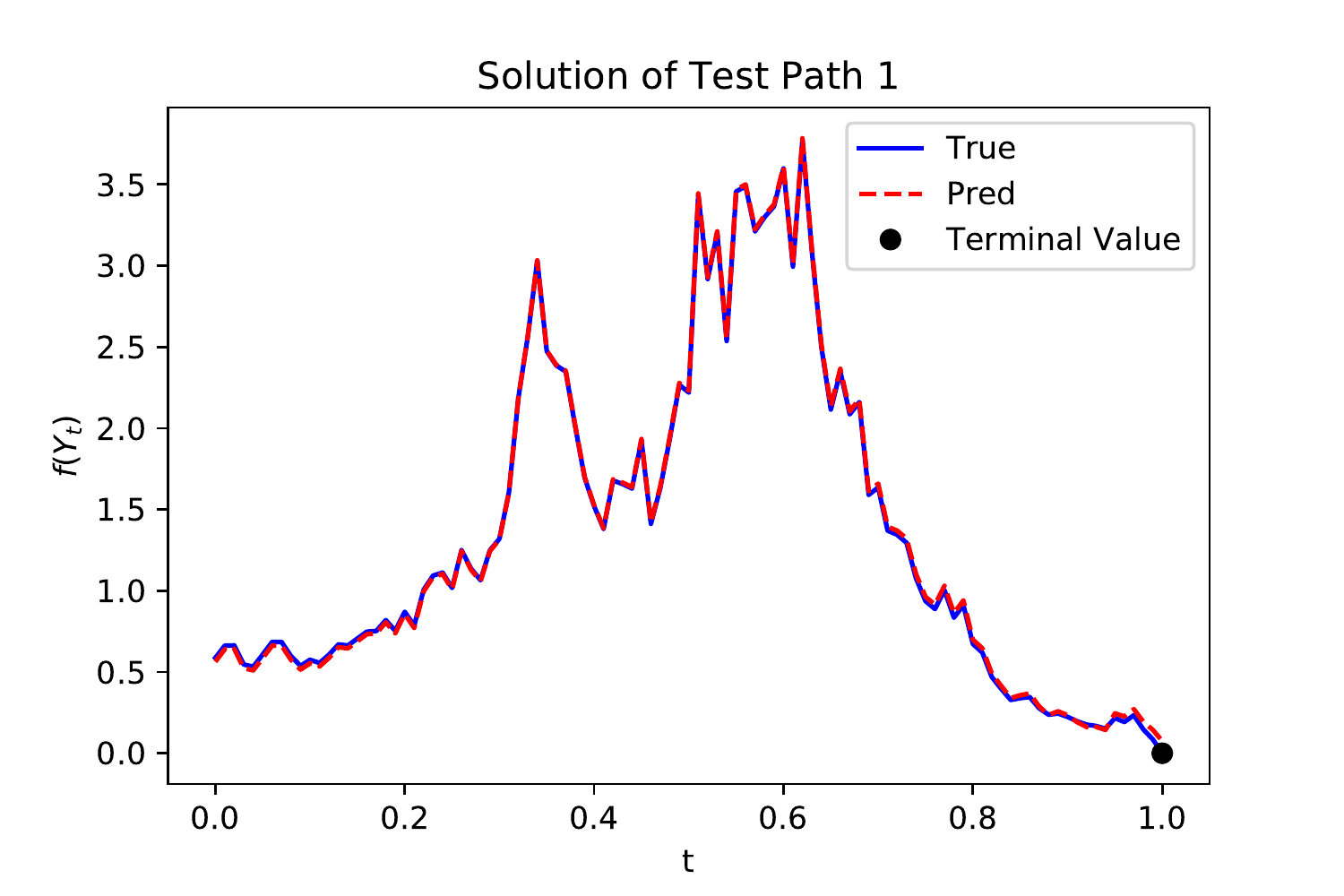}
  \includegraphics[width = 0.45\textwidth, height = 5cm]{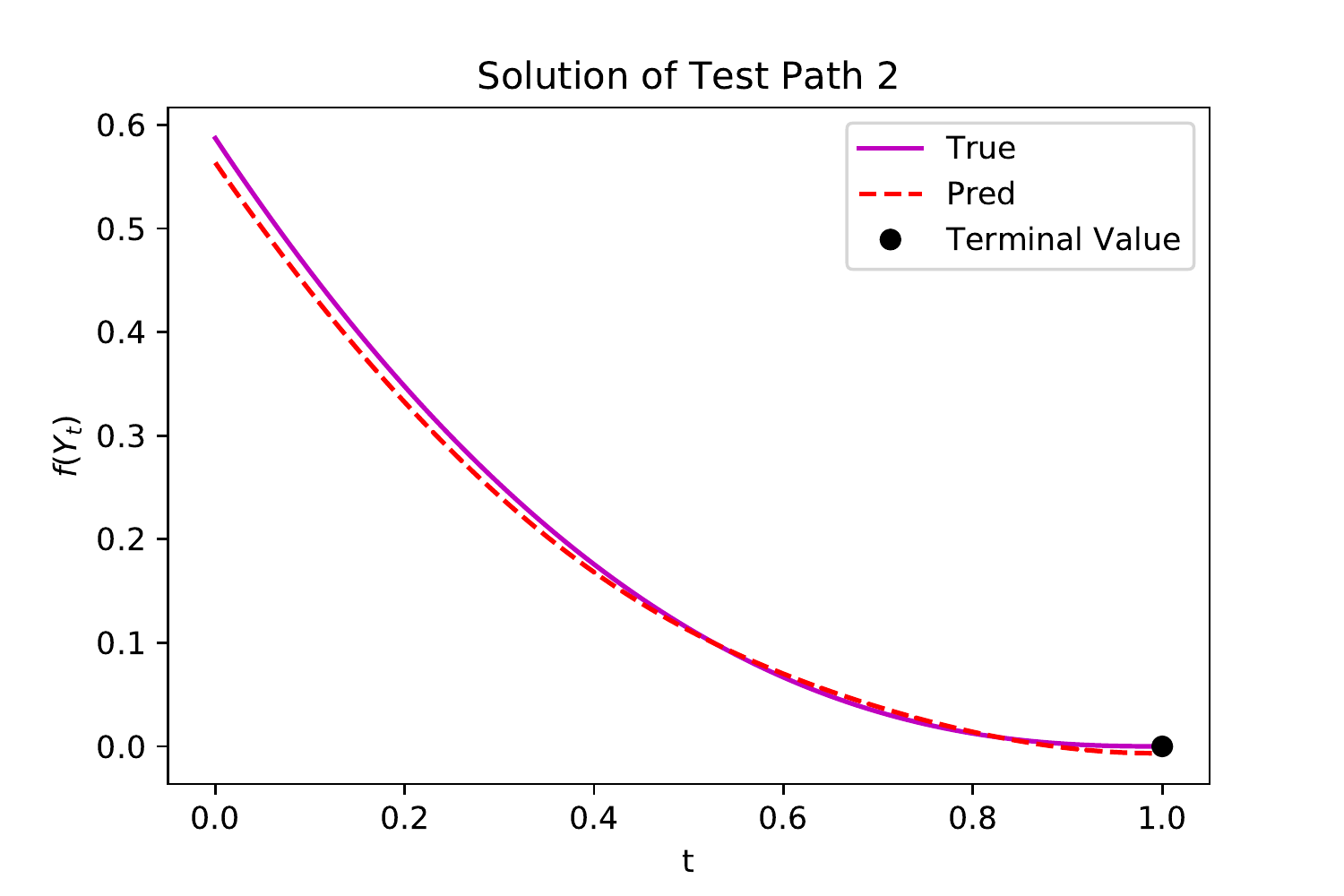}  \includegraphics[width = 0.45\textwidth, height = 5cm]{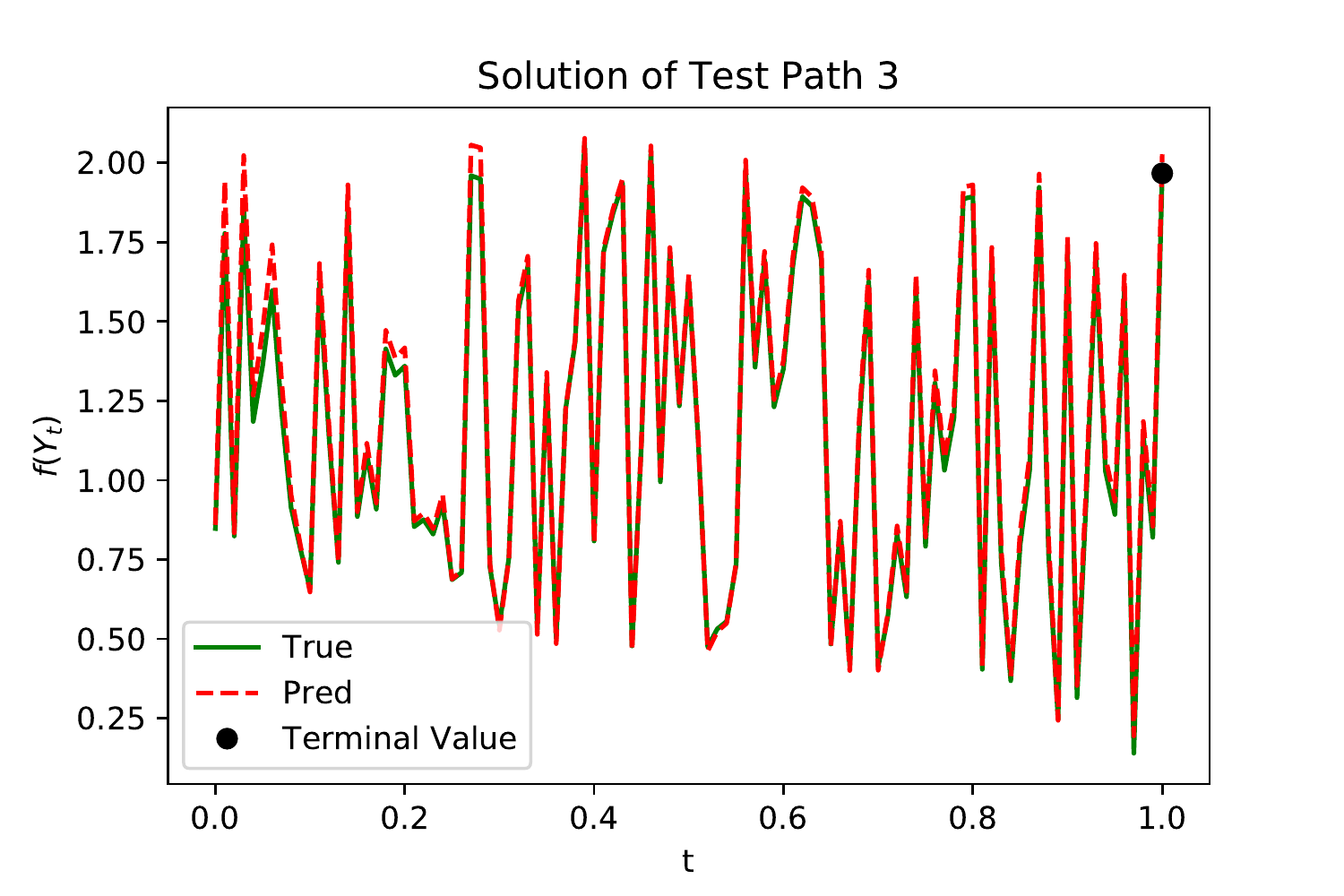}
  \caption{Three representative paths with corresponding solutions for the lookback option.}
  \label{fig:lookback}
    \end{center}
\end{figure}

A lookback call option with floating strike is given by the payoff
$$g(Y_T) =  y_T - \inf_{0 \leq t \leq T} y_t.$$
If we denote $m_t = \inf_{0 \leq u \leq t} y_u$, the closed-form solution, assuming $q=0$, for the price of this option can be written as
\begin{align*}
f(Y_t) = y_t \Phi(a_1) - m_t e^{-r(T-t)} \Phi(a_2) - y_t \frac{\sigma^2}{2r} \left( \Phi(-a_1) - e^{-r(T-t)} \left(\frac{m_t}{y_t}\right)^{2r/\sigma^2} \Phi(-a_3)\right),   
\end{align*}
where
\begin{align*}
a_1 = \frac{\log(y_t/m_t) + (r + \sigma^2/2)(T-t)}{\sigma \sqrt{T-t}}, \quad a_2 = a_1 - \sigma \sqrt{T-t} \mbox{ and } a_3 = a_1 - \frac{2r}{\sigma} \sqrt{T-t}.
\end{align*}
 Figures \ref{fig:lookback} shows three representative test paths with corresponding closed-form solutions. Our algorithm shows a promising result.

\subsubsection{Barrier option}\label{sec:barrier}

There are several types of barrier options, see for instance, \cite{barrier} and \cite{multiscale_fouque_new_book}. Here, we will focus on the case of down-and-out call options. More precisely, the option becomes worthless whether the spot value crosses a down barrier $B < S_0$. Otherwise, the payoff is a call with strike $K \geq B$. The payoff functional can then be written as
$$g(Y_T) = (y_T - K)^+ 1_{\left\{\substack{\inf y_t \\ \scriptscriptstyle{0 \leq t \leq T}} \ > \ B \right\}}.$$
In addition, the solution should also satisfy the boundary condition $f(Y_t) = 0$ if the barrier $B$ was crossed by the path $Y_t$. A closed-form solution is available:
\begin{align}
f(Y_t) = \begin{cases}
f_{do}(y_t, T-t), \mbox{ if } \ds \inf_{0 \leq u \leq t} y_u > B,\\ \\
0, \mbox{ if } \ds \inf_{0 \leq u \leq t} y_u \leq B,
\end{cases}
\end{align}
where
$$f_{do}(y_t, T-t) = C_{BS}(y_t, T - t) - \left( \frac{y_t}{B}\right)^{1-\lambda} C_{BS}\left(\frac{B^2}{y_t}, T-t\right),$$
$C_{BS}(y_t, T-t)$ is the price of a call option with strike $K$ and maturity $T$ at $(t,y_t)$ and
$$\lambda = \frac{2(r-q)}{\sigma^2}.$$

Due to the fact that the option become valueless when the stock price crosses the barrier, we need to slightly modify the loss function in our algorithm. In this case, the loss for a given sample path $j$ at time $t_i$ is 
$$J^{(j)}_{t_i} (\theta) = 
\begin{cases}
|u(Y^{(j)}_{t_i};\theta) - 0| & \mbox{ if } \inf_{0 \leq i' \leq i} Y^{(j)}_{t_{i'}} < B,\\
\left(\Delta_{t} u(Y^{(j)}_{t_i};\theta) + \cL u(Y^{(j)}_{t_i};\theta)\right)^2 & \mbox{ otherwise.}\\
\end{cases}$$
The total loss is calculated as 
$$J_{N,M}(\theta) = \frac{1}{M} \frac{1}{N}\sum_{j=1}^M \sum_{i=0}^N J^{(j)}_{t_i} (\theta) +  \frac{1}{M}\sum_{j=1}^M  \left[\left(u(Y^{(j)}_{t_N}; \theta) - g(Y_{t_N}^{(j)})  1_{\left\{\substack{\inf y_{t_i} \\ \scriptscriptstyle{0 \leq i \leq N}} \ > \ B \right\}} \right)^2 + |u(Y^{(j)}_{t_N}; \theta) - 0| 1_{\left\{\substack{\inf y_{t_i} \\ \scriptscriptstyle{0 \leq i \leq N}} \ < \ B \right\}} \right].$$
We then minimize the above loss objective using stochastic gradient descent algorithm and update parameter $\theta$. 

\begin{figure}[htbp]
\begin{center}
  \includegraphics[width = 0.45\textwidth, height = 5cm]{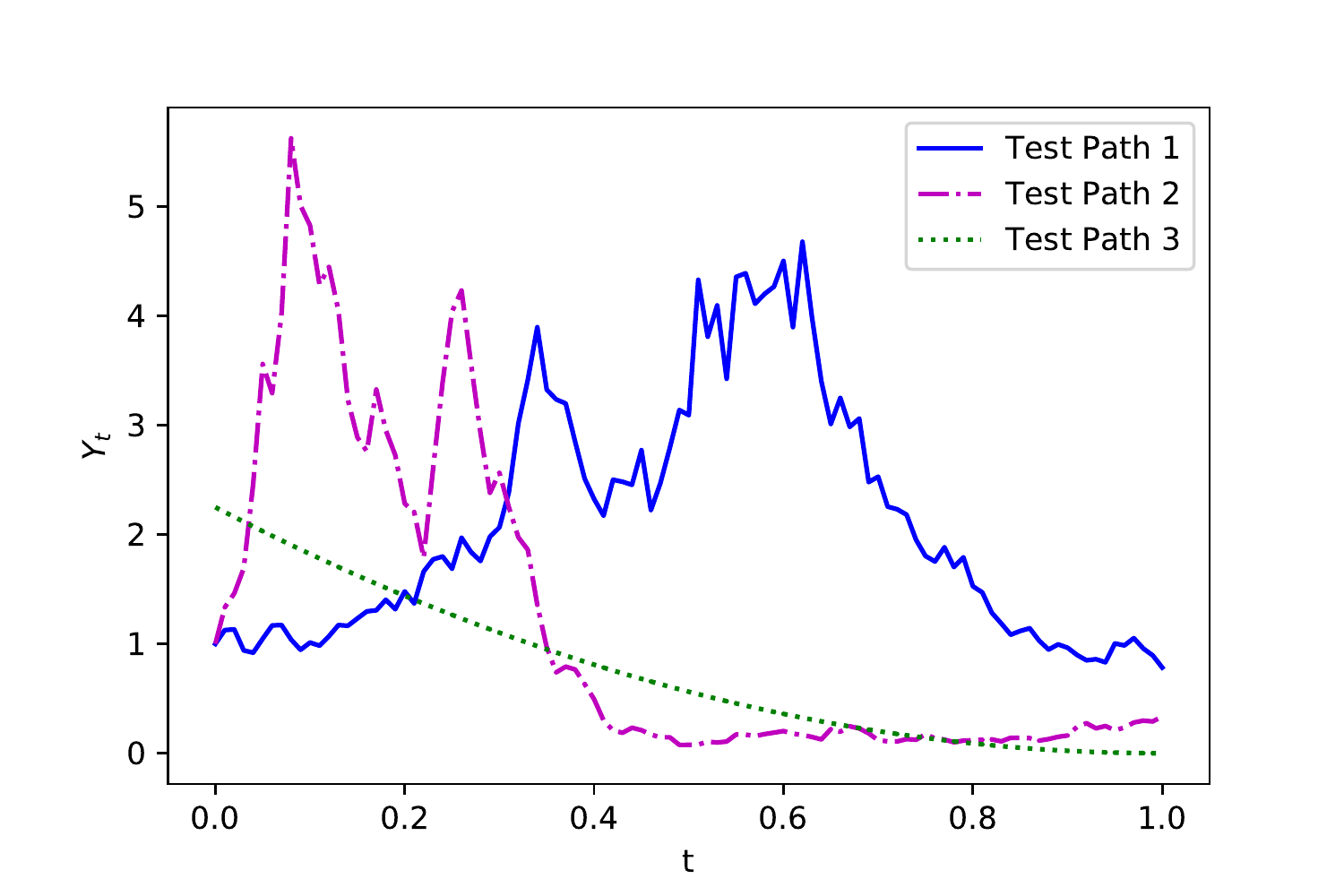}
  \includegraphics[width = 0.45\textwidth, height = 5cm]{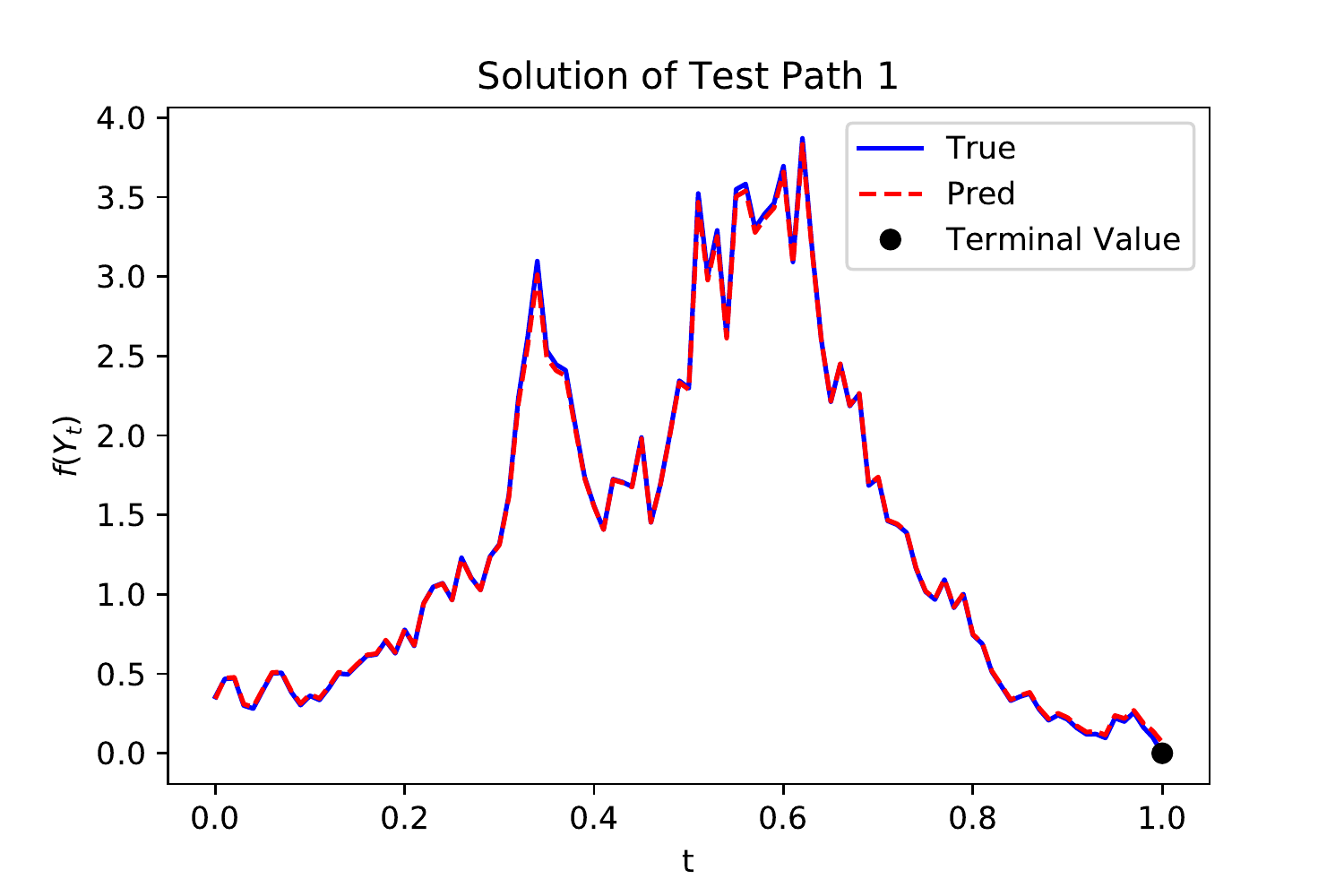}
  \includegraphics[width = 0.45\textwidth, height = 5cm]{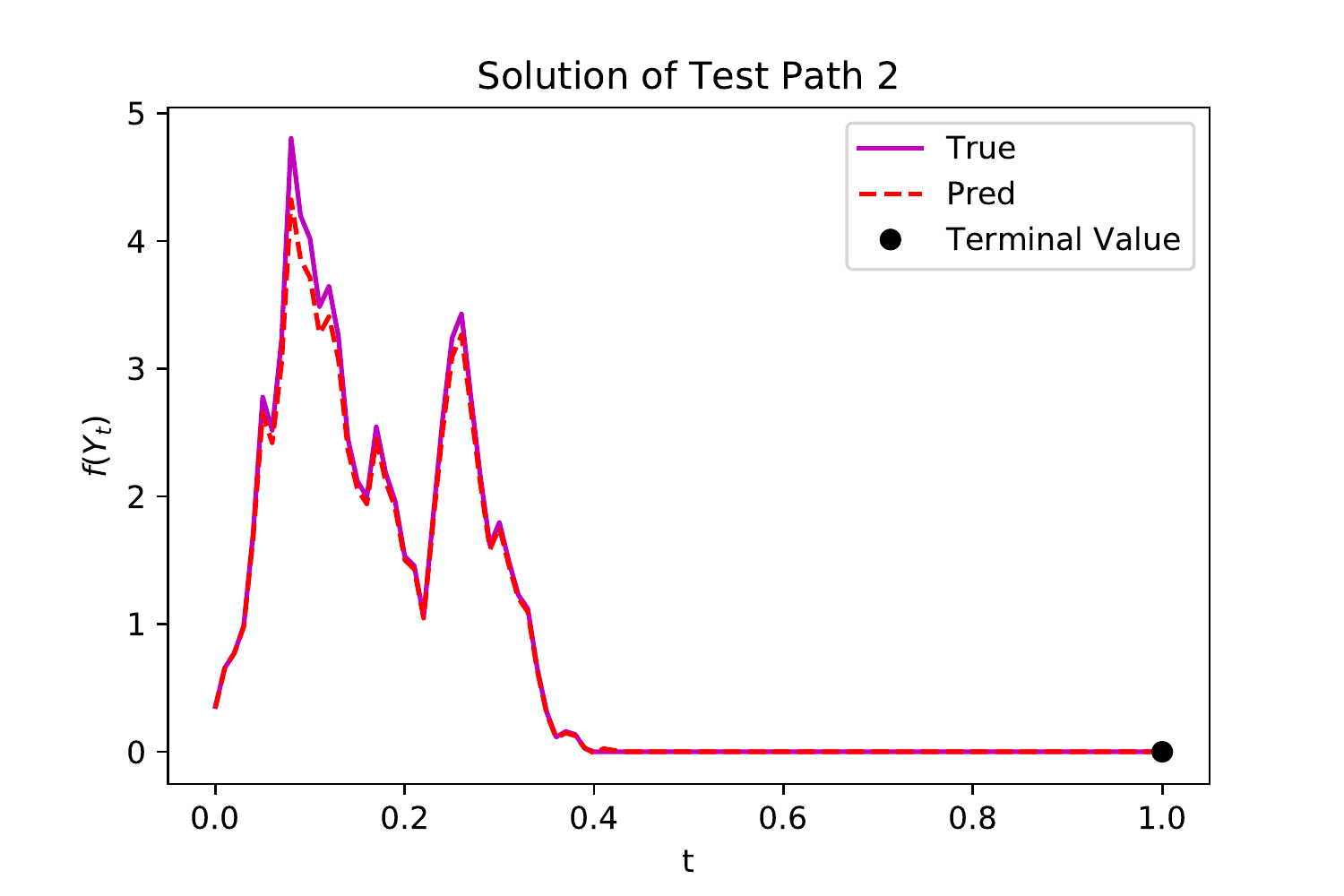}  \includegraphics[width = 0.45\textwidth, height = 5cm]{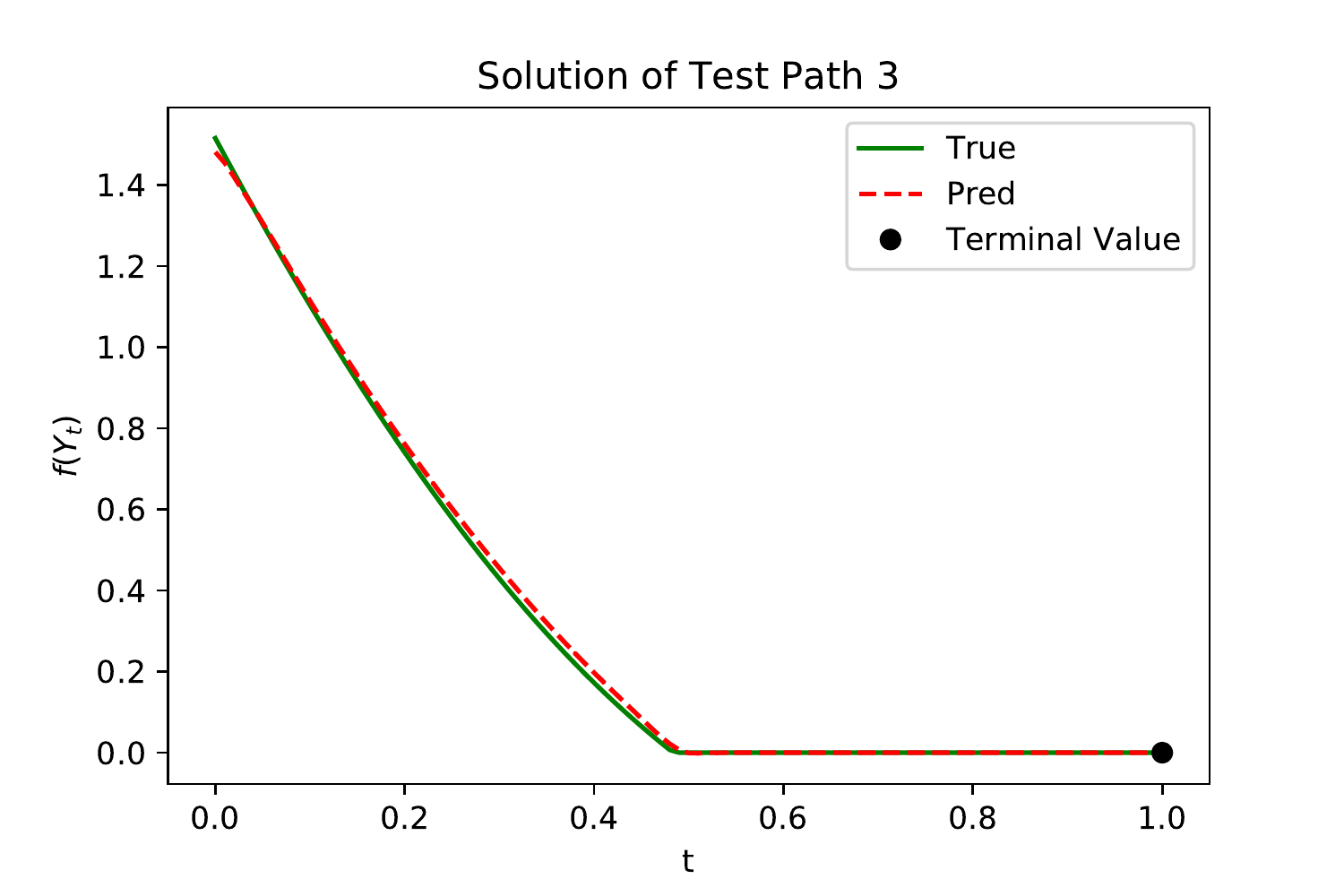}
  \caption{Three representative paths with corresponding solutions for the down-and-out call option.}
  \label{fig:barrier}
    \end{center}
\end{figure}

In the numerical implementation, we choose $B = 0.6$ and $K = 0.8$.
Figure \ref{fig:barrier} plots three representative paths and the corresponding solutions.  Ttest path 1 is a geometric Brownian motion with the parameters described above. Note this path does not cross the barrier. Test path 2 is another geometric Brownian motion but with $\sigma  = 2$. This path down crosses the barrier around $t = 0.4$. The third test path is a smooth path $y_t = 2.25(1-t)^2$. As a result, the predicted solutions and the true solutions are approximately the same.

\begin{remark}\label{rmk:barrier}
In our numerical experiment, we simulate paths $Y$ at time $t_i$ equally spaced by $\delta t$. When verifying whether the barrier was crossed, we have available these discretized values $y_{t_i}$. It would be possible to have $\inf_{i=0,\ldots,n-1} y_{t_i} > B$, but $\inf_{0\leq t \leq T} y_t \leq B$. This problem vanishes when $\delta t$ goes to 0. One approach would be then to consider a sufficiently small $\delta t$ in order to diminish this issue. This concern appears in the usual Monte Carlo methods to price barrier options and the methods used there might be adapted to assist us here, see \cite{barrier_MC}.
\end{remark}

\subsubsection{Exotic Option in Stochastic Volatility Models}\label{sec:heston}

A more complex model we could consider is the well-known Heston model:
$$\begin{cases}
dx_t = (r - q)x_tdt + \sqrt{v_t} x_t dw_t,\\
dv_t = \kappa(m - v_t)dt + \xi \sqrt{v_t} dw^*_t,\\
dw_t dw^*_t = \rho dt
\end{cases}
$$
The price at time $t$ of a general path-dependent option with maturity $T$ and payoff $g: \Lambda_T \longrightarrow \bR$ can be written as the functional $f(Y_t, v)$ and solves the PPDE
\begin{align}
\begin{cases}
\Delta_t f(Y_t, v) + (r - q) y_t \Delta_x f(Y_t, v) + \displaystyle \frac{1}{2}v y_t^2 \Delta_{xx} f(Y_t, v) - r f(Y_t,v) \\[10pt]
+\kappa(m - v) \partial_v f(Y_t, v) + \frac{1}{2} \xi^2 v \partial_{vv} f(Y_t, v) + \rho \xi v y_t \Delta_x \partial_v f(Y_t,v) = 0\\[10pt]
f(Y_T, v) = g(Y_T).
\end{cases}
\end{align}

\begin{figure}[htbp]
\begin{center}
  \includegraphics[width = 0.45\textwidth, height = 5cm]{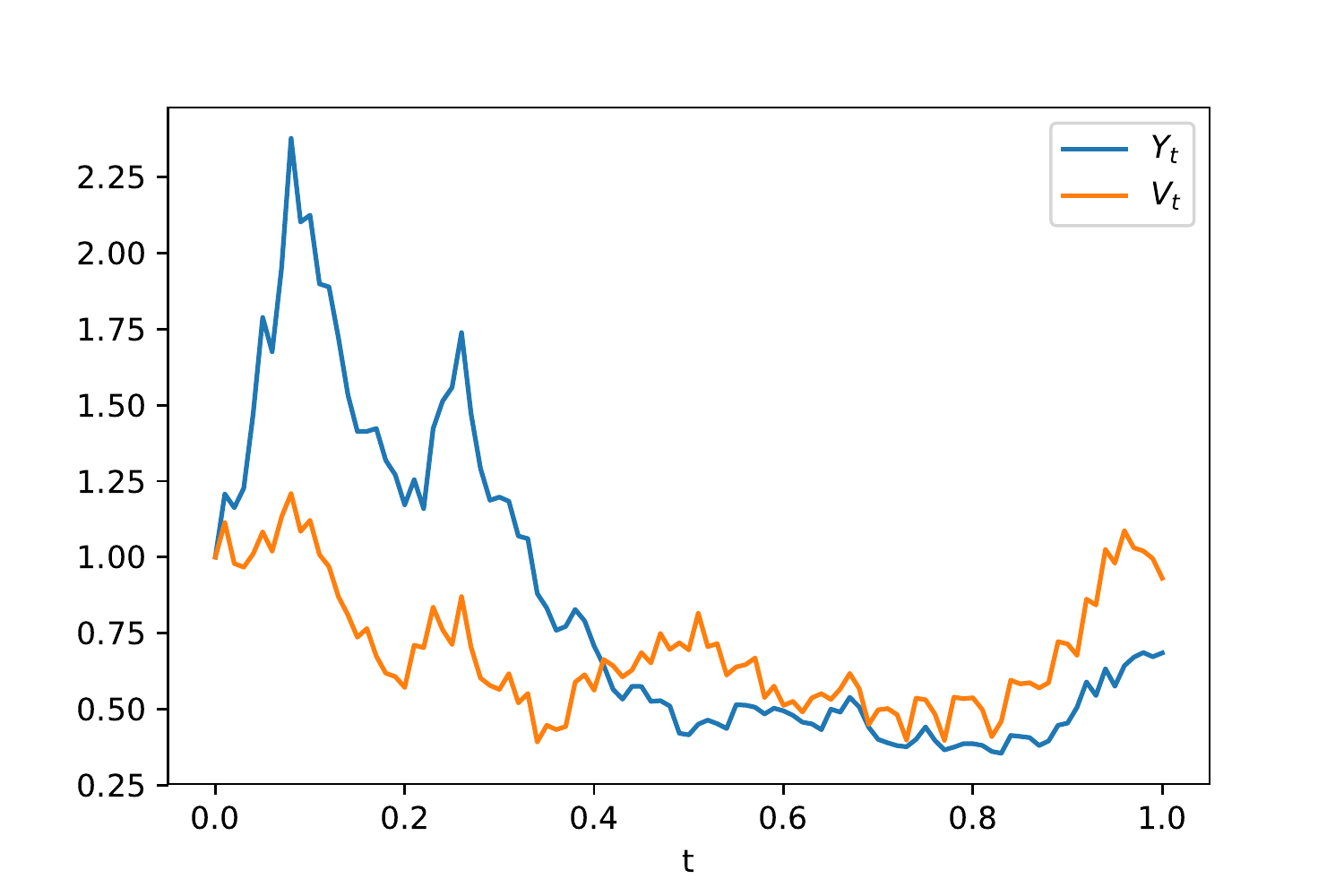}
  \includegraphics[width = 0.45\textwidth, height = 5cm]{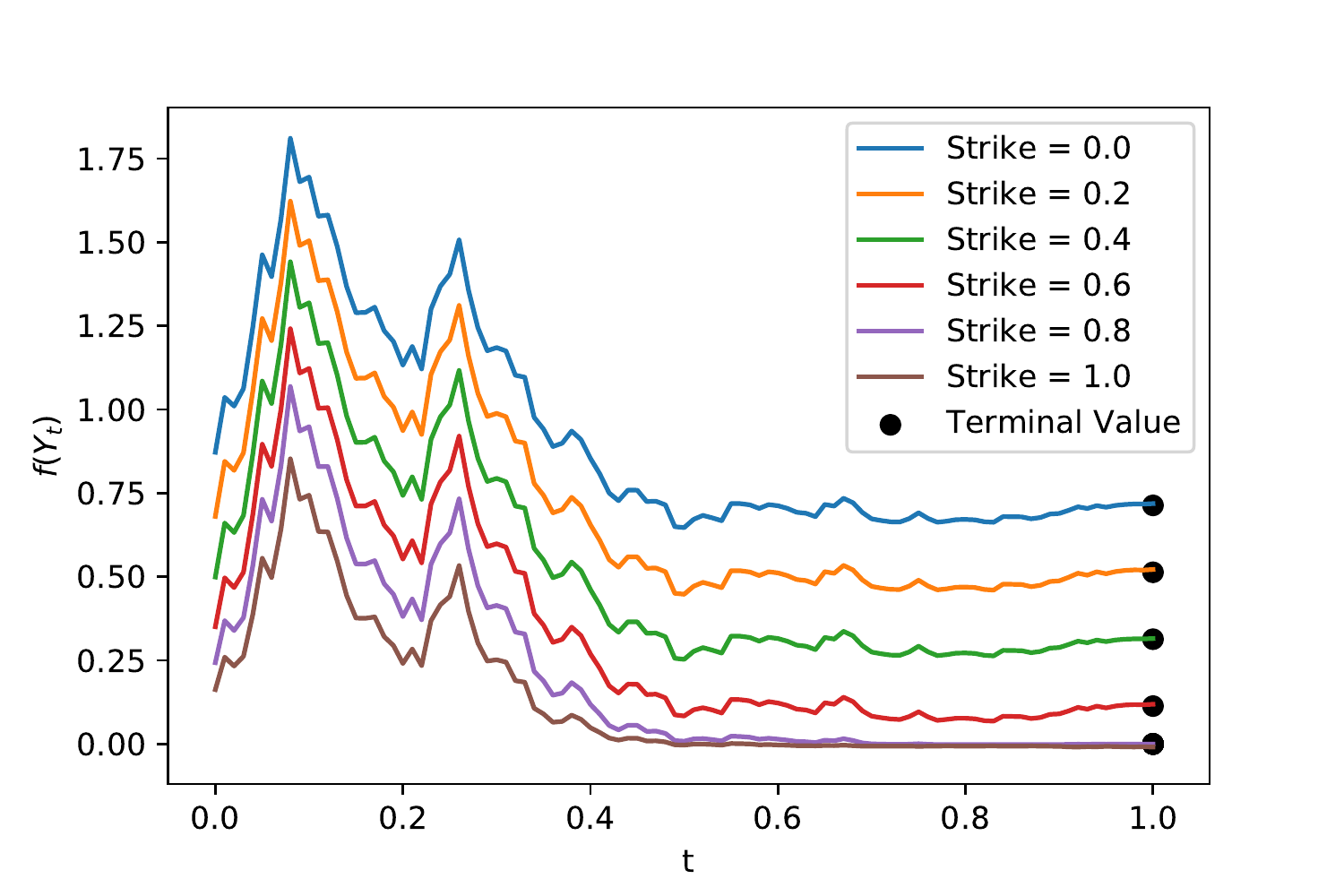}
  \caption{Solutions to the Heston model given a pair of paths of $(Y_t, V_t)$ (on the left) by varying the strike prices.}
  \label{fig:sto_vol}
    \end{center}
\end{figure}

\begin{figure}[htbp]
\begin{center}
  \includegraphics[width = 0.45\textwidth, height = 5cm]{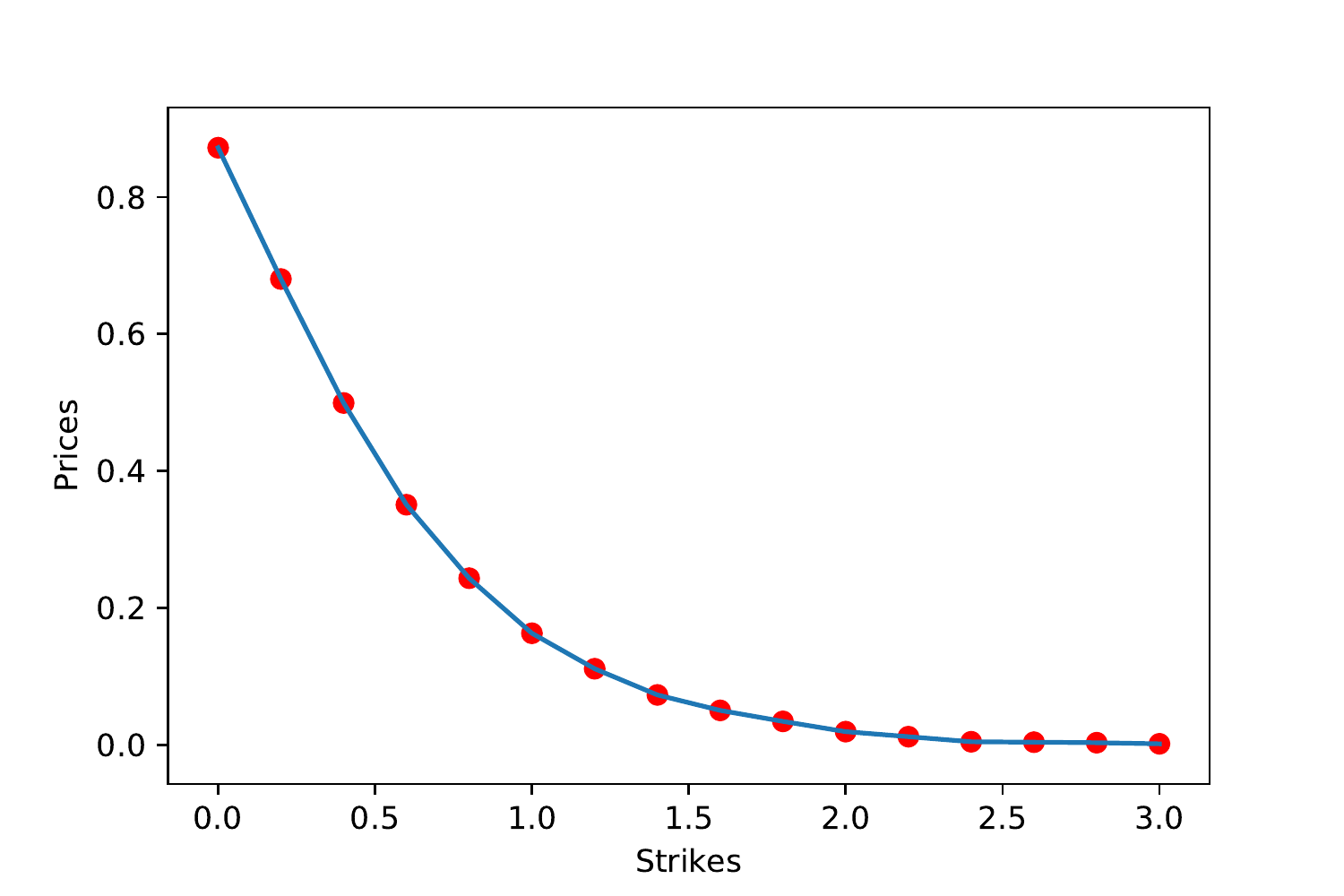}
    \includegraphics[width = 0.45\textwidth, height = 5cm]{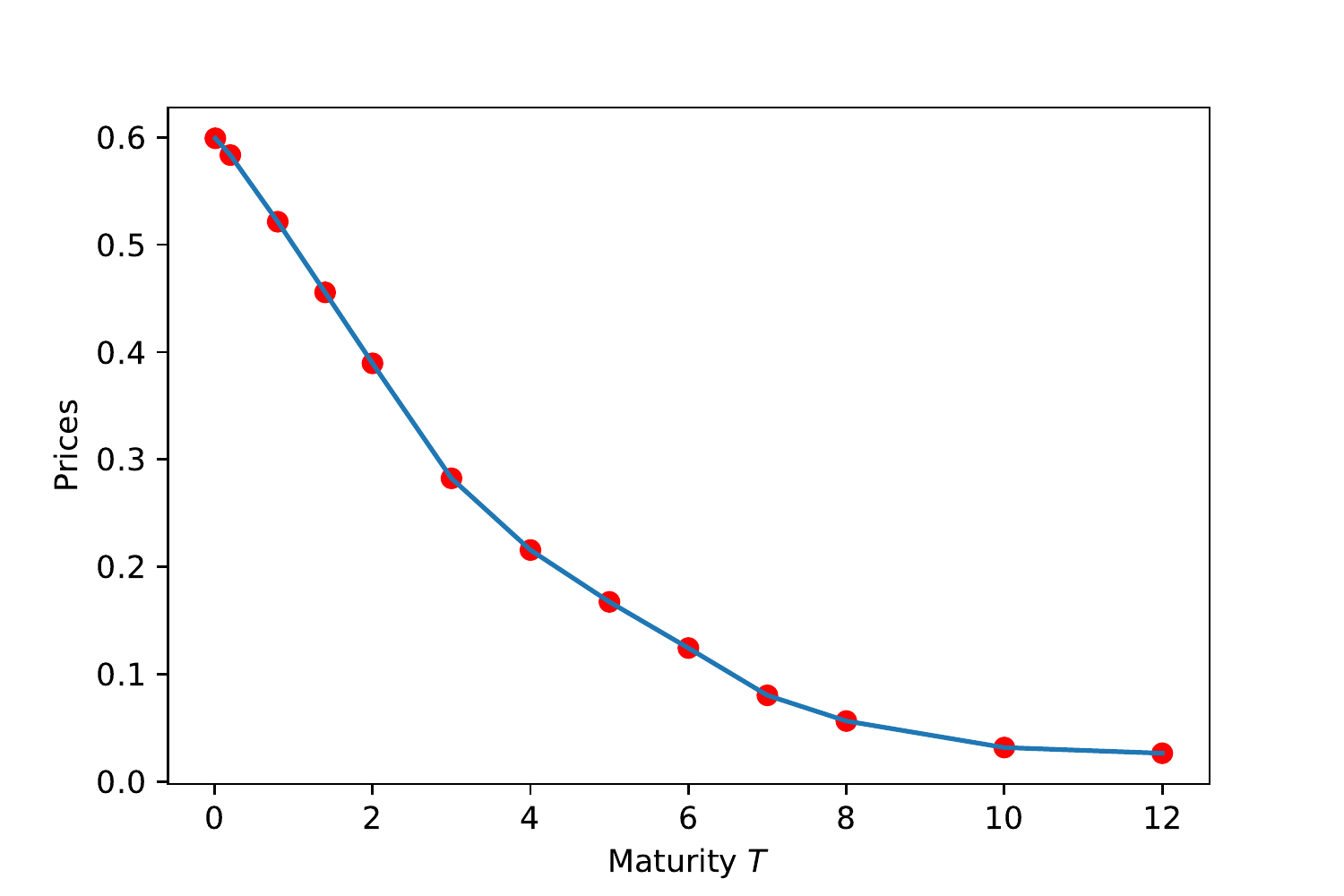}
  \caption{On the left: prices vs strike prices $K$. On the Right: prices vs maturity times $T$.}
  \label{fig:sto_vol_prices}
    \end{center}
\end{figure}

The generalization of our algorithm to this multidimensional case is straightforward. We will consider the geometric Asian option as in Section \ref{sec:geo_asian}. For the numerical implementation, we specify $r = 0.03, q = 0.01, \kappa = 3, m = 1, \xi = 1, \rho = 0.6, x_0 = v_0 = 1$. For a fixed maturity time $T = 1$, Figure  \ref{fig:sto_vol} plots a pair of stock prices path realization and volatility path realization on the left-hand side. Solutions predicted from our algorithm are shown on the right by with different strikes from 0 to 1. On the left of Figure \ref{fig:sto_vol_prices}, we plot the patterns of option prices versus strikes with fixed maturity $T = 1$,  and on the right we plot the option prices versus different maturities with fixed strike price $K = 0.4$.

\vspace{2cm}

\subsection{Non-linear PPDE}

This is the last numerical example. We consider a non-linear PPDE with closed-form solution. This example was studied in \cite{monotone_ppde}.

The closed-formula solution is given by $f(Y_t) = \cos(y_t + I_t)$, where $I_t = \int_0^t y_u du$ is the running integral, and the PPDE being considered is
$$\begin{cases}
\Delta_t f(Y_t) + (\underline{\mu} 1_{\{\Delta_x f(Y_t) > 0\}} + \overline{\mu} 1_{\{\Delta_x f(Y_t) < 0\}})\Delta_x f(Y_t) \\ \\
\ds + \frac{1}{2}(\underline{\sigma}^2 1_{\{\Delta_{xx} f(Y_t) < 0\}} + \overline{\sigma}^2 1_{\{\Delta_{xx} f(Y_t) > 0\}})\Delta_{xx}  f(Y_t) + \phi(Y_t) = 0,\\ \\
f(Y_T) = g(Y_T).
\end{cases}$$
We want to find the suitable source $\phi$ so $f$ satisfies the PPDE above. Notice that
$$\Delta_t f(Y_t) = -\sin (y_t + I_t) y_t$$
$$\Delta_x f(Y_t) = -\sin (y_t + I_t)$$
$$\Delta_{xx} f(Y_t) = -\cos (y_t + I_t)$$
Plugging them into the PPDE, it yields 
\begin{align*}
&-\sin (y_t + I_t) y_t + (\underline{\mu} 1_{\{\Delta_x f(Y_t) > 0\}} + \overline{\mu} 1_{\{\Delta_x f(Y_t) < 0\}}) [-\sin (y_t + I_t)]\\
&+ \frac{1}{2}(\underline{\sigma}^2 1_{\{\Delta_{xx} f(Y_t) < 0\}} + \overline{\sigma}^2 1_{\{\Delta_{xx} f(Y_t) > 0\}}) [-\cos (y_t + I_t)] + \phi(Y_t)= 0.
\end{align*}
Rearranging
\begin{align*}
    &-\sin (y_t + I_t)(y_t + \underline{\mu} 1_{\{\Delta_x f(Y_t) > 0\}} + \overline{\mu} 1_{\{\Delta_x f(Y_t) < 0\}})\\
    &- \frac{1}{2} \cos (y_t + I_t) (\underline{\sigma}^2 1_{\{\Delta_{xx} f(Y_t) < 0\}} + \overline{\sigma}^2 1_{\{\Delta_{xx} f(Y_t) > 0\}}) + \phi(Y_t) = 0.
\end{align*}
Then,
\begin{align*}
    \phi(Y_t) &= (y_t + \underline{\mu}) \min\left(\sin (y_t + I_t), 0 \right) + (y_t + \overline{\mu}) \max\left(\sin (y_t + I_t), 0\right)\\
    &+ \frac{\underline{\sigma}^2}{2}\max\left(\cos (y_t + I_t), 0\right) 
    + \frac{\overline{\sigma}^2}{2}\min\left(\cos (y_t + I_t), 0\right).
\end{align*}
Moreover
$$g(Y_T) = \cos(y_T + I_T).$$

\begin{figure}[htbp]
\begin{center}
  \includegraphics[width = 0.45\textwidth, height = 5cm]{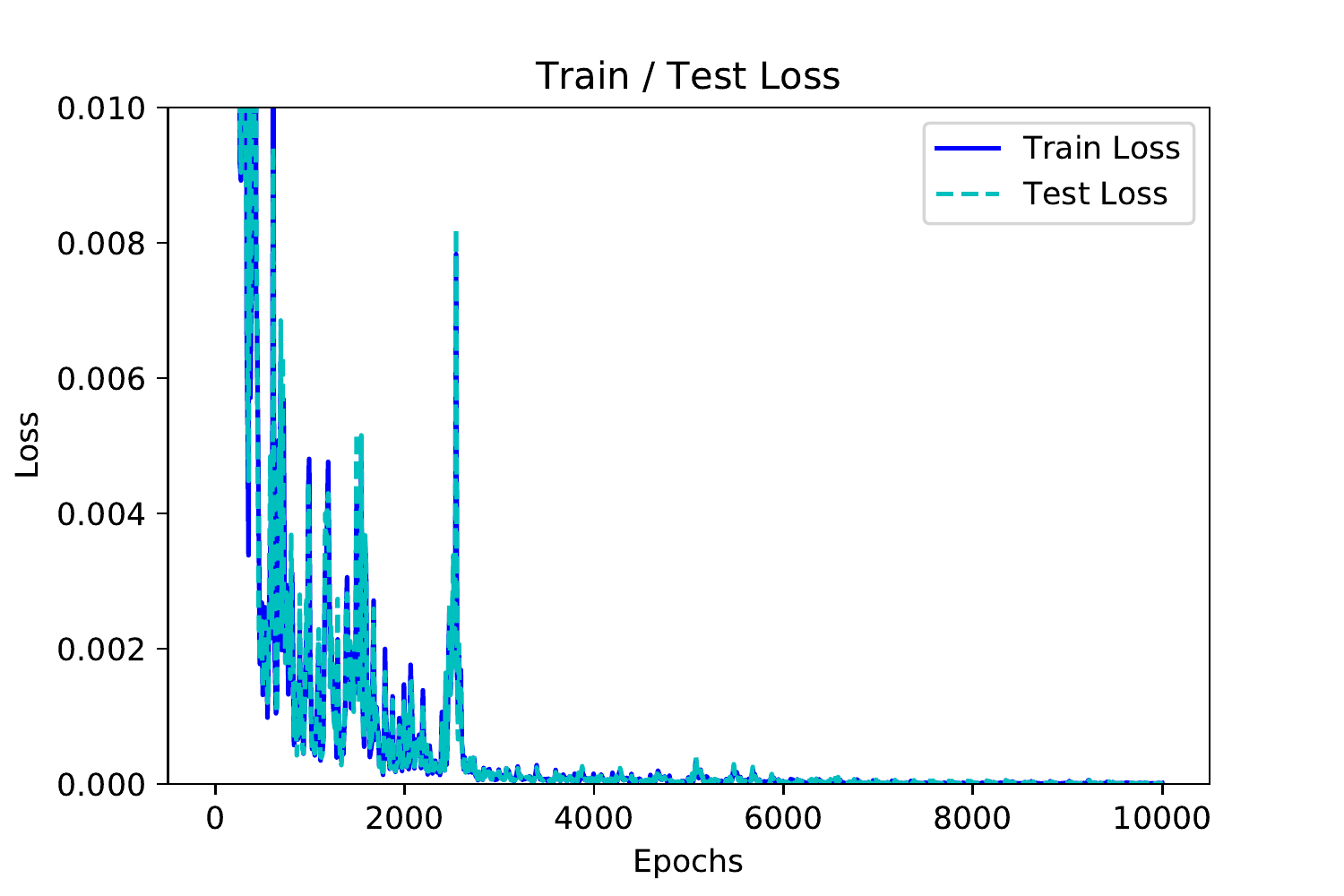}

  \caption{Train and test losses for the path-independent example.}
  \label{fig:nonlinear_loss}
    \end{center}
\end{figure}

\begin{figure}[htbp]
\begin{center}
  \includegraphics[width = 0.45\textwidth, height = 5cm]{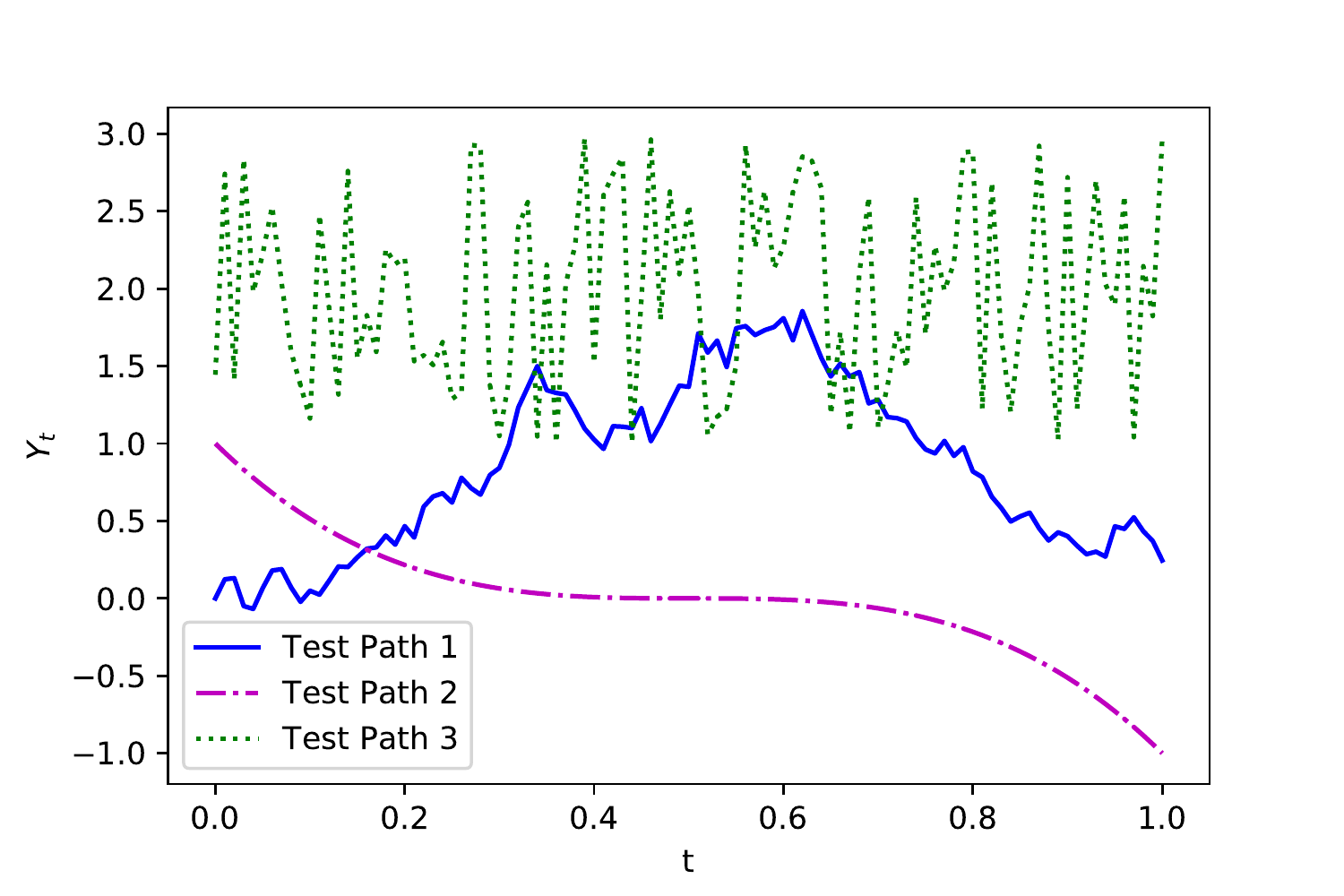}
  \includegraphics[width = 0.45\textwidth, height = 5cm]{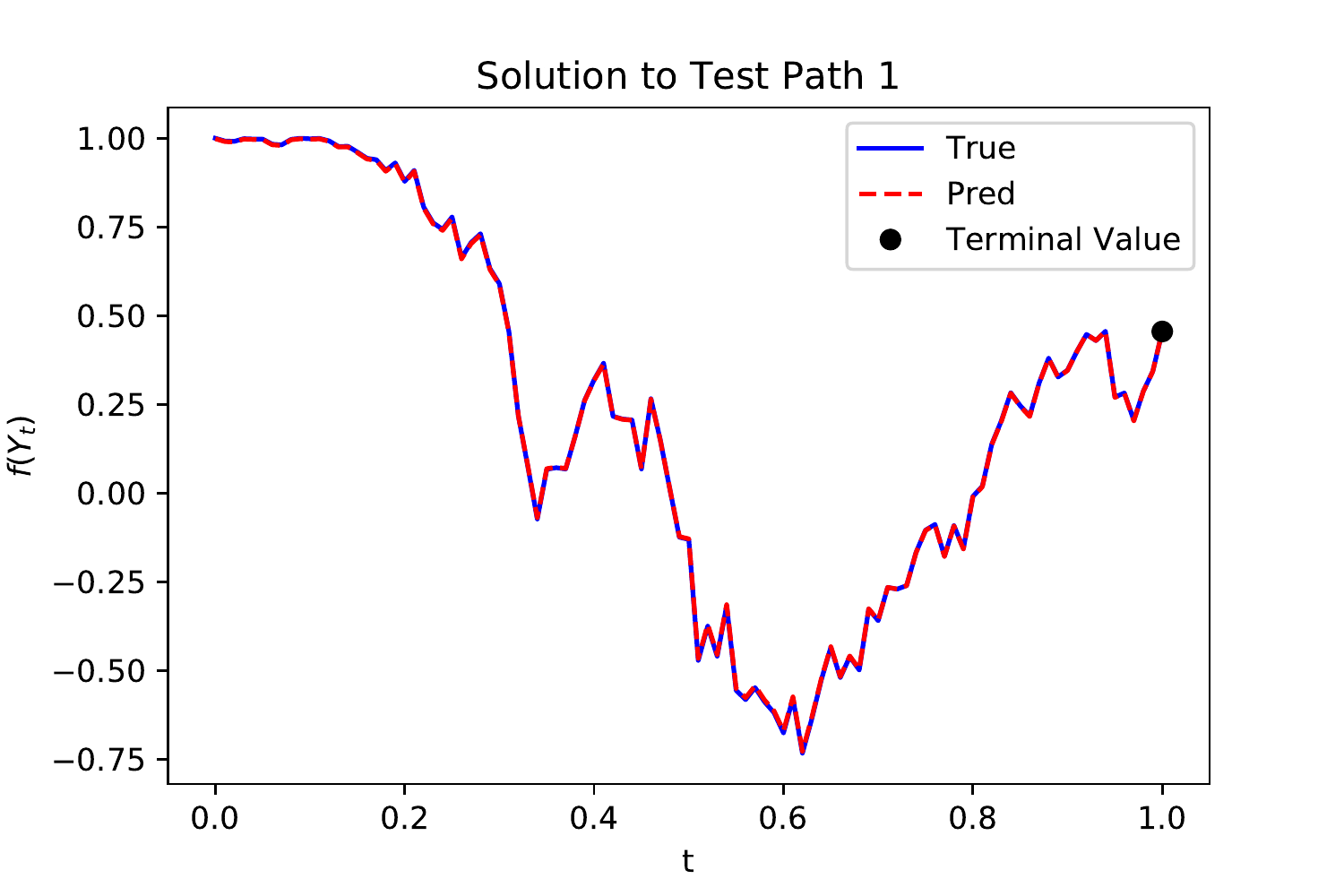}
  \includegraphics[width = 0.45\textwidth, height = 5cm]{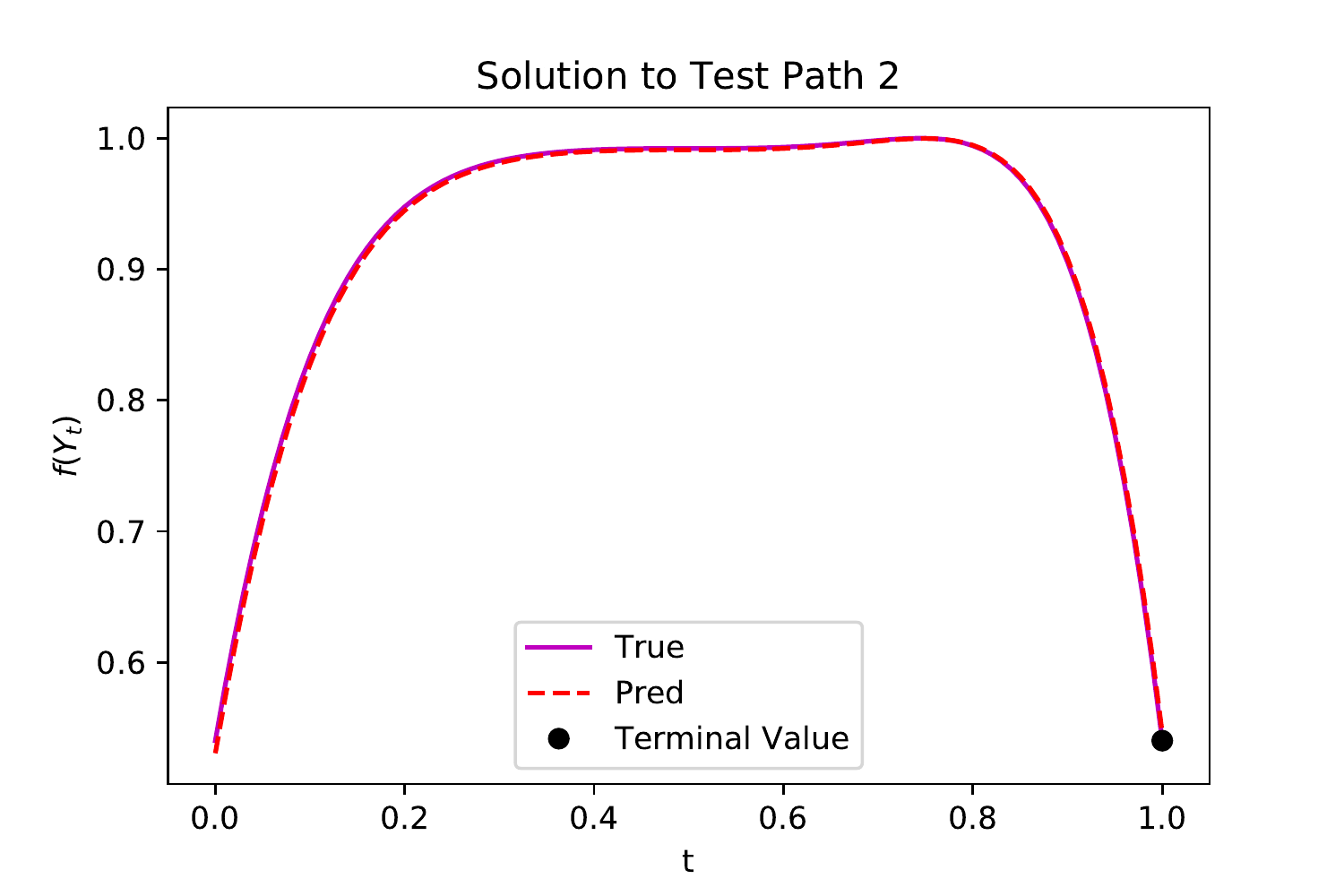}
  \includegraphics[width = 0.45\textwidth, height = 5cm]{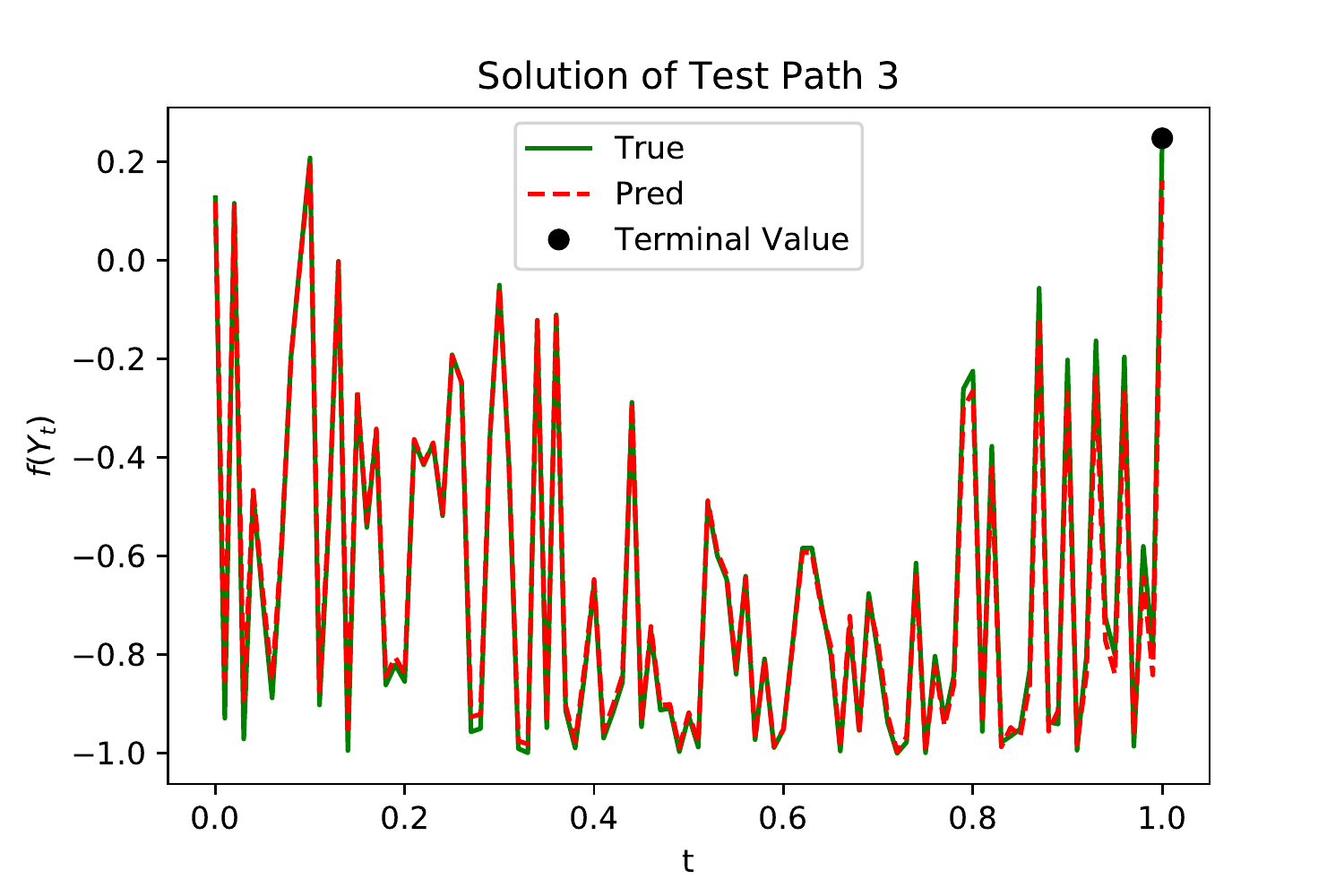}
  \caption{Three representative paths with corresponding solutions for the non-linear PPDE.}
  \label{fig:nonlinear}
    \end{center}
\end{figure}

The motivation for this problem is the following stochastic differential game:
$$u_0 = \inf_{\mu \in [\underline{\mu}, \overline{\mu}]} \sup_{\sigma \in [\underline{\sigma}, \overline{\sigma}]} \bE\left[g(X_T^{\mu, \sigma}) + \int_0^T \phi(X_t^{\mu, \sigma}) dt\right],$$
where
\begin{align*}
dx_t^{\mu, \sigma} = \mu_t dt + \sigma_t dW_t \mbox{ with } x_0^{\mu,\sigma} = 0.
\end{align*}
We use standard Brownian motion paths to train the neural network. We specify the coefficients to be $\underline{\mu} = -0.2, \overline{\mu} = 0.2, \underline{\sigma} = 0.2, \mbox{and } \overline{\sigma} = 0.3$. While keeping track of the signs of spatial derivatives, our algorithm works in the same way as in the other examples. Loss reaches around $5.9 \times 10^{-6}$ after 15000 epochs, and loss is plotted in Figure \ref{fig:nonlinear_loss}. Three representative paths with their corresponding solutions are presented in Figure \ref{fig:nonlinear}. Test path 1 is a realization of standard Brownian motion path. Test path 2 is a smooth path $y_t = (1-2t)^3$. Test path 3 is is $y_{t_i} \sim U(1,3), \ i \in \{1, \dots, 100\}$.  

\section{Conclusion and Future Work}

We have proposed a new method to solve PPDEs based on neural networks, called Path-Dependent Deep Galerking Method (PDGM). A novel network architecture was developed in order to deal with the objects from the functional It\^o calculus. There are very few methods available to solve these equations; for a discussion about them, see \cite{monotone_ppde} and references therein. We then showed the vast capabilities of the PDGM in various examples.

Future work could be divided between two main avenues. Firstly, one could study theoretical questions regarding the PDGM method as its consistency, speed of convergence and stability. Secondly, one could apply the method to more complex situations. As mentioned in the introduction, one could also extend PDGM to the different family of PPDEs originated from \cite{fito_zhang_fractional}.

Additionally, the notion of monotone numerical schemes was generalized to the PPDE setting in \cite{monotone_ppde}. As future research one could study if the proposed method here is monotonic as defined in aforesaid reference.

\bibliography{references}

\end{document}